\documentclass[11pt,a4paper]{article}
\usepackage{jheppub}
\usepackage{lipsum}
\usepackage{braket}
\usepackage{tensor}
\usepackage{dsfont}
\usepackage{caption}
\usepackage[dvipsnames*,svgnames]{xcolor}
\usepackage{subcaption}
\usepackage{tikz}
\usepackage{pgfplots}
\usetikzlibrary{datavisualization}
\usetikzlibrary{datavisualization.formats.functions}
\usetikzlibrary{decorations.pathreplacing,calligraphy}
\usetikzlibrary{decorations.pathreplacing}
\usetikzlibrary{decorations.pathmorphing, patterns,shapes}
\usepackage{bm}

\newmuskip\pFqmuskip
\newcommand*\pFq[6][8]{%
  \begingroup % only local assignments
  \pFqmuskip=#1mu\relax
  \mathchardef\normalcomma=\mathcode`,
  \mathcode`\,=\string"8000
  \begingroup\lccode`\~=`\,
  \lowercase{\endgroup\let~}\pFqcomma
  {}_{#2}F_{#3}{\left[\left.\genfrac..{0pt}{}{#4}{#5}\right| #6\right]}%
  \endgroup
}
\newcommand{\pFqcomma}{{\normalcomma}\mskip\pFqmuskip}
\newcommand{\EE}{\varepsilon_3}
\newcommand{\PP}{e^{i\mathcal{P}}}
\newcommand{\invPP}{e^{-i\mathcal{P}}}

\title{Long-range to the Rescue of Yang-Baxter}

\preprint{DESY-24-113, ZMP-HH/24-16}

\author[a]{Deniz N Bozkurt,}
\author[a]{Juan Miguel Nieto Garc\'ia,}
\author[b]{Elli Pomoni}

\affiliation[a]{II. Institut f\"ur Theoretische Physik, Universit\"at Hamburg, Luruper Chaussee 149,
22607 Hamburg, Germany}
\affiliation[b]{Deutsches Elektronen-Synchrotron DESY, Notkestr. 85, 22607 Hamburg, Germany}

\emailAdd{deniz.bozkurt@desy.de}
\emailAdd{juan.miguel.nieto.garcia@desy.de}
\emailAdd{elli.pomoni@desy.de}

\abstract{ 

\vspace*{0.5cm}

We study the spin chain model which captures the one-loop spectral problem of a prototypical example of an $\mathcal{N}=2$ SCFT in four dimensions.
Up to date, this spin chain model remains unfathomable; 
the coordinate Bethe Ansatz does not lead to a solution from three magnons on, as
the Yang-Baxter equation is not satisfied by the two-magnon scattering coefficients.
In this paper, we find a long-range solution to the eigenvalue problem 
for three magnons.
 Remarkably, the scattering coefficients of our solution together with the position-dependent corrections, 
 obey an infinite tower of Yang-Baxter equations.
Our method of solving the three-magnon problem is interesting in its own right and generalizes the coordinate Bethe Ansatz approach to cases where the permutation symmetry is broken.

}

\makeatletter
\gdef\@fpheader{}
\makeatother

\begin{document} 
\maketitle
\flushbottom

\section{Introduction}
\label{sec:intro}

Integrability in the context of the AdS/CFT correspondence has been an exciting and fertile field of research for more than twenty years \cite{Beisert:2010jr,Minahan:2002ve}.
A major open problem in this strand of investigation is the question of whether four-dimensional gauge theories in the planar limit
can still be integrable when supersymmetry is less than maximal \cite{Korchemsky:2010kj,Zoubos:2010kh,Pomoni:2019oib}.

For generic gauge theories, it has been believed that  integrability is lost because the  Yang-Baxter equation (YBE) is not satisfied \cite{Gadde:2010zi}.\footnote{Only for a
 sparse set of CFTs \cite{Beisert:2005he,Beisert:2005if,Gurdogan:2015csr}, and special subsectors \cite{Pomoni:2013poa} integrability is still believed to persist.}
 Even though this way of thinking  is naive \cite{Gadde:2010zi,Pomoni:2021pbj} 
and traces of a richer structure have been seen, we do not have enough evidence for a solid conclusion yet \cite{Pomoni:2021pbj,Hanno}.
The YBE is a well-known criterion for integrability. It is a necessary but not sufficient condition for the integrability of rational and trigonometric models.
For elliptic models, the YBE is only satisfied on a specific choice of basis \cite{Baxter:1982zz,Belavin:1981ix}.
For other choices of basis, a modified form, also known as the dynamical  Yang-Baxter equation (DYBE) is  the
criterion for integrability \cite{Felder:1994be,Felder:1996xym,konnobook}.\footnote{For the choice of basis where the YBE is satisfied \cite{Baxter:1982zz,Belavin:1981ix}, the Sklyanin algebra captures the symmetry of the problem \cite{Sklyanin:1982tf}. Due to its particular form, it does not allow for representations with
 highest weight states.
On the other hand, generically SCFTs have BPS operators which correspond to the highest weight states. Thus naturally the Hilbert space of SCFTs is not organised in the basis where the YBE is satisfied.}

The simplest context in which the  YBE arises is the study of the three-magnon coordinate Bethe ansatz (CBA), where it captures the fact that the scattering of three magnons factorizes into the product of two magnon scatterings.
For spin chains capturing the spectral problem of gauge theories with lower supersymmetry, a solution for the three-magnon problem is not known up to date, rendering the Yang-Baxter discussion premature. With this paper, we break this impasse by explicitly constructing such a solution.  

For concreteness, we consider what is arguably the simplest example of a half-maximal $\mathcal{N}=2$ superconformal field theory (SCFT) \cite{Douglas:1996sw,Gadde:2009dj}.
This theory is obtained from
the maximally supersymmetric theory in four dimensions $\mathcal{N}=4$ super Yang-Mills (SYM)
via orbifolding (with the simplest discrete subgroup $\mathbb Z_2\subset SU(2) \subset SU(4)_R$  R-symmetry) and then moving along its conformal manifold away from the special ``orbifold point''. The conformal manifold of this theory is two-dimensional, with coordinates the Yang-Mills marginal couplings $g_1,g_2$, and the ``orbifold point'' is the one-dimensional sub-manifold  $g_1=g_2$.

The spin chain model capturing the one-loop spectral problem was constructed in \cite{Gadde:2010zi,Liendo:2011xb}. In this paper, we consider a scalar subsector that descends from an $SU(2)$ of the mother $\mathcal{N}=4$ SYM, such that the  $SU(2)$ symmetry of the spin chain corresponds to a subgroup of the R-symmetry of $\mathcal{N}=4$ SYM. This symmetry is broken as we descend to the $\mathcal{N}=2$ SCFT via orbifolding. The study of this subsector is enough to demonstrate the phenomena we wish to study. The generalization to the full scalar subsector \cite{Gadde:2010zi} as well as the complete one-loop spin chain \cite{Liendo:2011xb} should be straightforward once this ``broken $SU(2)$ subsector'' is understood.

The Hamiltonian of the spin chain model we study was found to be nearest-neighbor, naturally capturing the fact that it arises from a one-loop computation. What is special about this  spin chain  is that the Hilbert space is not a tensor product  space %$\otimes_{\ell=1}^L V_\ell$ 
but a certain projection of the tensor product space. The projection arises from the fact that in the planar limit the allowed color contractions of the fields that define the single-trace state space %$V_\ell$
 restricts the total space,  as not all states in the tensor product are allowed. This Hilbert space is better described mathematically as a dynamical spin chain  \cite{Pomoni:2021pbj}.

A consequence of the dynamical nature of the three-magnon Hilbert space
is that the permutation symmetry of the wave function is broken. Hence, even though the spin chain Hamiltonian is nearest-neighbor, the three-magnon eigenvector has to be a long-range wave function.
If there was ordinary permutation symmetry, it would be possible to write a solution in terms of the usual Bethe Ansatz.
The absence of it implies that corrections are needed to solve the eigenvalue problem with additive energy eigenvalue. Precisely the long-range corrections to the usual Bethe Ansatz allow for the YBE to be reestablished.

\bigskip

This paper is organized as follows:

\begin{itemize}
    \item In Section \ref{sec:intro-model} we will begin by introducing the spin chain model under study, reviewing all the necessary details, for a self-contained paper.
    
     \item In Section \ref{sec:CBA} we review how the usual CBA allows for the one- and two-magnon problem solutions and explain how it fails to give a solution for three magnons precisely because the YBE is not satisfied.
We briefly discuss why the insertion of contact terms is not enough to obtain a solution (away from a special kinematic limit), postponing the details to the Appendix \ref{sec:contact-terms}.

  \item In Section \ref{sec:infinite} we introduce the long-range ansatz that we employ. We derive the equations that capture the eigenvalue problem with additive energies and finally show how to solve them. We discover that the general solution still contains an infinite number of coefficients that remain unfixed capturing the fact that we have not imposed any boundary condition to the infinitely long chain solution.

    \item In Section \ref{sec:special} we present three special solutions
    (with no remaining unfixed parameters)
    that capture the symmetries and the physics of the problem. Furthermore, we present the details of the resulting generating functions under these special choices in Appendix \ref{sec:G123}.
     
    \item In Section \ref{sec:Yang} we show that
the scattering coefficients of our solutions, labeled by two integers, obey an infinite tower
of YBEs.

\end{itemize}
Finally, in Section \ref{sec:conclusions}, we conclude and present some interesting future directions. 
 
\section{The Spin Chain Model}
\label{sec:intro-model}

\subsection{The Gauge Theory}

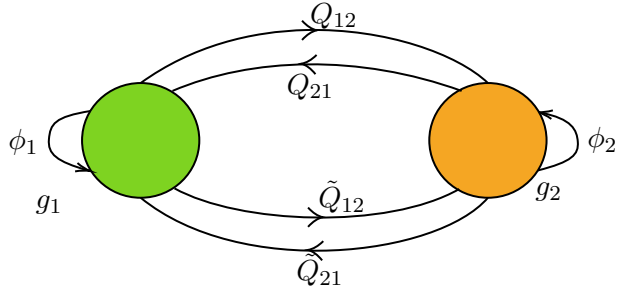
\begin{figure}[t]
\begin{center}

\tikzset{every picture/.style={line width=0.75pt}} %set default line width to 0.75pt        

\begin{tikzpicture}[x=0.5pt,y=0.5pt,yscale=-1,xscale=1]
%uncomment if require: \path (0,321); %set diagram left start at 0, and has height of 321

%Shape: Ellipse [id:dp7361143573255293] 
\draw  [fill={rgb, 255:red, 126; green, 211; blue, 33 }  ,fill opacity=1 ] (161.3,162.32) .. controls (161.3,138.42) and (180.95,119.04) .. (205.2,119.04) .. controls (229.44,119.04) and (249.1,138.42) .. (249.1,162.32) .. controls (249.1,186.22) and (229.44,205.6) .. (205.2,205.6) .. controls (180.95,205.6) and (161.3,186.22) .. (161.3,162.32) -- cycle ;
%Curve Lines [id:da3437910296394796] 
\draw    (167.9,140.1) .. controls (140.5,144.5) and (137.5,150) .. (136.5,161.5) .. controls (135.53,172.71) and (142.16,178.22) .. (165.66,184.98) ;
\draw [shift={(167.5,185.5)}, rotate = 195.64] [color={rgb, 255:red, 0; green, 0; blue, 0 }  ][line width=0.75]    (10.93,-3.29) .. controls (6.95,-1.4) and (3.31,-0.3) .. (0,0) .. controls (3.31,0.3) and (6.95,1.4) .. (10.93,3.29)   ;
%Curve Lines [id:da4636084656304217] 
\draw    (504.74,141.42) .. controls (528.23,146.02) and (531.84,154.42) .. (532.5,163.5) .. controls (533.18,172.86) and (528.5,178.5) .. (502.1,185.1) ;
\draw [shift={(502.5,141)}, rotate = 10.01] [color={rgb, 255:red, 0; green, 0; blue, 0 }  ][line width=0.75]    (10.93,-3.29) .. controls (6.95,-1.4) and (3.31,-0.3) .. (0,0) .. controls (3.31,0.3) and (6.95,1.4) .. (10.93,3.29)   ;
%Shape: Ellipse [id:dp48152619377057015] 
\draw  [fill={rgb, 255:red, 245; green, 166; blue, 35 }  ,fill opacity=1 ] (421.3,162.32) .. controls (421.3,138.42) and (440.95,119.04) .. (465.2,119.04) .. controls (489.44,119.04) and (509.1,138.42) .. (509.1,162.32) .. controls (509.1,186.22) and (489.44,205.6) .. (465.2,205.6) .. controls (440.95,205.6) and (421.3,186.22) .. (421.3,162.32) -- cycle ;
%Curve Lines [id:da21356486707384936] 
\draw    (229,125) .. controls (288.5,99.5) and (368.5,97) .. (445.5,125) ;
%Curve Lines [id:da899440931806935] 
\draw    (230.5,198.5) .. controls (279,225.5) and (389.5,230) .. (443.7,199.04) ;
%Curve Lines [id:da3186288459146429] 
\draw    (205.2,205.6) .. controls (280,267.05) and (427,250.05) .. (465.2,205.6) ;
%Curve Lines [id:da3353680149824555] 
\draw    (205.2,119.04) .. controls (243.5,91.15) and (287.88,79.96) .. (329.52,79.05) .. controls (388.77,77.75) and (442.47,97.26) .. (465.2,119.04) ;
\draw   (326.02,72) .. controls (329.68,76.04) and (333.34,78.46) .. (337,79.27) .. controls (333.34,80.08) and (329.68,82.51) .. (326.02,86.55) ;
\draw   (336.06,112.36) .. controls (332.65,108.11) and (329.14,105.48) .. (325.53,104.45) .. controls (329.23,103.86) and (333.03,101.65) .. (336.92,97.84) ;
\draw   (330.52,213) .. controls (334.18,217.04) and (337.84,219.46) .. (341.5,220.27) .. controls (337.84,221.08) and (334.18,223.51) .. (330.52,227.55) ;
\draw   (341.06,252.86) .. controls (337.65,248.61) and (334.14,245.98) .. (330.53,244.95) .. controls (334.23,244.36) and (338.03,242.15) .. (341.92,238.34) ;

% Text Node
\draw (103.9,151.5) node [anchor=north west][inner sep=0.75pt]    {$\phi _{1}$};
% Text Node
\draw (124.4,202) node [anchor=north west][inner sep=0.75pt]    {$g_{1}$};
% Text Node
\draw (537.5,150.2) node [anchor=north west][inner sep=0.75pt]    {$\phi _{2}$};
% Text Node
\draw (498.6,192.1) node [anchor=north west][inner sep=0.75pt]    {$g_{2}$};
% Text Node
\draw (329.9,56.4) node [anchor=north west][inner sep=0.75pt]    {$Q_{12}$};
% Text Node
\draw (312.4,110) node [anchor=north west][inner sep=0.75pt]    {$Q_{21}$};
% Text Node
\draw (336,190.5) node [anchor=north west][inner sep=0.75pt]    {$\tilde{Q}_{12}$};
% Text Node
\draw (319.5,245.6) node [anchor=north west][inner sep=0.75pt]    {$\tilde{Q}_{21}$};

\end{tikzpicture}

\end{center}
\caption{\textit{The $\mathbb Z_2$ orbifold theory with $SU(N)_1\times SU(N)_2$}.}
\label{fig:guiver}
\end{figure}

In this paper we study the spin chain of the one-parameter family of $\mathcal{N}=2$ SCFTs arising from marginally deforming the $\mathbb Z_2$ orbifold of $\mathcal{N}=4$ SYM  away from the ``orbifold point'' \cite{Gadde:2009dj,Gadde:2010zi}.
The $\mathcal{N}=2$ daughter SCFT has  $SU(N)_1\times SU(N)_2$ gauge group and  $SU(2)_R\times U(1)_r$ R-symmetry.
The field content is captured by the quiver diagram depicted in Figure \ref{fig:guiver} and includes two bifundamental hypermultiplets as well as two adjoint $\mathcal{N}=2$ vector multiplets. 
Following the notation of \cite{Pomoni:2021pbj},
these fields descend from the three complex scalars of $\mathcal{N}=4$ in the following manner,
\begin{align}
 X\rightarrow\begin{pmatrix}
        &Q_{12}\\Q_{21}&
    \end{pmatrix}\;,\quad\quad\quad Y\rightarrow\begin{pmatrix}
        &\tilde Q_{12}\\\tilde Q_{21}&
    \end{pmatrix}\;,\quad\quad\quad Z\rightarrow    \begin{pmatrix}
        \phi_1&\\&\phi_2
    \end{pmatrix}\;.\label{eq:fields}
\end{align}
The scalars $\phi_1$ and $\phi_2$ transform in the adjoint representation of one of the $SU(N)_{i=1,2}$ gauge groups, respectively. The two hypermultiples, include the bifundamental scalars $Q_{12},\bar{\tilde Q}_{12}$ and $Q_{21},\bar{\tilde Q}_{21}$, where the first index refers to the gauge node for which the field is in the fundamental representation, and the second index refers to the gauge node for which the scalar field is in the anti-fundamental representation. 
Marginally deforming away from the orbifold point with the parameter $\kappa=g_2/g_1$ gives the theory with two distinct coupling constants,
\begin{align}
    \mathcal{W}_{\mathbb Z_2}=ig_1 \text{tr}_2\left(\tilde Q_{21}\phi_1Q_{12}-Q_{21}\phi_1\tilde Q_{12}\right)-ig_2 \text{tr}_1\left( Q_{12}\phi_2\tilde Q_{21}-\tilde Q_{12}\phi_2Q_{21}\right)\;,
    \label{eq:Z2-superpotential}
\end{align}
with the parameter $\kappa\in \left[0,1\right]$. In addition to the $SU(2)_R\times U(1)_r$ R-symmetry, this theory has an extra $SU(2)_L$ global symmetry. Also, the theory interpolates between the orbifold point theory for $\kappa=1$ and the $\mathcal{N}=2$ superconformal QCD (SCQCD) with the $U(N_f)=U(2N)$ flavor symmetry for $\kappa=0$ \cite{Gadde:2009dj}. 

\subsection{The Spin Chain}
\label{subsec:SpinChain}

To construct the Hilbert space of the spin chain, we consider the single-trace operators built from only two of  the three scalar fields, $\phi_i$ and $Q_{ij}$ (not $\tilde{Q}_{ij}$) which can be understood as forming two doublets ($i=1,2$) of the broken $SU(2) \subset SU(4)_R$ \cite{Hanno}.
Using the $\mathcal{N}=4$ SYM language from which it descends, this sector can be refereed to as the $XZ$ sector.
The fact that these fields are in different representations of the gauge group restricts the allowed single-trace operators in the planar limit.  Single-trace operators can only include an even number of bifundamental fields and these two bifundamental fields, $Q_{12}$ and $Q_{21}$, should always appear in an alternating fashion with or without $\phi$s in between. Examples of the possible configurations include,
\begin{align}
    \text{tr}_1\left(\phi_1^L\right)\;,\quad\quad\text{tr}\left(\phi_1^{L-A}Q_{12}\phi_2^AQ_{21}\right)\;,\quad\quad\text{tr}\left(\phi_1^{L-A-B-C}Q_{12}\phi_2^AQ_{21}\phi_1^{B}Q_{12}\phi_2^CQ_{21}\right)\;.\label{eq:ex-single-trace}
\end{align}
%When we transfer these single-trace operators to the spin chain picture, to have a unified description of the model for any number of excitations, 
As usual, to study the spectral problem of this spin chain
we first consider infinitely long open spin chains.
The two (degenerate) vacua constructed by the adjoint scalars of the theory are\footnote{
$\mathcal{N}=2$ SCFTs have two types of BPS operators which could be used as vacua around which we study the $n$-magnon problem. These are operators either with $\Delta=\pm r$ with $r$ being the $U(1)_r$ symmetry charge or $\Delta= 2R$ with $R$ being the $SU(2)_R$ symmetry Cartan eigenvalue.
In this paper, we consider the vacua with $\Delta=\pm r$ which we also refer to as $\phi$-vacua. In \cite{Pomoni:2021pbj}
the study of the $Q$-vacuum was pursued.}
 \begin{align}
\ket{\text{vac}}_{i}=\ket{\cdots\phi_i\phi_i\phi_i\cdots}\;,\quad\quad\text{for}\quad i=1,2\;.\label{eq:vacuum}
 \end{align}
 
 On these vacua, we have two types of magnon excitations, given by $Q_{12}$ and $Q_{21}$. Only the states with an even number of magnons admit periodic or twisted periodic boundary conditions, and can be identified with single-trace operators, for example,
 \begin{align}
\ket{{\phi_1}^{a_1}_{a_2} {\phi_1}^{a_2} _{a_3} \cdots {\phi_1}^{a_{L-1}}_{a_1}} &\longleftrightarrow \text{tr}\left(
{\phi_1}^L\right)\;,\label{eq:gaugeop1} \\
\ket{{\phi_1}^{a_1}_{a_2} {Q_{12}^{a_2}}_{\Bar{a}_1} {\phi_2}^{\Bar{a}_1}_{\Bar{a}_2} \cdots{\phi_2}^{\Bar{a}_{L-1}}_{\Bar{a}_L}}& \longleftrightarrow \text{(not closable)} \;,\label{eq:gaugeop2} \\
\ket{{\phi_1}^{a_1}_{a_2} {Q_{12}^{a_2}}_{\Bar{a}_1} {\phi_2}^{\Bar{a}_1}_{\Bar{a}_2} \cdots{Q_{21}}^{\Bar{a}_{L-1}}_{a_1}} &\longleftrightarrow \text{tr}\left(\phi_1Q_{12}\phi_2^{L-4}Q_{21}\right)\;.\label{eq:gaugeop3} 
\end{align}
Therefore, the states with an odd number of excitations cannot be closed and by considering infinitely long open spin chains states we avoid the discussion on the boundary conditions.

 Although the states with an odd number of magnons do not belong to the physical Hilbert space that corresponds to single-trace operators of the gauge theory, they are crucial for the process of diagonalization of the Hamiltonian. In this paper, we study infinitely long open spin chains and leave the discussion of the physical spectrum of the gauge theory to the upcoming work \cite{fourmagnon}.

The Hamiltonian of the ``broken $SU(2)$ sector'' is nearest-neighbor type and can be taken from \cite{Gadde:2010zi},
\begin{align}
        \mathcal{H}=\sum_l \mathcal{H}_{l,l+1}\;,\label{eq:total-hamiltonian}
\end{align}
where
    {\scriptsize{
			\begin{eqnarray}
				\label{Hamiltonian}
				& \mathcal{H}_{\ell,\ell+1}=
				& \bordermatrix{
					&\phi_1 \phi_1& \phi_1Q_{12} & Q_{12}\phi_2 &Q_{12}Q_{21} & \phi_2\phi_2 & \phi_2Q_{21} &Q_{21}\phi_1 &Q_{21}Q_{12} \cr
					&&&& \cr
					\phi_1 \phi_1& 0 & 0 & 0 & 0 & 0 & 0 & 0 & 0
					\cr
					\phi_1Q_{12} & 0  & 1/\kappa & -1 & 0 & 0 & 0 & 0 & 0\cr
					Q_{12}\phi_2 & 0 & -1 &\kappa &0 & 0 & 0 & 0 & 0\cr
					Q_{12}Q_{21} & 0 & 0 & 0  &0 & 0 & 0 & 0 & 0\cr
					\phi_2\phi_2 & 0 & 0 & 0 & 0 & 0 &0& 0 & 0\cr
					\phi_2Q_{21} & 0 & 0 & 0 & 0 & 0 & \kappa & -1 & 0\cr
					Q_{21}\phi_1 & 0 & 0 & 0 & 0 & 0 & -1 & 1/\kappa & 0\cr
					Q_{21}Q_{12}  & 0 & 0 & 0 & 0 & 0 & 0 & 0 & 0} \;.
	\end{eqnarray}}}

 The fact that we do not have a tensor product Hilbert space is evident from the nearest-neighbor Hamiltonian, \eqref{Hamiltonian}, which acts on an eight-dimensional vector space instead of a sixteen-dimensional one. Moreover, by organizing the two-site basis vectors into two sub-sets that are related to each other under the exchange of color indices $i,j$ of the fundamental fields, $\phi_i$ and $Q_{ij}$, we can represent the Hamiltonian in a block diagonal form. Under the exchange of color indices, together with inverting $\kappa$ to $1/\kappa$, we can map these two blocks of the Hamiltonian to each other. 
This $\mathbb Z_2$ symmetry of the Hamiltonian corresponds to the $\mathbb Z_2$ symmetry of the quiver diagram in Figure \ref{fig:guiver}. More explicitly, the transformation,
\begin{align}
    \mathbb Z_2: \begin{pmatrix}
        \phi_1\\ Q_{12}
    \end{pmatrix}\leftrightarrow\begin{pmatrix}
        \phi_2\\ Q_{21}
    \end{pmatrix}\quad\text{together with}\quad \kappa\leftrightarrow 1/\kappa\;,\label{eq:Z2map}
\end{align}
 leaves the Hamiltonian invariant. This will play a crucial role when we start constructing eigenvectors.

Additionally, each one of the blocks we see inside the Hamiltonian resembles a Temperly-Lieb Hamiltonian with quantum parameter $\kappa$ or $\frac{1}{\kappa}$. However, since the basis of the Hilbert space is highly non-trivial, this model is not equivalent to two copies of the Temperly-Lieb model. The constraints coming from the color contraction rules, together with the $\mathbb Z_2$ symmetry, make this model a dynamical spin chain model \cite{Pomoni:2021pbj}. Nevertheless, taking $\kappa\to1$ reduces the Hamiltonian given in \eqref{Hamiltonian} to two copies of the Heisenberg XXX model,\footnote{We can also obtain the Heisenberg XXX model for open infinite spin chains by taking the quantum parameter of a Temperly-Lieb model to 1.} which is integrable as discovered by \cite{Beisert:2005he}. This limit corresponds to the orbifold point limit from the gauge theory side. We expect our solution to recover the eigenvalues and eigenvectors of these two XXX models as we take the $\kappa\to1$ limit. We will show that this is indeed the case.

 In the following section, we present the Coordinate Bethe Ansatz (CBA) method and various ways to generalize the CBA in the cases that it fails to capture the dynamical nature of the spin chains.

\section{Coordinate Bethe Ansatz}
\label{sec:CBA}

\subsection{One-Magnon Problem}

We consider one excited state in a sea of $\phi$s. Depending on the color indices we start with, we obtain two distinct one-magnon states. The CBA for one-magnon is given by the state,
\begin{align}
    \ket{\Psi(p)}_{12}=\sum_{l=-\infty}^\infty e^{ip l}\ket{\dots \phi_1 \phi_1 Q_{12}(l) \phi_2 \phi_2 \dots }\;,\label{eq:one-mag12}\\
    \ket{\Psi(p)}_{21}=\sum_{l=-\infty}^\infty e^{ip l}\ket{\dots \phi_2 \phi_2 Q_{21}(l) \phi_1 \phi_1 \dots }\;,\label{eq:one-mag21}
\end{align}
where the field $Q_{12}$ sits in position $l$. Here, we represent the position vector by the coordinate of the magnon. We dress every position state with a plane wave factor and sum over all possible positions that the excited state can occupy to obtain the CBA for the one-magnon state. To distinguish between the two types of magnons, we label the momentum space states with the color indices of the magnon.

We derive the dispersion relation of the model by solving the eigenvalue equation,
\begin{align}
    \mathcal{H}\ket{\Psi(p)}_{ij}=E(p)\ket{\Psi(p)}_{ij}\;,\quad\text{for}\quad ij\in\{12,21\}\;.
\end{align}
The dispersion relation is identical for both states,
\begin{align}
    E_1(p)=\left(\sqrt{\kappa}-\frac{1}{\sqrt{\kappa}}\right)^2+4\sin^2\left(\frac{p}{2}\right)\label{eq:disprel}\;.
\end{align}
The agreement between the dispersion relations of two states is a consequence of a $\mathbb Z_2$ symmetry of the Hamiltonian \eqref{eq:Z2map}. This operator maps distinct one-magnon states to each other,
\begin{align}
    \mathbb Z_2\ket{\Psi(p)}_{ij}=\ket{\Psi(p)}_{ji}\;,\quad\quad\text{for}\quad\quad ij\in\{12,21\}\;.\label{eq:Z2action}
\end{align}
The action of the $\mathbb Z_2$ operator generalizes to any number of excitations. We will make extensive use of this symmetry in this paper.

It is also worth mentioning that the dispersion relation \eqref{eq:disprel} coincides with the dispersion relation of a Heisenberg XXZ spin chain with the anisotropy parameter, $\Delta=\left(\kappa+1/\kappa\right)/2$. Indeed, when we take $\kappa\to1$, the mass gap closes and we obtain the dispersion relation for the Heisenberg XXX model. At the orbifolding point, we obtain two copies of the Heisenberg XXX model, which is integrable for infinite open spin chains.

Another important transformation in the context of spin chains is parity, $\mathcal{P}$, which reverses the ordering of the states of our chain (supplemented in our case with the transformation $Q_{12} \leftrightarrow Q_{21}$ for consistency in the color indices)
   \begin{align}
       \mathcal{P}\ket{\Psi(p)}_{12}&=\sum_{l=-\infty}^\infty e^{ip l}\mathcal{P}\ket{\cdots\phi_1Q_{12}(l)\phi_2\cdots}\nonumber\\
       &=\sum_{l=-\infty}^\infty e^{ip l}\ket{\cdots\phi_2Q_{21}(-l) \phi_1\cdots}\nonumber\\
       &=\ket{\Psi(-p)}_{21}\;.\label{eq:parity-onemagnon}
   \end{align}
At first glance, the parity of the model seems broken due to the distinction between the magnons, $Q_{12}$ and $Q_{21}$.\footnote{Although it still commutes with Hamiltonian, parity maps the distinct one-magnon states to each other instead of just reversing the sign of the momentum parameter. Therefore, it is not broken as a symmetry of the spin chain model.}  However, by employing the $\mathbb Z_2$ symmetry of the Hamiltonian, parity is restored,
   \begin{align}
       \mathbb Z_2 \mathcal{P}\ket{\Psi(p)}_{ij}=\ket{\Psi(-p)}_{ij}\;,\label{eq:restoredparity}
   \end{align}
   
   This is a property of our spin chains which we will use to characterize the three-magnon solution better. We first consider the two-magnon eigenvalue problem.

\subsection{Two-Magnon Problem}
\label{subsec:two}

We proceed by solving the eigenvalue problem for two-magnon spin chain states,
\begin{align}
    \mathcal{H}\ket{\Psi(p_1,p_2)}=E_2(p_1,p_2)\ket{\Psi(p_1,p_2)}\;.\label{eq:twoeig}
\end{align}
The eigenvalue of the two-magnon states is given by summing over the one-magnon energy eigenvalue, \eqref{eq:disprel} for the independent momentum variables,
\begin{align}
E_2(p_1,p_2)=E_1(p_1)+E_1(p_2)\;.\label{eq:twoeigenvalue}
\end{align}
The CBA for two magnons is given by summing over all possible configurations of the magnons with a fixed ordering. The form of the eigenvalues, \eqref{eq:twoeigenvalue} can be derived by the asymptotic version of the eigenvalue problem.\footnote{By asymptotic we mean to project the two-magnon eigenvalue problem to position space by contracting with a state, $\braket{l_1,l_2|\mathcal{H}-E_2|\Psi(p_1,p_2)}$ such that the magnons are far apart, $l_2>l_1+1$. This expression only vanishes with the particular CBA, \eqref{eq:two-magnonCBA} when the eigenvalue is the sum of the dispersion relations of two momentum variables.} Depending on the order of the magnons on the spin chain we can have two distinct eigenvectors,
\begin{align}
    \cdots\phi_1Q_{12}\phi_2\cdots\phi_2Q_{21}\phi_1\cdots\;,\quad\quad\quad \cdots\phi_2Q_{21}\phi_1\cdots\phi_1Q_{12}\phi_2\cdots\;.\nonumber
\end{align}
As with the one-magnon eigenvectors, these two configurations are labeled by the color indices of the first magnon on the spin chain. We introduce a wave function for each state in position space, incorporating the two possible ways the magnons can carry the momentum, $p_1$ and $p_2$. Initially, the CBA for two excited states takes the form
\begin{align}
    \ket{\Psi(p_1,p_2)}_{ij}=\sum_{l_1<l_2}\psi_{ij}(p_1,p_2;l_1,l_2)\ket{Q_{ij}(l_1),Q_{ji}(l_2)}\label{eq:two-magnonCBA}\;,
\end{align}
such that the wave function under a particular choice of normalization is expressed as
\begin{align}
    \psi_{ij}(p_1,p_2;l_1,l_2)=(e^{il_1p_1+il_2p_2}+S_{ij}(p_1,p_2)e^{il_1p_2+il_2p_1})\;.\label{eq:two-wave}
\end{align}
for $ij\in\{12,21\}$. Here, we inserted a phase between these two plane wave contributions, which is the \emph{scattering coefficient}. The contribution of the scattering coefficient will ensure that the eigenvalue equation, \eqref{eq:twoeig} holds for any configuration on the position space, including the \emph{interacting magnons} given as $\ket{Q_{12}(l),Q_{21}(l+1)}$ . Therefore, we solve the eigenvalue problem \eqref{eq:twoeig} to obtain the scattering coefficients. It turns out that the scattering coefficients are described in terms of
\begin{align}
    S_\kappa(p_1,p_2)=-\frac{1+e^{ip_1+ip_2}-2\kappa e^{ip_2}}{1+e^{ip_1+ip_2}-2\kappa e^{ip_1}} \; , \label{eq:scatdef}
\end{align}
such that the scattering coefficients of the two eigenvectors are related to each other by $\kappa\to1/\kappa$ transformation,
\begin{align}
	S_{12}=S_\kappa(p_1,p_2)\;,\label{eq:scat12}
\end{align}
whereas
\begin{align}
	S_{21}=S_{1/\kappa}(p_1,p_2)\;.\label{eq:scat21}
\end{align}
This is a consequence of the $\mathbb Z_2$ symmetry. Notably, the distinction between this model and two copies of the Heisenberg XXZ spin chain becomes apparent only at the two-magnon level. We observe that the anisotropy parameter $\Delta=(\kappa+1/\kappa)/2$ splits into two components and these two pieces parameterize the scattering coefficients. This splitting highlights the unique features of the model and underscores the role of $\mathbb Z_2$ symmetry in differentiating it from the Heisenberg XXZ spin chain. Additionally, taking the orbifold limit would make the distinct scattering coefficients \eqref{eq:scat12} and \eqref{eq:scat21}, coincide and we would obtain the result for two copies of the Heisenberg XXX chain as anticipated. Other than $\kappa=\pm1$, the scattering coefficients also coincide under the kinematic limit, $p_1+p_2=\pi$,
\begin{align}
    S_\kappa(p_1,\pi-p_1)=S_{1/\kappa}(p_1,\pi-p_1)\;,\label{eq:pi-relation}
\end{align}
for arbitrary $p_1$ and $\kappa$.
This turns out to be crucial for the three-magnon CBA with contract terms, which is presented in Appendix \ref{sec:contact-terms}.

Furthermore, the parity operator acts by reversing the sign of the momentum variables and does not need to be combined by the $\mathbb Z_2$ operator. This phenomenon is deeply connected with the fact that two magnon states, as well as all other states with an even number of excitations, allow periodic boundary conditions. Since the states at the far left and far right of the spin chain agree, we observe,
\begin{align}
    \mathcal{P}\ket{\Psi(p_1,p_2)}_{ij}=S_{ij}(p_1,p_2)\ket{\Psi(-p_1,-p_2)}_{ij}\;.\label{eq:parity-twobdy}
\end{align}

The periodic boundary conditions imply the following Bethe equations, which were analyzed in \cite{Gadde:2010zi},
\begin{align}
    e^{ip_iL}=S_{ji}(p_i,p_j)\quad\quad\text{for}\quad\quad i,j=1,2\;\; \text{and}\;\; i\neq j\;. \label{eq:betheeq2}
\end{align}

As a preparation for the next section, we take a detour to consider deforming the two-body CBA. We insert \emph{contact terms} to the wave function which only plays a role when the magnons are interacting,  \cite{Pomoni:2021pbj,Bibikov2016},
\begin{align}
    \ket{\Phi(p_1,p_2)}_{12}=&\sum_{l_1<l_2}\Big(\left(A_{12}+D_{12}\;\delta(l_2,l_1+1)\right)e^{ip_1l_1+ip_2l_2}\nonumber\\&\quad\quad\quad+\left(A_{21}+D_{21}\;\delta(l_2,l_1+1)\right)e^{ip_2l_1+ip_1l_2}\Big)\ket{l_1,l_2}_{12}\label{eq:twobody-cont}\;.
\end{align}
As scattering coefficients, contact terms are functions of momentum variables. The eigenvalue problem imposes the following relations among scattering coefficients and contact terms,
\begin{align}
    \frac{A_{21}}{A_{12}}=S_\kappa(p_1,p_2)\;,\quad\quad\quad D_{21}=-e^{ip_2-ip_1}D_{12}\;.\label{eq: two-contact-rel}
\end{align}

Since the contact terms only become non-trivial for $l_2=l_1+1$, we observe that the relation given above implies that the contribution from the contact terms vanishes. Although we can generalize the idea of the contact terms by replacing $D_{ij}\delta(l_2,l_1+1)$ with an infinite sum $\sum_nD^n_{ij}\delta(l_2,l_1+n)$, the solution of the eigenvalue problem would still wash out the contact term contributions. Therefore, we state that the two-magnon eigenvector cannot be generalized by the contribution of the contact terms. We will revisit the idea of inserting contact terms for the three-magnon problem.

\subsection{How CBA Fails for the Three-Magnon Problem} \label{subsec:threemagnon}
The eigenvalue problem for three magnons is a natural extension of the two-magnon case. We consider the eigenvalue problem,
    \begin{align}
       \mathcal{H}\ket{\Psi(p_1,p_2,p_3)}=E_3(p_1,p_2,p_3)\ket{\Psi(p_1,p_2,p_3)}\label{eq:eigthree}\;,
    \end{align}
    such that the eigenvalue of the three-magnon state is the sum of the energy eigenvalues for each momentum variable,
    \begin{align}
        E_3(p_1,p_2,p_3)=E_1(p_1)+E_1(p_2)+E_1(p_3)\;.\label{eq:threeeigv}
    \end{align}
    The natural extension of the CBA to states with three excitations should naively have the following form, 
	\begin{align}
		\ket{\varphi(p_1,p_2,p_3)}_{12}&=\sum_{l_1<l_2<l_3}\varphi_{12}(p_1,p_2,p_3;l_1,l_2,l_3)\ket{l_1,l_2,l_3}_{12}\label{eq:naiveansatz}\;,\\
        \varphi_{12}(p_1,p_2,p_3;l_1,l_2,l_3)&=\left(\sum_{\sigma\in\mathcal{S}_3} A_\sigma\;e^{ip_{\sigma(1)}l_1+ip_{\sigma(2)}l_2+ip_{\sigma(3)}l_3}\right)\;.\label{eq:naivewave}
	\end{align}
    Here, the wave function includes a sum over all elements of the symmetric group, $\mathcal{S}_3$. The group elements can be represented as the possible permutations of the three momentum parameters. As in the case of one and two magnons, the momentum space vector is labeled by the indices of the excitation at $l_1$ and the rest of the magnons are fully characterized by this piece of information. Additionally, there exists another state, $\ket{\varphi(p_1,p_2,p_3)}_{21}$, corresponding to the image of the state given in \eqref{eq:naiveansatz} under the action of $Z_2$. 

    By using the CBA given in \eqref{eq:naiveansatz} we attempt to solve the three-magnon eigenvalue problem. We observe that for well-separated magnons the eigenvalue problem is immediately satisfied, 
	\begin{align}
		\braket{l_1,l_2,l_3|\mathcal{H}-E_3|\varphi(p_1,p_2,p_3)}=0\;,\quad\text{for all}\quad l_2-l_1>1, \;l_3-l_2>1\label{eq:naive-non-int}\;.
	\end{align}
    When two or more magnons are next to each other, we obtain a finite set of linear equations for the scattering coefficients $A_\sigma$. Here we consider the configurations that have only two interacting magnons and the third one is far away. The equations associated with the configurations with three interacting magnons are a linear combination of the former. Depending on whether the pair of interacting magnons is located at the left or right of the third one,
   \begin{align}
        &\cdots \phi_1Q_{12}Q_{21}\phi_1\cdots\phi_1Q_{12}\phi_2\cdots \; , \qquad \qquad 
        &\cdots \phi_1Q_{12}\phi_2\cdots\phi_2Q_{21}Q_{12}\phi_2\cdots\nonumber \; ,
    \end{align}
    the equations we obtain are different,
	\begin{align}
		&\braket{l_1,l_1+1,l_3|\mathcal{H}-E_3|\varphi(p_1,p_2,p_3)}_{12}=e^{il_1(p_1+p_2+p_3)}\nonumber\\&\quad\quad\quad\times\sum_{ij\in\{12,13,23\}}\left(A_{jik}a(p_i,p_j,\kappa)+A_{ijk}a(p_j,a_i,\kappa)\right)e^{i(l_3-l_1-1)p_k}\;,\label{eq:twobodyl-naive}\\
  &\braket{l_1,l_3-1,l_3|\mathcal{H}-E_3|\varphi(p_1,p_2,p_3)}_{12}=e^{il_3(p_1+p_2+p_3)}\nonumber\\&\quad\quad\times\sum_{ij\in\{12,13,23\}}\left(A_{kji}a(p_i,p_j,\kappa^{-1})+A_{kij}a(p_j,a_i,\kappa^{-1})\right)e^{i(l_1-l_3+1)(p_i+p_j)}\;,\label{eq:twobodyr-naive}
	\end{align}
  The function
  \begin{align}
      a(p_1,p_2,\kappa)=1+e^{ip_1+ip_2}-2\kappa e^{ip_1}\;, \label{eq:scatnum}
  \end{align}
  is the denominator of the scattering coefficient given in \eqref{eq:scatdef}.
    We assume that the scattering coefficients are independent of the coordinates of the magnons, meaning that each of the three terms of the sums in the previous equations have to separately vanish. We write these six equations in the matrix form,
    \begin{align}
	\resizebox{0.9\textwidth}{!}{$\begin{pmatrix}
		a(p_3,p_2;1/\kappa)&a(p_2,p_3;1/\kappa)&0&0&0&0\\
        a(p_2,p_1;\kappa)&0&a(p_1,p_2;\kappa)&0&0&0\\
        0&a(p_3,p_1;1/\kappa)&0&a(p_1,p_3;1/\kappa)&0&0\\
        0&0&a(p_3,p_1;\kappa)&0&a(p_1,p_3;\kappa)&0\\
        0&0&0&a(p_3,p_2;\kappa)&0&a(p_2,p_3;\kappa)\\
		0&0&0&0&a(p_2,p_1;1/\kappa)&a(p_1,p_2;1/\kappa)
	\end{pmatrix}\begin{pmatrix}
	A_{123}\\A_{132}\\A_{213}\\A_{231}\\A_{312}\\A_{321}
	\end{pmatrix}=0$\;.}\label{eq:homlineq}
\end{align}
    We can check that the determinant of the matrix of coefficients does not vanish, implying that only the trivial solution exists. Therefore, the CBA given in \eqref{eq:naiveansatz} fails to provide a solution for the eigenvectors of the three-body eigenvalue problem, \eqref{eq:eigthree}.

    We can understand the failure of the CBA from a different perspective. On one hand, solving the equations \eqref{eq:twobodyl-naive} implies that the scattering coefficients have to fulfill the relations
    \begin{align}
       \frac{A_{213}}{A_{123}}=S_\kappa(p_1,p_2)\;,\quad\quad\frac{A_{312}}{A_{132}}=S_\kappa(p_1,p_3)\;,\quad\quad\frac{A_{321}}{A_{231}}=S_\kappa(p_2,p_3)\;.\label{eq:naive-sol-left}
    \end{align}
    On the other hand, we obtain the following relations between the scattering coefficients when we only solve \eqref{eq:twobodyr-naive},
    \begin{align}
       \quad\quad\quad\frac{A_{321}}{A_{312}}=S_{1/\kappa}(p_1,p_2)\;,\quad\quad\frac{A_{231}}{A_{213}}=S_{1/\kappa}(p_1,p_3)\;,\quad\quad \frac{A_{132}}{A_{123}}=S_{1/\kappa}(p_2,p_3)\;.\label{eq:naive-sol-right}
    \end{align}
    When we combine these results, the failure of the CBA can be mapped to the fact that the scattering coefficients of the CBA do not fulfill the YBE
	\begin{equation}
		S_\kappa(p_2,p_3) S_{1/\kappa}(p_1,p_3)S_\kappa(p_1,p_2) \neq S_{1/\kappa}(p_1,p_2)S_\kappa(p_1,p_3) S_{1/\kappa}(p_2,p_3) \;. \label{eq:neq}
	\end{equation}
    This observation was initially made in \cite{Gadde:2010zi}.  It becomes evident that this issue recurs for all higher numbers of magnons due to the permutation symmetric, factorizable nature of the CBA. It is clear that as long as we use a permutationally symmetric form for the eigenvector ansatz, the YBE will arise. The goal of this paper is to present a systematic way to upgrade the factorization of the scattering coefficients which does not satisfy YBE so that they satisfy a form of DYBE. In this case, the mismatch between the parameters $\kappa$ and $1/\kappa$ would be cured by the shifts of a dynamical parameter \cite{Felder:1994be,Felder:1996xym}.

\subsubsection{Contact Terms Do Not Solve the Three-Magnon Problem}

    We first attempt to find a solution for the three magnon problem by adding contact terms, thus effectively allowing the scattering coefficients of the model to have two distinct forms. 
    When magnons are interacting on the position space, the contact terms  change the asymptotic form of the scattering coefficients to its interacting form. Therefore, we deform the two-magnon interacting equations of the three-magnon CBA and try to lift the inconsistency presented in the previous subsection. After solving all the equations coming from the three-magnon eigenvalue problem with contact terms, we observe that the contributions of the contact terms still effectively vanish due to relations coming from the non-interacting equations. We already observe this phenomenon for the two-magnon problem, as discussed at the end of Section \ref{subsec:two}.

    However, a non-trivial solution for the three-magnon contact term ansatz emerges upon imposing momentum constraints. Under these constraints, certain conditions similar to \eqref{eq: two-contact-rel} that typically nullify the overall contribution of contact terms, are relaxed. This yields a \emph{partial solution} to the three-magnon problem for a certain \emph{kinematic limit}. This is not entirely satisfactory, as it only grants access to a limited portion of the spin chain model's spectrum. We present the details of this partial solution of the contact term ansatz for three magnons in Appendix \ref{sec:contact-terms}. If we consider inserting more contact terms for finitely many configurations of the three magnons in the position space, we run into the same problem. To overcome this, we introduce position-dependent corrections for every configuration in the position space, ultimately leading to a fully satisfactory solution. It becomes evident that achieving a comprehensive solution to the three-magnon problem for the long-range ansatz requires the incorporation of infinitely many corrections to capture the entire spectrum of the model.

\section{Three-Body Long-Range Solution}
\label{sec:infinite}
\begin{figure}[h]
       \centering
       \includegraphics[scale=0.25]{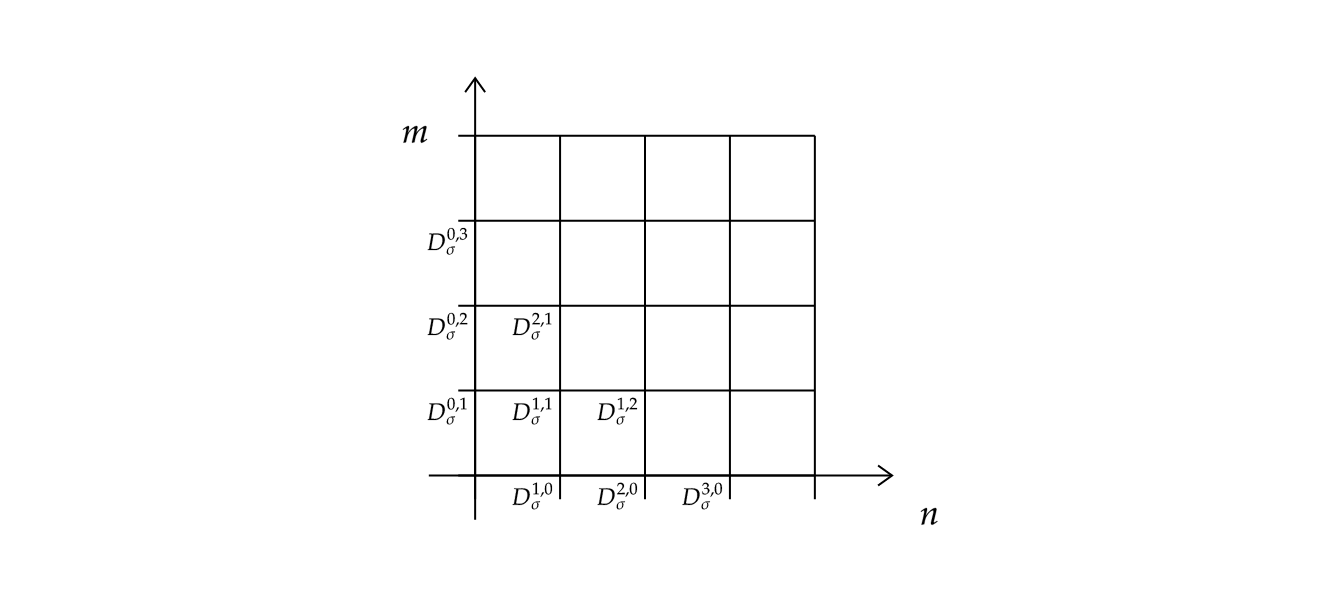}
       \caption{\it The position-dependent corrections $D^{n,m}_\sigma$ are labeled by two integers, the distances between the three magnons, thus naturally live on a two-dimensional lattice. They are also labeled by  the permutations $\sigma$.
       }
       \label{fig:lattice}
   \end{figure}
Generalizing the CBA with contact terms, we introduce an ansatz with an infinite number of correction terms,
\begin{align}
       \ket{\Psi(p_1,p_2,p_3)}_{12} = \sum_{l_1<l_2<l_3} \Psi(p_1,p_2,p_3;l_1,l_2,l_3) \ket{l_1,l_2,l_3}_{12} \; ,\label{eq:infcontansatz}
         \end{align}
         with
    \begin{align}
\Psi(p_1,p_2,p_3;l_1,l_2,l_3)=\left(\sum_{\sigma\in S_3} \left(A_\sigma+D^{l_2-l_1-1,l_3-l_2-1}_\sigma\right)e^{i\Vec{p}_{\sigma}\cdot\Vec{l}}\right) \; .\label{eq:infocontwave}
   \end{align}
   We label all the scattering coefficients and the position-dependent corrections by permutations, $\sigma$, according to the order of the momentum variables that appear in the plane wave contributions. In this ansatz, we require the position-dependent corrections to be functions of the momenta, the parameter $\kappa$, and the relative position of the excitations. Therefore,  our ansatz includes infinite undetermined coefficients for each permutation which can be enumerated on a two-dimensional lattice for each permutation, together with six scattering coefficients $A_\sigma$. Since we consider infinite open spin chains without specifying boundary conditions, the corresponding two-dimensional lattice is $\mathbb Z^{\geq0}\times \mathbb Z^{\geq0}$, which is depicted in Figure \ref{fig:lattice}. Notice also that the scattering coefficients can always be absorbed into the definition of the position-dependent corrections. We have instead decided to separate them and set the coefficients $D^{0,0}_\sigma=0$ to make the connection with the case of two magnons more clear. Additionally, the translational invariance of the wave function is equivalent to vanishing total momenta as it happens for the usual CBA, since the position-dependent corrections only depend on the relative distance.

    \subsection{Equations of Three-Magnon Problem}

   When we substitute the ansatz \eqref{eq:infocontwave}  into the eigenvalue problem, we find three distinct types of equations based on the relative positions of the excitations. As discussed in Section \ref{subsec:threemagnon}, these correspond to the cases where all the magnons are well separated, where only two magnons are interacting, and where all three magnons are interacting simultaneously. Unlike the usual CBA equations, the configurations where the magnons are well separated do not vanish immediately but instead provide an important constraint on the position-dependent coefficients, $D^{n,m}_\sigma$. Consequently, the separation of the three-magnon dispersion relation into the sum of three independent one-magnon dispersion relations \eqref{eq:threeeigv} does not naturally arise from this particular ansatz. Instead, we demand the eigenvalue to have an additive form and obtain a family of solutions for the eigenvector with the position-dependent corrections.
   
   When the magnons are well separated, the eigenvalue problem is given as
   \begin{align}
       \gamma(n,m)=\braket{l_1,l_2,l_3|\mathcal{H}-E_3|\Psi(p)}_{12}=0\;,\label{eq:non-int}
    \end{align}
    where $n=l_2-l_1-1>0$ and $m=l_3-l_2-1>0$. The explicit form of these \emph{non-interacting equations} given by the wave function is, \eqref{eq:infocontwave},
    \begin{align}
       \gamma(n,m)&=\left(3\kappa+\frac{3}{\kappa}-E_3\right)\Psi(p_1,p_2,p_3;l_1,l_2,l_3)-\Psi(p_1,p_2,p_3;l_1\pm1,l_2,l_3)\nonumber\\&\quad\quad\quad-\Psi(p_1,p_2,p_3;l_1,l_2\pm1,l_3)-\Psi(p_1,p_2,p_3;l_1,l_2,l_3\pm1) \; . \label{eq:non-int-braket}
   \end{align}
   It is useful to separate the coefficients that accompany the plane waves to distinguish the position dependence arising from the position-dependent corrections from that of the plane waves. Thus, the non-interacting equations \eqref{eq:non-int-braket} split as
   \begin{align}
   \label{eqn:defgamma}
      \gamma(n,m)=\sum_{\sigma\in\mathcal{S}_3}\gamma_{\sigma}(p_1,p_2,p_3,n,m)\;e^{i(n+1)(p_{\sigma(2)}+p_{\sigma(3)})+i(m+1)p_{\sigma(3)}}
   \end{align}
  where we omit the overall plane wave contribution, $\exp(il_1(p_1+p_2+p_3))$. The term associated with the identity permutation takes the form
  \begin{align}
       \gamma_{123}(n,m)=-e^{i p_1} D_{123}^{n-1,m}-e^{-i p_1} D_{123}^{n+1,m}-e^{-i p_3} D_{123}^{n,m-1}\nonumber\\+\EE D_{123}^{n,m}-e^{ i p_3} D_{123}^{n,m+1}-e^{i p_2} D_{123}^{n+1,m-1}-e^{-ip_2}D_{123}^{n-1,m+1}
       \, .\label{eq:gamma123}
   \end{align}
 The remaining terms can be obtained from this expression by permuting the labels of the position-dependent corrections and momenta. In addition, we have introduced the symbol
  \begin{align}
     \EE=3\left(\kappa+\frac{1}{\kappa}\right)-E_3=\sum_{j=1}^3 \left( e^{i p_j} +e^{-i p_j} \right) \; ,\label{eq:tot-energy}
 \end{align}
which is closely related to the total energy of the three-magnon state, as it will appear at several points in our computations.

Similarly to the eigenvalue equations coming from the CBA, we impose that each $\gamma_\sigma$ vanishes independently of the plane wave contributions. Therefore, for each permutation, we obtain a relation between the coefficients $D_\sigma^{n,m}$ with shifted position coordinates. These relations will be the base of the permutation symmetries of our solution.

   \bigskip

When only two magnons are interacting, we obtain the \emph{two-magnon interaction} equations. Then we must distinguish whether the interacting pair is positioned to the left or right of the third magnon. For the case where the pair of interacting magnons are in the left of the spin chain, we have
    \begin{equation}
        \beta_l(m)=\braket{l_1,l_1+1,l_3|\mathcal{H}-E_3|\Psi(p_1,p_2,p_3)}_{12}=0\;,\label{eq:two-int-left}
    \end{equation}
    for all $m=l_3-l_2-1>0$. In terms of the wave functions the projection of the eigenvalue problem to the position space takes the following form,
    \begin{align}
        \beta_l(m)&=\left(\kappa+\frac{3}{\kappa}-E_3\right)\Psi(p_1,p_2,p_3;l_1,l_1+1,l_3)-\Psi(p_1,p_2,p_3;l_1-1,l_1+1,l_3)\nonumber\\&\quad\quad\quad\quad\quad\quad\quad-\Psi(p_1,p_2,p_3;l_1,l_1+2,l_3)-\Psi(p_1,p_2,p_3;l_1,l_1+1,l_3\pm1)\label{eq:two-int-l-braket}\;.
    \end{align}
    For the case where the pair of interacting magnons are on the right, we have
    \begin{equation}
        \beta_r(n)=\braket{l_1,l_3-1,l_3|\mathcal{H}-E_3|\Psi(p_1,p_2,p_3)}_{12}=0\;,\label{eq:two-int-right}
    \end{equation}
    for all $n=l_2-l_1-1>0$. Similarly to the previous case, this expression takes the form,
    \begin{align}
        \beta_r(n)&=\left(3\kappa+\frac{1}{\kappa}-E_3\right)\Psi(p_1,p_2,p_3;l_1,l_3-1,l_3)-\Psi(p_1,p_2,p_3;l_1\pm1,l_3-1,l_3)\nonumber\\&\quad\quad\quad\quad\quad\quad\quad-\Psi(p_1,p_2,p_3;l_1,l_3-2,l_3)-\Psi(p_1,p_2,p_3;l_1,l_3-1,l_3+1)\;.\label{eq:two-int-r-braket}
    \end{align}
    When we compare the left two-magnon interacting equations to the right, we can observe that they are related by the $\mathbb Z_2$ symmetry \eqref{eq:Z2map}. Recall that the action of $\mathbb Z_2$ symmetry exchanges $\kappa$ with $\frac{1}{\kappa}$ as well as $Q_{12}$ with $Q_{21}$. Looking at the pair of interacting magnons, the $\mathbb Z_2$ action would map the left two-magnon interacting pair to a right pair, but with the third magnon in the wrong position. By combining $\mathbb Z_2$ transformation with the action of parity operator, we can explicitly obtain \eqref{eq:two-int-right} from \eqref{eq:two-int-left} and vice versa. This is closely related to the observation about the restored parity invariance of one-magnon states, \eqref{eq:restoredparity}. After presenting all the equations, we will come back to this point.

    We now follow the same steps as for the non-interacting equations. We substitute the ansatz \eqref{eq:infocontwave}, decompose these equations into separate terms distinguished by the plane wave factors, and require that each term vanishes independently. In contrast to the non-interacting equations, one of the relative distances between magnons is set to zero, which implies that these equations can be split into three contributions instead of six. In particular, we have
    \begin{align}
      \beta_l(m)=&\beta_{l,(12)}(m)e^{i(m+2)p_3}+\beta_{l,(13)}(m)e^{i(m+2)p_2}+\beta_{l,(23)}(m)e^{i(m+2)p_1}\label{eq:defbetal}\;,
   \end{align}
   where we omit the overall plane wave contribution, $\exp(il_1(p_1+p_2+p_3))$ and the first coefficient in the sum is given as
   \begin{align}
       \beta_{l,(12)}(m)=&A_{123}\; a(p_2,p_1,\kappa)+A_{213}\;a(p_1,p_2,\kappa)\nonumber\\&-e^{-i p_3} \left(e^{ip_1}D_{213}^{0,m-1}+e^{ip_2}D_{123}^{0,m-1}\right)-e^{i p_3} \left(e^{ip_1}D_{213}^{0,m+1}+e^{ip_2}D_{123}^{0,m+1}\right)\nonumber\\&-e^{-ip_1-ip_2}\left(e^{2ip_1}D_{213}^{1,m}+e^{2ip_2}D_{123}^{1,m}\right) - \left(e^{2ip_1}D_{213}^{1,m-1}+e^{2ip_2}D_{123}^{1,m-1}\right)\nonumber\\
       &+\left(\EE-2\kappa\right)\left(e^{ip_1}D_{213}^{0,m}+e^{ip_2}D_{123}^{0,m}\right)\, ,
       \label{eq:betal12}
   \end{align}
   where the function $a(p_1,p_2,\kappa)=1+e^{ip_1+ip_2}-2\kappa e^{ip_1}$ is defined in equation \eqref{eq:scatnum}.
    Here we label our equations using $(ij)\in\{(12),(13),(23)\}$, as we only have to keep track of these transpositions. The remaining expressions coming from the left two-magnon interacting equation can be obtained by permuting the labels of momentum variables as well as the undetermined coefficients. Concerning these transpositions, both the scattering coefficients and the position-dependent corrections contribute in pairs.

    Similarly, the two-magnon interacting equation from the right splits as   
   \begin{align}
       \beta_r(n)=&\beta_{r,(12)}(m)e^{i(p_1+p_2)(n+1)}+\beta_{r,(13)}(n)e^{i(p_1+p_3)(n+1)}+\beta_{r,(23)}(n)e^{i(p_2+p_3)(n+1)}\;\label{eq:defbetar}
   \end{align}
 with
   \begin{align}
       \beta_{r,(12)}(n)=&A_{312}\;a(p_2,p_1,\kappa^{-1})+A_{321}\;a(p_1,p_2,\kappa^{-1})\nonumber\\& -e^{i p_3} \left(e^{ip_1}D_{321}^{n-1,0}+e^{ip_2}D_{312}^{n-1,0}\right)-e^{-i p_3} \left(e^{ip_1}D_{321}^{n+1,0}+e^{ip_2}D_{312}^{n+1,0}\right)\nonumber\\&- \left(e^{2ip_1}D_{321}^{n,1}+e^{2ip_2}D_{312}^{n,1}\right) -e^{- i p_1-ip_2} \left(e^{2ip_1}D_{321}^{n-1,1}+e^{2ip_2}D_{312}^{n-1,1}\right) \nonumber\\
       &+\left(\EE-2\kappa^{-1}\right)\left(e^{ip_1}D_{321}^{n,0}+e^{ip_2}D_{312}^{n,0}\right)\;,\label{eq:betar12}
   \end{align}
   where we omit the overall plane wave contribution as before. Compared to the two-body interaction equations derived from the usual CBA, given in \eqref{eq:naive-sol-left} and \eqref{eq:naive-sol-right}, these equations are corrected by the contributions of the position-dependent corrections. In contrast with the case of the CBA with contact terms, discussed in detail in Appendix~\ref{sec:contact-terms}, this infinite tower of corrections cannot be fully washed out. This strongly indicates that the failure of the YBE, as presented in \eqref{eq:neq}, will not be an obstruction to obtaining a solution.

\bigskip
   
  In the case where the three magnons are next to each other, we have an expression that is free of position variables,
   \begin{equation}
        \alpha=\braket{l_1,l_1+1,l_1+2|\mathcal{H}-E_3|\Psi(p_1,p_2,p_3)}_{12}=0\;,\label{eq:threeint0}
    \end{equation}
    where
   \begin{align}
        &\braket{l_1,l_1+1,l_1+2|\mathcal{H}-E_3|\Psi(p_1,p_2,p_3)}_{12} =\left(\kappa+\frac{1}{\kappa}-E_3\right)\Psi(p_1,p_2,p_3;l_1,l_1+1,l_1+2)\nonumber\\
        &-\Psi(p_1,p_2,p_3;l_1-1,l_1+1,l_1+2)-\Psi(p_1,p_2,p_3;l_1,l_1+1,l_1+3)\; .\label{eq:threeint}
    \end{align}
    The explicit form of the \emph{three-magnon interacting equation} up to an overall plane wave contribution, $\exp(il_1(p_1+p_2+p_3))$ is given by
   \begin{align}
       \alpha(p_1,p_2,p_3) &=\sum_{\sigma\in S_3} \big[(\kappa+\frac{1}{\kappa}-E_3-e^{i p_{\sigma(3)}} - e^{-i p_{\sigma(1)}})A_\sigma \nonumber\\
       & -e^{-ip_{\sigma(1)}}D^{1,0}_\sigma -e^{ip_{\sigma(3)}}D^{0,1}_\sigma\big] e^{i (p_{\sigma(2)}+ 2p_{\sigma(3)})} \;.\label{eq:alpha}
   \end{align}
   This equation completes the list of the equations that are coming from the eigenvalue problem.

   Because we are working with open infinitely long spin chains, these three types of equations imply an infinite dimensional linear system of equations. Since the interacting equations mix the position-dependent corrections with different permutations, we have a single highly interwoven system of equations. To reveal the underlying structure, we pay close attention to the symmetries of the spin chain model. After analyzing the consequences of parity and $\mathbb{Z}_2$ symmetry, we solve the eigenvalue problem in a specific order to capture the residual permutation symmetry of our long-range ansatz.

\subsection{Parity and $\mathbb{Z}_2$ Symmetry}
\label{subsec:symmetry}
   
   We study the conditions imposed on the long-range ansatz for it to satisfy the analogous parity relation with the one-magnon eigenstate \eqref{eq:restoredparity}. Although we only discuss the equations coming from the eigenvalue problem for one of the two possible three-magnon states, $\ket{\Psi(p_1,p_2,p_3)}_{12}$, the $\mathbb Z_2$ pair of this state exists and it satisfies the eigenvalue equation \eqref{eq:eigthree},
   \begin{align}
       \mathbb Z_2\ket{\Psi(p_1,p_2,p_3)}_{12}=\ket{\Psi(p_1,p_2,p_3)}_{21}\;.\label{eq:z2forthree}
   \end{align}
   This implies that the entire wave function of the eigenvector, $\ket{\Psi}_{21}$ should be the $\kappa\to1/\kappa$ version of the wave function of the state, $\ket{\Psi}_{12}$, as given in \eqref{eq:infcontansatz}. Indeed, this can be realized by looking at the eigenvalue problem for the $\mathbb Z_2$ pair. We observe that the form of the non-interacting equations is the same for both sets of position-dependent corrections because the expressions given in \eqref{eq:gamma123} do not include explicit dependence on the $\kappa$ parameter. Additionally, the three-magnon interacting equation given in \eqref{eq:alpha} is invariant under the $\kappa\to1/\kappa$ transformation, therefore the coefficients of the eigenstate $\ket{\Psi(p_1,p_2,p_3)}_{21}$ also satisfy the same three-magnon interacting equation. However, the two-magnon interacting equations given explicitly in equations \eqref{eq:betal12} and \eqref{eq:betar12} significantly change under the $\mathbb Z_2$ transformation. Hence, both scattering coefficients and the position-dependent corrections of the $\mathbb Z_2$ pair state, have the form,
   \begin{align}
       \braket{l_1,l_2,l_3|\Psi(p_1,p_2,p_3)}_{21}= \sum_{\sigma\in S_3} \left(A_\sigma(p,1/\kappa)+D^{l_2-l_1-1,l_3-l_2-1}_\sigma(p,1/\kappa)\right)e^{i\Vec{p}_{\sigma}\cdot\Vec{l}}\;.\label{eq:z2-wavefunction}
   \end{align}
   Moreover, when we consider the action of the parity, we obtain
   \begin{align}  &\mathcal{P}\ket{\Psi(p_1,p_2,p_3)}_{12}\nonumber\\&\quad=\sum_{-l_3<-l_2<-l_1} \left(\sum_{\sigma\in S_3} \left(A_\sigma(p,\kappa)+D^{l_2-l_1-1,l_3-l_2-1}_\sigma(p,\kappa)\right)e^{i\Vec{p}_{\sigma}\cdot\Vec{l}}\right) \ket{-l_3,-l_2,-l_1}_{21}\;. \label{eq:parity-action3}
   \end{align}
   Although the state that is obtained by the action of the parity operator is not necessarily an eigenvector of the Hamiltonian, we can restore parity as for the one-magnon state using the action of $\mathbb Z_2$. The parity symmetry, which is severely broken, can only be restored if there exists a three-magnon eigenstate whose wave function includes the scattering coefficients $\hat{A}_\sigma(p,\kappa)$ and position-dependent corrections $\hat{D}_\sigma^{n,m}(p,\kappa)$ such that the coefficients of the initial eigenstate $\ket{\Psi}_{12}$ satisfy the following relations,
   \begin{align}
       A_\sigma(p,\kappa)=e^{i\varphi(p,\kappa)}\hat{A}_{\sigma^r}(-p,1/\kappa)\;,\quad\quad D_\sigma^{n,m}(p,\kappa)=e^{i\varphi(p,\kappa)}\hat{D}_{\sigma^r}^{m,n}(-p,1/\kappa)\;,\label{eq:parity-identity}
   \end{align}
   where for $\sigma=(ijk)$ the reflection of the permutation index is denoted as $\sigma^r=(kji)$. We explicitly show the existence of two eigenvectors that satisfy these relations in Section \ref{sec:special}. The overall phase factor $e^{i\varphi(p,\kappa)}$ is incorporated into these relations to avoid over-restricting the solution space for the coefficients of the three-magnon state. The two eigenvectors whose coefficients satisfy the given relations, \eqref{eq:parity-identity} coincide then we obtain,
   \begin{align}
       \mathbb{Z}_2\mathcal{P}\ket{\Psi(p_1,p_2,p_3)}_{12}=e^{i\varphi(p,\kappa)}\ket{\Psi(-p_1,-p_2,-p_3)}_{12}\label{eq:restoreparity3}\;.
   \end{align}
   Furthermore, applying parity twice should give us back the original state, so the phase has to satisfy $e^{i\varphi(p,\kappa)} e^{i\varphi(-p,\kappa^{-1})}=1$.

   \subsection{The Solution}
   \label{sec:solution}

   \begin{table}[t!]

       \begin{center}

\begin{tabular}{ |c|c|c| } 
 \hline  & Equations & Undetermined coefficients\\ \hline
First Class  & $\alpha=0$, $\beta_{l,(ij)}(1)=0$, $\beta_{r,(ij)}(1)=0$ & $A_\sigma$, $D^{1,0}_\sigma$, $D^{0,1}_\sigma$, $D^{2,0}_\sigma$, $D^{1,1}_\sigma$, $D^{0,2}_\sigma$ \\ \hline
  & $\gamma_\sigma (n,m)=0$ & $D^{n,m}_\sigma$ for $n,m \geq 2$ \\ 
 Second Class & $\beta_{l,(ij)}(m+1)-\beta_{l,(ij)}(m)=0$ & $D^{n,m}_\sigma$ for $m \geq 2$ and $n=0,1$ \\ 
  & $\beta_{r,(ij)}(nm+1)-\beta_{r,(ij)}(n)=0$ & $D^{n,m}_\sigma$ for $m=0,1$ and $n \geq 2$ \\ \hline
\end{tabular}
\end{center}
       \caption{\it Summary of the method, we list the equations coming from the eigenvalue problem and the undetermined coefficients that appear in these equations.
      }
    \label{fig:method}
   \end{table}

   Our strategy to solve these equations is depicted in Table \ref{fig:method} and is as follows. We separate the expressions coming from the eigenvalue problem into two classes. The first class of equations does not involve any free position variables, and we use them to obtain the scattering coefficients of the model. Therefore, the first class includes the interacting equations with minimum separation, listed as
   \begin{align}
       \alpha(p_1,p_2,p_3)=0\;,\quad\quad\beta_{l,(ij)}(1)=0\;,\quad\quad\beta_{r,(ij)}(1)=0 \;,\label{eq:firstclass}
   \end{align}
   for all transpositions of $\mathcal{S}_3$, denoted as $(ij)$. These seven equations are mainly used to determine the scattering coefficients, $A_\sigma$, up to an overall normalization which is unspecified in the form of the coefficient $A_{123}$. Together with the five scattering coefficients, we should fix at least two of the position-dependent corrections to ensure that all seven equations are solved.

    The remaining equations constitute the second class of equations. They give us recurrence relations for the position-dependent corrections, $D^{n,m}_\sigma$. We can separate the eigenvalue problem for position-dependent corrections from the scattering coefficients by considering the difference of two successive two-body equations. Thus, for $n,m\geq1$, we will solve the equations
   \begin{align}
       \gamma_\sigma(n,m)=0\;,\quad\quad \beta_{l,(ij)}(n+1)-\beta_{l,(ij)}(n)=0 \;,\quad\quad \beta_{r,(ij)}(n+1)-\beta_{r,(ij)}(n)=0\;,  \label{eq:secondclass}
   \end{align}
   such that $\sigma\in\mathcal{S}_3$ and $(ij)$ are the distinct transpositions. These equations, together with \eqref{eq:firstclass}, span the infinite linear system of equations for the scattering coefficients and the corrections. By dividing the set of two-magnon interacting equations into two sub-sets, which we refer to as the first and second class, we achieve an upper-triangular form for the infinite-dimensional system of equations which include all the two-magnon interacting eigenvalue equations and this will simplify the process of solving the two-magnon interacting equations drastically.
   
   A careful consideration of the set of equations given in \eqref{eq:firstclass} and \eqref{eq:secondclass} reveals that the solution of the position-dependent corrections coming from the first class of expressions is the initial conditions for the recursion relations in the second class. Furthermore, the relations coming from the two-body interacting equations are the initial conditions of the non-interacting type equations. Therefore, we start solving the eigenvalue problem from the non-interacting equations. After acquiring the most general set of position-dependent corrections that solve the non-interacting equations, we move on to the two-body interacting recurrence relations and incorporate these relations into the non-interacting ones. Finally, we present the most general way to solve the first class of equations before taking advantage of the large set of undetermined coefficients and obtaining distinct subfamilies of the three-magnon eigenvectors, which admit interesting physical interpretations.
   
   \subsubsection{Solving the Non-interacting Equations}
   \label{subsubsec:non-int}

\begin{figure}[t]
       \centering
       \includegraphics[scale=0.25]{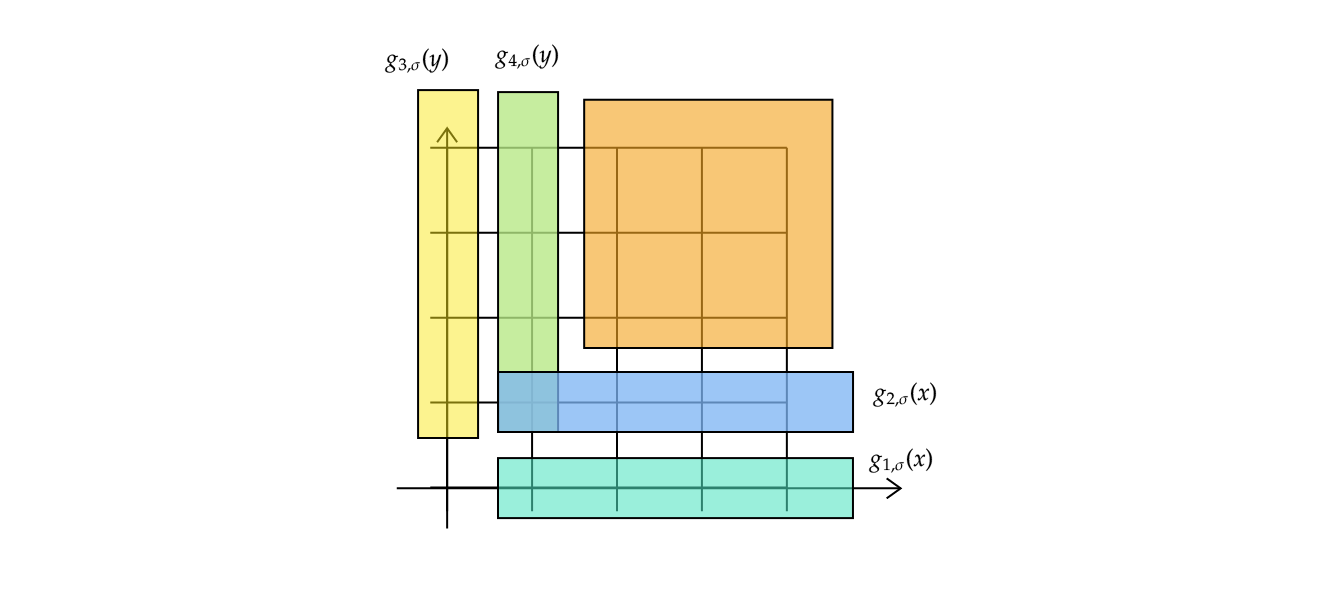}      
        \caption{\textit{Imposing non-interacting recurrence relations gives us all position-dependent corrections colored with orange, in terms of the rest. The analyticity condition further gives the solution of the position-dependent corrections. Depending on the choice of parametrization either $g_{2,\sigma}$ or $g_{4,\sigma}$ is given by analyticity.}}
       \label{fig:non-interactingsol}
   \end{figure}

    We start from the non-interacting equations, which we solve via the generating function method. We introduce a complex two-variable function whose Taylor expansion around $(0,0)$ gives us the position-dependent corrections for each permutation, 
   \begin{align}
       G_{\sigma}(x,y)=\sum_{n,m=1}^{\infty} D_{\sigma}^{n,m}(p_1,p_2,p_3;\kappa) x^ny^m\;. \label{eq:GDefinition}
   \end{align}
   The non-interacting equations, $\gamma_\sigma(n,m)=0$, can then be used to manipulate the initial form of the generating function. First, we refine the position-dependent corrections to get rid of a universal factor in all eigenvalue problem expressions, 
   \begin{align}
       \tilde D_{\sigma}^{n,m}=e^{-inp_{\sigma(1)}+imp_{\sigma(3)}}D_{\sigma}^{n,m}\;.\label{eq:scaledDs}
   \end{align}
   The solution of the non-interacting equations for each permutation is encoded in the closed form of the generating function,
   \begin{align}
   G_\sigma(e^{-ip_{\sigma(1)}}x,e^{ip_{\sigma(3)}}y)&=g_{1,\sigma}(x)+g_{3,\sigma}(x)-\frac{\invPP x^2 y \tilde D^{0,1}_\sigma+\PP x y^2 \tilde D^{1,0}_{\sigma} }{Q(x,y)}  \notag \\
       &+\frac{\PP y^2\left(1 +\invPP x \right) }{Q(x,y)} g_{1,\sigma}(x)+\frac{\invPP x^2\left( 1 +\PP y\right) }{Q(x,y)} g_{3,\sigma}(y)\nonumber \\
       &-\frac{xy\left(1+\invPP x\right)}{Q(x,y)} g_{2,\sigma}(x)-\frac{xy\left(1+\PP y\right)}{Q(x,y)} g_{4,\sigma}(y)\label{eq:generating-non-int}\; ,
   \end{align}
   where the common denominator $Q(x,y)$ is a third-order polynomial of two variables without any constant term,
   \begin{align}
   \label{eq:ellipticCurve}
       Q(x,y)=\EE x y-\invPP x^2-\PP y^2-x-y-x^2 y- y^2 x\;.
   \end{align}
   Vanishing of this polynomial defines an elliptic curve which we use to understand the solution of the non-interacting equations better.\footnote{This can be checked by bringing it into its Weierstrass form, which has non-vanishing discriminant and J-invariant for generic values of $\EE$ and $\PP$.} The coefficients that appear in the generating function are $\PP=\exp\left(i(p_1+p_2+p_3)\right)$, which is the exponential of the total momentum, and $\EE$, which is closely related to total energy, \eqref{eq:tot-energy}. The functions $g_{i,\sigma}$ are generating functions that consist of a particular sub-set of position-dependent corrections. These are not fixed by the non-interacting equations and they are defined as follows,
   \begin{align}
       &g_{1,\sigma}(x)=G_{\sigma}(e^{-ip_{\sigma(1)}}x,0)=\sum_{n=1}^\infty \tilde D_{\sigma}^{n,0} x^n\;,\label{eq:one-variable1}\\
       &g_{2,\sigma}(x)=\partial_yG_{\sigma}(e^{-ip_{\sigma(1)}}x,e^{ip_{\sigma(3)}}y)|_{y=0}=\sum_{n=1}^\infty \tilde D_{\sigma}^{n,1} x^n\;,\label{eq:one-variable2}\\
       &g_{3,\sigma}(y)=G_{\sigma}(0,e^{ip_{\sigma(3)}}y)=\sum_{m=1}^\infty \tilde D_{\sigma}^{0,m} y^m\;,\label{eq:one-variable3}\\
       & g_{4,\sigma}(y)=\partial_xG_{\sigma}(e^{-ip_{\sigma(1)}}x,e^{ip_{\sigma(3)}}y)|_{x=0}=\sum_{m=1}^\infty \tilde D_{\sigma}^{1,m} x^n\label{eq:one-variable4}\;.
   \end{align}  
   Thus, solving the non-interacting equations reduces the number of undetermined coefficients from a two-dimensional lattice parameterized by positive integers $\mathbb{Z}^{>0}\times\mathbb{Z}^{>0}$, to the boundaries $\{0,1\}\times\mathbb Z^{>0}$ and $\mathbb Z^{>0}\times\{0,1\}$ of that lattice for each permutation. This is depicted in Figure \ref{fig:non-interactingsol}. We should also stress that the scaling of the position-dependent corrections, given in \eqref{eq:scaledDs}, together with the re-scaling of the variables $x$ and $y$, allow us to write the generating functions $G_\sigma$ only in terms of energy $\EE$ and total momentum $\PP$. Consequently, the resulting form of the generating function \eqref{eq:generating-non-int} is exactly the same for each permutation, up to the undetermined $g_{i,\sigma}$. This implies that it is symmetric in terms of permutations of the indices.

    At first glance, the form of the generating function \eqref{eq:generating-non-int} implies that all one-variable generating functions defined in \eqref{eq:one-variable1}-\eqref{eq:one-variable4} are undetermined. The rest of the position-dependent corrections for $n,m\geq2$ can be obtained by taking the derivatives with respect to the generating function, 
    \begin{align}
       \tilde D_{\sigma}^{n,m}=\lim_{x,y\to0}\frac{1}{n!m!}\partial^{(n)}_x\partial^{(m)}_y  G_{\sigma}(e^{-ip_{\sigma(1)}}x,e^{ip_{\sigma(3)}}y) \;.\label{eq:derivative-cont}
   \end{align}
    
    After closer inspection, we observe that the functions \eqref{eq:one-variable1}-\eqref{eq:one-variable4} cannot be completely independent. The analyticity of the generating function \eqref{eq:generating-non-int} is not guaranteed for four independent functions, $g_{i,\sigma}$. Equivalently, we observe that the limit that appears in equation \eqref{eq:derivative-cont} does not exist from every direction on the complex plane for four independent one-variable generating functions.  As a consequence, we have to demand that the numerator generating function given in \eqref{eq:generating-non-int} should vanish on the algebraic curve defined by $Q(x,y)=0$, that is\footnote{We do not need to impose a stronger condition on the numerator because the denominator $Q(x,y)$ only generates a single pole at $(0,0)$.}
    \begin{align}
       &y(1+\invPP x) \left( \PP y g_{1,\sigma}(x) -  x g_{2,\sigma}(x) \right)\nonumber\\&\quad\quad +x(1+\PP y) \left( \invPP x g_{3,\sigma}(y) - y g_{4,\sigma}(y) \right)\nonumber\\&\quad\quad\quad\quad\quad\quad\quad-\left. xy\left(\invPP x \tilde D^{0,1}_{\sigma}+\PP y \tilde D^{1,0}_{\sigma} \right) \right|_{Q(x,y)=0}=0 \; . \label{eq:analyticityofG}
    \end{align}
    Once we ensure that this condition of the elliptic curve that includes the origin, $(x,y)=(0,0)$ is satisfied, the limit \eqref{eq:derivative-cont} exists on a disc around $(0,0)$ and we can extract the position-dependent corrections. To explicitly state the analyticity condition over the four one-variable generating functions given in \eqref{eq:one-variable1}-\eqref{eq:one-variable4} we factorize the elliptic curve,
    \begin{align}
        Q(x,y)=(x-\rho_+(y,p))(x-\rho_-(y,p))\;,
    \end{align}
   such that
  \begin{align}
       \rho_\pm(y,p)=\frac{-\PP\left( y^2-\EE  y+1\right)\pm\sqrt{e^{2i\mathcal{P}}\left(y^2-\EE  y+1\right)^2-4\PP y (1+\PP y)^2}}{2 (1+\PP y)}\;.\label{eq:elpara}
   \end{align}
   Here, $p$ denotes the dependence on the momentum variables, $p_i$ all together. The function $\rho_\pm$ parametrizes the two different sheets of the elliptic curve on the complex plane. We explicitly show the dependence of $\rho_\pm$ on the momentum variables because we want to make use of the property,
%   labelled the function, $\rho_\pm$ by the dummy complex variable, $y$, and the momentum variables to make use of the observation,
   \begin{align}
       Q(x,y)=(x-\rho_+(y,p))(x-\rho_-(y,p))=(y-\rho_+(x,-p))(y-\rho_-(x,-p))\;.\label{eq:factorizedEll}
   \end{align}
%   This is a consequence of the action of parity on the generating function. The explicit form of the action is spelled out in expression, \eqref{eq:two-body-rel}. 
Employing the symmetric form of the non-interacting equations, we will obtain a highly symmetric solution of the two-magnon interacting equations. Since we are only interested in the Taylor expansion around the origin, we only consider the relation coming from the curve $\rho_+$, as the curve $(\rho_{-}(y,p),y)$ does not contain the origin.\footnote{This statement is highly dependent on which branch we choose. For $-\pi<\sum_i p_i<\pi$ we use the positive branch of the square root. Essentially, it is sufficient to consider one of the curves $(\rho_\pm(y,p),y)$ depending on the choice of the branch and this can be checked by comparing the analyticity relation, \eqref{eq:analyticityofG} to the existence of the limit conditions for any $n,m$ given in \eqref{eq:derivative-cont}.} Hence, we can derive an explicit relation from the analyticity condition, \eqref{eq:analyticityofG},\footnote{When we restate this equation in terms of position-dependent corrections, we can see that each coefficient of the Taylor expansion in $y$ is a linear combination of the non-interacting equations $\gamma_\sigma(n,m)$. Thus, the analyticity condition \eqref{eq:analyticityofG} is not a spurious equation, but it captures a linear combination of the non-interacting equations that are not naively captured in $G_\sigma(x,y)$.}
   \begin{align}
       g_{4,\sigma}(y)=&\frac{\rho_+(y,p)}{\PP y} g_{3,\sigma}(y)+\frac{1+\invPP \rho_+(y,p)}{1+\PP y} \left[ \frac{\PP y}{\rho_+(y,p)} g_{1,\sigma}(\rho_+(y,p)) -   g_{2,\sigma}(\rho_+(y,p)) \right] \notag \\
       -&\frac{ \invPP \rho_+(y,p)}{1+\PP y} \tilde D^{0,1}_{\sigma}-\frac{\PP y}{1+\PP y} \tilde D^{1,0}_{\sigma}\;.\label{eq:sol-analyticity}
   \end{align}
   This expression reduces the number of undetermined position-dependent corrections given in the form of four one-variable generating functions, \eqref{eq:one-variable1}-\eqref{eq:one-variable4} by one for each permutation. Therefore, the final form of the generating function is a meromorphic function of two complex variables, $x,y\in\mathbb C$ given as

   \begin{align}
       G_\sigma^\gamma(e^{-ip_{\sigma(1)}}&x,e^{ip_{\sigma(3)}}y)=-\frac{\invPP x y(x-\rho_+(y,p)) \tilde D^{0,1}_\sigma}{Q(x,y)}  \notag \\
       &+\frac{\EE xy - \invPP x^2 -x -y - x^2 y }{Q(x,y)} g_{1,\sigma}(x)-\frac{xy^2\left(1+\PP \rho_+(y,p)^{-1}\right)}{Q(x,y)}g_{1,\sigma}(\rho_+(y,p))\nonumber \\
       &+\frac{\EE xy - \PP y^2 -x -y - y^2 x-\invPP x \rho_+(y,p)\left(1+\PP y\right) }{Q(x,y)} g_{3,\sigma}(y)\nonumber \\
       &-\frac{xy\left(1+\invPP x\right)}{Q(x,y)} g_{2,\sigma}(x)+\frac{xy\left(1+\invPP \rho_+(y,p)\right)}{Q(x,y)} g_{2,\sigma}(\rho_+(y,p))\;. \label{eq:g-gamma}
   \end{align}
   On the other hand, by using the second parametrization given in \eqref{eq:factorizedEll}, we could write the analyticity relation as 
   \begin{align}
       g_{2,\sigma}(x) =&\frac{\rho_+(x,-p)}{\invPP x}  g_{1,\sigma}(x)+\frac{1+\PP \rho_+(x,-p)}{1+\invPP x} \left[\frac{\invPP x }{\rho_+(x,-p)}g_{3,\sigma}(\rho_+(x,-p))-  \vphantom{\frac{\invPP x }{\rho_+(x,-p)}} g_{4,\sigma}(\rho_+(x,-p)) \right]\nonumber\\&  -\frac{\invPP x}{1+\invPP x} \tilde D^{0,1}_{\sigma}-\frac{\PP \rho_+(x,-p)}{(1+\invPP x)} \tilde D^{1,0}_{\sigma} \; .\label{eq:sol-analyticity2}
    \end{align}
   Then, the form of the generating function after imposing the analyticity condition would slightly differ, but it is equivalent to the one presented in \eqref{eq:g-gamma},
   \begin{align}
   G^\gamma_\sigma(&e^{-ip_{\sigma(1)}}x,e^{ip_{\sigma(3)}}y)=-\frac{\PP x y\left(y-\rho_+(x,-p)\right) \tilde D^{1,0}_{\sigma} }{Q(x,y)}  \notag \\
   &+\frac{\EE xy - \invPP x^2 -x -y - x^2 y-\PP\rho_+(x,-p)y\left(1+\invPP x\right)}{Q(x,y)} g_{1,\sigma}(x)\nonumber\\
   &+\frac{\EE xy-\PP y^2-x-y-y^2x }{Q(x,y)} g_{3,\sigma}(y)-\frac{x^2y\left(1+\invPP\rho_+(x,-p)^{-1}\right)}{Q(x,y)}g_{3,\sigma}(\rho_+(x,-p))\nonumber \\
    &-\frac{xy\left(1+\PP y\right)}{Q(x,y)} g_{4,\sigma}(y)+\frac{xy\left(1+\PP\rho_+(x,-p)\right)}{Q(x,y)}g_{4,\sigma}(\rho_+(x,-p))\; ,\label{eq:g-gamma-2}
   \end{align}
   This is exactly the parity transform of the \eqref{eq:g-gamma}. Since we are only solving the non-interacting equations which do not have any explicit $\kappa$ dependence, the action of parity, as given in \eqref{eq:parity-action3}, transforms to the following map in the level of generating functions,
   \begin{align}
       \{D_\sigma^{n,m},x,y,p\}\leftrightarrow \{D_\sigma^{m,n},y,x,-p\}\;. \label{eq:non-int-rel}
   \end{align}
   Similarly, the two parametrizations of the elliptic curve given in \eqref{eq:factorizedEll} are a result of this relation. Furthermore, this is a subset of the operations in which we need to map the left two-magnon interacting equations to the right two-magnon interacting equations. This is given in \eqref{eq:two-body-rel}. In addition to the $\kappa\to1/\kappa$ transformation, we also do not see the reflection of the permutations, as the generating functions that solve the non-interacting equations are invariant under the permutation of momentum indices.

   The resulting form of the generating function, \eqref{eq:g-gamma} and \eqref{eq:g-gamma-2} will be altered once more when we solve the two-magnon interacting equations. At this point, solution of any position-dependent corrections, $D^{n,m}_\sigma$ for $n,m\geq 2$ are given as a linear combination of undetermined coefficients, encoded in $g_{i,\sigma}$ for $i=1,2,3$ or $i=1,3,4$ and the factors in front of the undetermined position-dependent corrections are only powers and linear combinations of $\PP$ and $\EE$. Furthermore, the generating function satisfies a very simple partial differential equation which remains to be explored
   \begin{align}
       \partial_x^2\partial_y^2\left(Q(x,y)G_\sigma^\gamma(x,y)\right)=0\;.\label{eq:partialeq}
   \end{align}
   for each permutation, $\sigma$.

   \subsubsection{Solving the Two-Magnon Interacting Equations}
   \label{subsubsec:two-magnons}

   We move on to solve the two-body interacting equations given in the form of difference equations,
   \begin{align}
       \beta_{l,(ij)}(m+1)-\beta_{l,(ij)}(m)=0\;,\label{eq:two-body-rec1}\\\beta_{r,(ij)}(n+1)-\beta_{r,(ij)}(n)=0\;,\label{eq:two-body-rec2}
   \end{align}
   for all $n,m\geq 1$ and all transpositions of $\mathcal{S}_3$. The generating functions associated with these equations take the form of algebraic equations involving the undetermined position-dependent coefficients $g_{i,\sigma}$ associated with two different permutations. We can recast all three of the two-magnon interacting recursion relations from the left, \eqref{eq:two-body-rec1}, into the following expression
   \begin{align}
       &\left(e^{ip_1}g_{3,213}(y)+e^{ip_2}g_{3,123}(y)\right)\left(2\kappa - \EE+ \frac{1}{y} +y\right)\nonumber\\&\quad\quad\quad\quad+\left(e^{ip_1}g_{4,213}(y)+e^{ip_2}g_{4,123}(y)\right) \left(1+\PP y\right)=C_{12}(y)\label{eq:two-body-rec12}\;,
   \end{align}
   such that the position-dependent corrections that are left out of the one-variable generating functions are absorbed into the following function
   \begin{align}
       &C_{12}(y)=\left(1-\frac{y(\EE - 2\kappa)}{1-e^{ip_3}y}\right)\left(e^{ip_1}\tilde D_{213}^{0,1}+e^{ip_2}\tilde D_{123}^{0,1}\right)+ \frac{y}{1-e^{ip_3}y} \left[\left(e^{ip_1}\tilde D_{213}^{0,2}+e^{ip_2}\tilde D_{123}^{0,2}\right)\right.\nonumber\\&\quad\quad\quad\quad\quad\quad\quad\left.+\left(e^{ip_1}\tilde D_{213}^{1,1}+e^{ip_2}\tilde D_{123}^{1,1}\right)+ye^{ip_3}\PP\left(e^{ip_1}\tilde D_{213}^{1,0}+e^{ip_2}\tilde D_{123}^{1,0}\right) \right]\label{eq:const12}\;.
   \end{align}
   These terms will be determined by the first class of equations, \eqref{eq:firstclass}. The rest of the relations coming from \eqref{eq:two-body-rec1} can be obtained by permuting the indices of this expression.% From the right version of the two-body interacting configuration we attain the following relation between the generating functions,

   Similarly, the two-body interacting equations from the right give us the following relation between the generating functions 
   \begin{align}
       &\left(e^{ip_1}g_{1,321}(x)+e^{ip_2}g_{1,312}(x)\right)\left(\frac{2}{\kappa} - \EE+ \frac{1}{x} +x\right)\nonumber\\&\quad\quad\quad\quad+\left(e^{ip_1}g_{2,321}(x)+e^{ip_2}g_{2,312}(x)\right)\left(1+\invPP x\right)=\tilde C_{12}(x)\label{eq:two-body-rec21}\;.
   \end{align}
   Here, we absorbed the position-dependent coefficients which do not fit into one of the one-variable generating functions into the following functions,
   \begin{align}
       &\tilde C_{12}(x)=\left(1- \frac{x(\EE - \frac{2}{\kappa})}{1-e^{-ip_3}x}\right)\left(e^{ip_1}\tilde D_{321}^{1,0}+e^{ip_2}\tilde D_{312}^{1,0}\right)+ \frac{x}{1-e^{-ip_3}x} \Big[\left(e^{ip_1}\tilde D_{321}^{2,0}+e^{ip_2}\tilde D_{312}^{2,0}\right)\nonumber\\&\quad\quad\quad\quad\quad\quad\quad+\left(e^{ip_1}\tilde D_{321}^{1,1}+e^{ip_2}\tilde D_{312}^{1,1}\right)+xe^{-ip_3}\invPP\left(e^{ip_1}\tilde D_{321}^{0,1}+e^{ip_2}\tilde D_{312}^{0,1}\right) \Big]\;.\label{eq:const21}
   \end{align}
   From \eqref{eq:two-body-rec21} and \eqref{eq:const21}, the remaining expressions can be easily obtained by permuting the indices as usual. The functions are given in \eqref{eq:const12} and \eqref{eq:const21} contain the particular position-dependent corrections that appear in the first class of equations, \eqref{eq:firstclass}.

   Note that we can map the two-body interacting equations from the left configuration to the right, and vice versa, using the given transformation
   \begin{align}
       \{p,x,y,\kappa,D_{ijk}^{n,m}\}\leftrightarrow\{-p,y,x,\kappa^{-1},D_{kji}^{m,n}\}\;.\label{eq:two-body-rel}
   \end{align}
   This is a mixed version of the parity and the $\mathbb Z_2$ transformation and is reminiscent of the symmetry $\mathbb Z_2\mathcal{P}$ as discussed in Section~\ref{subsec:symmetry}. As we pointed out in the previous section, under this set of transformations the non-interacting equations and the three-magnon equations stay invariant and the left and right two-magnon interacting equations transform to each other. It is also the extended version of the map that was given for the non-interacting equations, \eqref{eq:non-int-rel}.

   Combining the analyticity condition with the two-magnon interacting equations allows us to obtain half of the one-variable generating functions, which were introduced in \eqref{eq:one-variable1}-\eqref{eq:one-variable4}. The expressions of these generating functions capture the underlying permutation symmetry structure of the spin chain model. After a careful analysis of the two-magnon interacting equations, we conclude the right two-magnon equation which is presented in \eqref{eq:two-body-rec21} should be combined with the analyticity condition for the generating function $G_{312}$ with the $(x,\rho_+(x,-p))$ parametrization of the elliptic curve. This particular choice of parametrization gives us the analyticity condition as written in \eqref{eq:sol-analyticity2}. The solutions are given as,
   \begin{align}
       &g_{1,312}(x)=\frac{x \left(e^{-i p_2}\tilde C_{12}(x)+\invPP x\tilde D_{312}^{0,1}+\PP \rho_+(x,-p)\tilde D_{312}^{1,0}\right)}{\tilde h_{12}(x,p)}\nonumber\\&\quad-\frac{e^{ip_1-ip_2}(1+2x/\kappa-\EE x+x^2)}{\tilde h_{12}(x,p)}g_{1,321}(x)-\frac{e^{ip_1-ip_2}x(1+\invPP x)}{\tilde h_{12}(x,p)}g_{2,321}(x)\nonumber\\&\quad-\frac{x^2(1+\invPP/\rho_+(x,-p))}{\tilde h_{12}(x,p)}g_{3,312}(\rho_+(x,-p))+\frac{x(1+\PP\rho_+(x,-p))}{\tilde h_{12}(x,p)}g_{4,312}(\rho_+(x,-p))\;,\label{eq:sol-g1312}
    \end{align}
    and,
    \begin{align}
       &g_{2,312}(x)=\frac{e^{-ip_2}\PP \rho_+(x,-p)\tilde C_{12}(x)}{\tilde h_{12}(x,-p)}-\frac{e^{ip_1-ip_2}(\PP+x)\rho_+(x,-p)}{\tilde h_{12}(x,p)}g_{2,321}(x)\nonumber\\&-\frac{\left(1+x^2+2x/\kappa-\EE x\right)}{ \tilde h_{12}(x,-p)}\Bigg(\frac{x{\tilde D_{312}^{0,1}+e^{2i\mathcal{P}}\rho_+(x,-p) \tilde D_{312}^{1,0}}}{(\PP+x)}+\frac{e^{ip_1-ip_2}\PP \rho_+(x,-p)}{x} g_{1,321}(x)
       \nonumber\\&-\frac{1+\PP\rho_+(x,-p)}{1+\invPP x}\left(\frac{\invPP x}{\rho_+(x,-p)}g_{3,312}(\rho_+(x,-p))- g_{4,312}(\rho_+(x,-p))\right)\Bigg)\;.\label{eq:sol-g2312}
   \end{align}
   Although the expressions look complicated, it is remarkable that they admit a compact form in terms of the one-variable generating functions. These expressions are given in terms of relatively simple functions and they are analytic around the origin. The common denominator is given as
   \begin{align}
       \tilde h_{12}(x,p)=1+2x/\kappa-\EE x +x\rho_+(x,-p) +\PP \rho_+(x,-p)+x^2\;. 
   \end{align}
   The rest of the right two-magnon interacting equations can be obtained by permuting the indices to the right,
   \begin{align}
       g_{i,312}(x)\longrightarrow g_{i,231}(x)\longrightarrow g_{i,123}(x)\;,\quad\quad\text{for}\quad i=1,2\;.\label{eq:one-var-sol-right}
   \end{align}
   The relationship between the generating functions, connected through index rotation, reveals how the components of permutation symmetry manifest. For each one of these one-variable generating functions $g_{i,\sigma}$, obtaining the solution from the equations \eqref{eq:sol-g1312} and \eqref{eq:sol-g2312} requires mapping the indices of momentum variables and the position-dependent corrections accordingly, $(312)\to\sigma$.\footnote{If we encounter $C_{ij}$ or $\tilde C_{ij}$ for $i>j$ after mapping the indices, then we have to exchange the order of indices just for those functions.}
   
   \begin{figure}
       \centering

\tikzset{every picture/.style={line width=0.75pt}} %set default line width to 0.75pt        

\begin{tikzpicture}[x=0.75pt,y=0.75pt,yscale=-1,xscale=1]
%uncomment if require: \path (0,224); %set diagram left start at 0, and has height of 224

%Left Right Arrow [id:dp9749204013625127] 
\draw  [fill={rgb, 255:red, 184; green, 233; blue, 134 }  ,fill opacity=0.86 ] (161.2,51.4) -- (174.8,46.8) -- (174.8,49.1) -- (202,49.1) -- (202,46.8) -- (215.6,51.4) -- (202,56) -- (202,53.7) -- (174.8,53.7) -- (174.8,56) -- cycle ;
%Left Right Arrow [id:dp8466388098971935] 
\draw  [fill={rgb, 255:red, 184; green, 233; blue, 134 }  ,fill opacity=0.86 ] (321.2,51.4) -- (334.8,46.8) -- (334.8,49.1) -- (362,49.1) -- (362,46.8) -- (375.6,51.4) -- (362,56) -- (362,53.7) -- (334.8,53.7) -- (334.8,56) -- cycle ;
%Left Right Arrow [id:dp5203158577738819] 
\draw  [fill={rgb, 255:red, 184; green, 233; blue, 134 }  ,fill opacity=0.86 ] (161.2,171.4) -- (174.8,166.8) -- (174.8,169.1) -- (202,169.1) -- (202,166.8) -- (215.6,171.4) -- (202,176) -- (202,173.7) -- (174.8,173.7) -- (174.8,176) -- cycle ;
%Left Right Arrow [id:dp504980209573849] 
\draw  [fill={rgb, 255:red, 184; green, 233; blue, 134 }  ,fill opacity=0.86 ] (321.2,171.4) -- (334.8,166.8) -- (334.8,169.1) -- (362,169.1) -- (362,166.8) -- (375.6,171.4) -- (362,176) -- (362,173.7) -- (334.8,173.7) -- (334.8,176) -- cycle ;
%Up Arrow [id:dp5499288323349212] 
\draw  [fill={rgb, 255:red, 245; green, 166; blue, 35 }  ,fill opacity=0.64 ] (264,79.82) -- (270.75,70.5) -- (277.5,79.82) -- (274.13,79.82) -- (274.13,93.8) -- (267.38,93.8) -- (267.38,79.82) -- cycle ;
%Down Arrow [id:dp3148262908454418] 
\draw  [fill={rgb, 255:red, 245; green, 166; blue, 35 }  ,fill opacity=0.62 ] (264.58,144.4) -- (268.09,144.4) -- (268.09,130.3) -- (275.13,130.3) -- (275.13,144.4) -- (278.64,144.4) -- (271.61,153.8) -- cycle ;

% Text Node
\draw (84,39.4) node [anchor=north west][inner sep=0.75pt]    {$g_{i,312}( x)$};
% Text Node
\draw (244,39.4) node [anchor=north west][inner sep=0.75pt]    {$g_{i,231}( x)$};
% Text Node
\draw (404,39.4) node [anchor=north west][inner sep=0.75pt]    {$g_{i,123}( x)$};
% Text Node
\draw (504,39.4) node [anchor=north west][inner sep=0.75pt]    {$i=1,2$};
% Text Node
\draw (84,159.4) node [anchor=north west][inner sep=0.75pt]    {$g_{i,213}( y)$};
% Text Node
\draw (244,159.4) node [anchor=north west][inner sep=0.75pt]    {$g_{i,132}( y)$};
% Text Node
\draw (404,159.4) node [anchor=north west][inner sep=0.75pt]    {$g_{i,321}( y)$};
% Text Node
\draw (504,159.4) node [anchor=north west][inner sep=0.75pt]    {$i=3,4$};
% Text Node
\draw (172.13,102.7) node [anchor=north west][inner sep=0.75pt]  [font=\scriptsize]  {${\textstyle \left\{p,x,y,\kappa ,D_{ijk}^{n,m}\right\} \rightleftarrows \left\{-p,y,x,\frac{1}{\kappa } ,D_{kji}^{m,n}\right\}}$};

\end{tikzpicture}

       \caption{\textit{We collect the solutions of the one-variable generating functions coming from the analyticity conditions and two-magnon interacting equations for each permutation. The ones that appear on the same horizontal line are related to each other under the permutation of position-dependent corrections and momentum variables. The ones which appear on the same vertical line are related to each other by changing the labels of position-dependent corrections and changing the parameters according to the prescription in the middle. These relations are also given in Figure \ref{fig:nice-relations}.}}
       \label{fig:sol-one-gen}
   \end{figure}
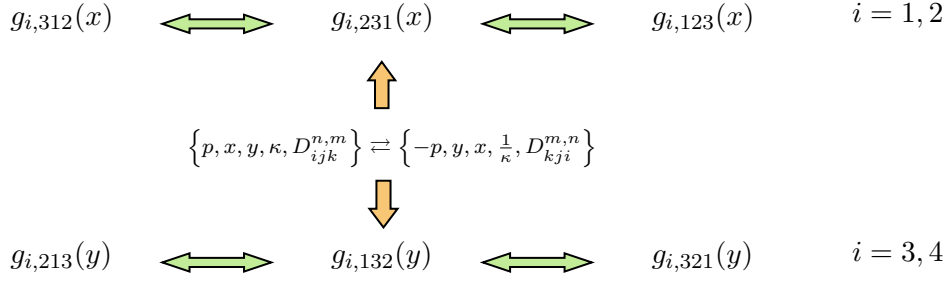
   
   Similarly, we combine the left two-magnon interacting equations with the analyticity condition for the remaining set of permutations. We choose to use the second parametrization of the elliptic curve given as $(\rho_+(y,p),y)$ together with the equation for the one-variable generating function \eqref{eq:sol-analyticity}. From those, we obtain,
   \begin{align}
       &g_{3,213}(y)=\frac{y \left(e^{-i p_1} C_{12}(y)+\PP y\tilde D_{213}^{1,0}+\invPP \rho_+(y,p)\tilde D_{213}^{0,1}\right)}{h_{12}(y,p)}\nonumber\\&\quad-\frac{e^{ip_2-ip_1}(1+2y\kappa-\EE y+y^2)}{h_{12}(y,p)}g_{3,123}(y)-\frac{e^{ip_2-ip_1}y(1+\PP y)}{h_{12}(y,p)}g_{4,123}(y)\nonumber\\&\quad-\frac{y^2(1+\PP/\rho_+(y,p))}{h_{12}(y,p)}g_{1,213}(\rho_+(y,p))+\frac{y(1+\invPP\rho_+(y,p))}{h_{12}(y,p)}g_{2,213}(\rho_+(y,p))\;,\label{eq:sol-g3213}
    \end{align}
    and,
    \begin{align}
       &g_{4,213}(y)=\frac{e^{-ip_1}\invPP \rho_+(y,p) C_{12}(y)}{ h_{12}(y,p) }-\frac{e^{ip_2-ip_1}(\invPP+y)\rho_+(y,p)}{h_{12}(y,p)}g_{4,123}(y)\nonumber\\
       &-\frac{1+y^2+2y\kappa-\EE y}{h_{12}(y,p)}\Bigg(\frac{y\tilde D_{213}^{1,0}+e^{-2i\mathcal{P}}\rho_+(y,p) \tilde D_{213}^{0,1}}{\invPP+y}+\frac{e^{ip_2-ip_1}\invPP \rho_+(y,p)}{y} g_{3,123}(y)\nonumber\\
       &-\frac{1+\invPP\rho_+(y,p)}{1+\PP y}\left(\frac{\PP y}{\rho_+(y,p)} g_{1,213}(\rho_+(y,p))-g_{2,213}(\rho_+(y,p))\right)\Bigg)\;.\label{eq:sol-g4213}
   \end{align}
   such that the common denominator is given as
   \begin{align}
       h_{12}(y,p)=1+2\kappa y-\EE y+y^2+\rho_+(y,p)\left(\invPP+y\right)\label{eq:h12}
   \end{align}
   And we can get the expressions for the solution of one-variable generating functions by permuting the indices of the expressions, \eqref{eq:sol-g3213} and \eqref{eq:sol-g4213},
   \begin{align}
       g_{i,213}(y)\longrightarrow g_{i,132}(y)\longrightarrow g_{i,321}(y)\;,\quad\quad\text{for}\quad i=3,4\;.\label{eq:one-var-sol-left}
   \end{align}

\begin{figure}
       \centering

\tikzset{every picture/.style={line width=0.75pt}} %set default line width to 0.75pt        

\begin{tikzpicture}[x=0.75pt,y=0.75pt,yscale=-1,xscale=1]
%uncomment if require: \path (0,320); %set diagram left start at 0, and has height of 320

%Straight Lines [id:da2300244569099148] 
\draw    (203.65,125.75) -- (204.46,171) ;
\draw [shift={(204.5,173)}, rotate = 268.97] [color={rgb, 255:red, 0; green, 0; blue, 0 }  ][line width=0.75]    (10.93,-3.29) .. controls (6.95,-1.4) and (3.31,-0.3) .. (0,0) .. controls (3.31,0.3) and (6.95,1.4) .. (10.93,3.29)   ;
%Straight Lines [id:da3288181446024635] 
\draw    (423.5,126.2) -- (424.46,171) ;
\draw [shift={(424.5,173)}, rotate = 268.78] [color={rgb, 255:red, 0; green, 0; blue, 0 }  ][line width=0.75]    (10.93,-3.29) .. controls (6.95,-1.4) and (3.31,-0.3) .. (0,0) .. controls (3.31,0.3) and (6.95,1.4) .. (10.93,3.29)   ;
%Straight Lines [id:da4721181436499069] 
\draw    (284.5,53) -- (230.67,88.6) ;
\draw [shift={(229,89.7)}, rotate = 326.52] [color={rgb, 255:red, 0; green, 0; blue, 0 }  ][line width=0.75]    (10.93,-3.29) .. controls (6.95,-1.4) and (3.31,-0.3) .. (0,0) .. controls (3.31,0.3) and (6.95,1.4) .. (10.93,3.29)   ;
%Straight Lines [id:da830084250402578] 
\draw    (344,53) -- (397.34,89.08) ;
\draw [shift={(399,90.2)}, rotate = 214.07] [color={rgb, 255:red, 0; green, 0; blue, 0 }  ][line width=0.75]    (10.93,-3.29) .. controls (6.95,-1.4) and (3.31,-0.3) .. (0,0) .. controls (3.31,0.3) and (6.95,1.4) .. (10.93,3.29)   ;
%Straight Lines [id:da0898994052762554] 
\draw    (404,213) -- (350.66,249.08) ;
\draw [shift={(349,250.2)}, rotate = 325.93] [color={rgb, 255:red, 0; green, 0; blue, 0 }  ][line width=0.75]    (10.93,-3.29) .. controls (6.95,-1.4) and (3.31,-0.3) .. (0,0) .. controls (3.31,0.3) and (6.95,1.4) .. (10.93,3.29)   ;
%Straight Lines [id:da7909404103918112] 
\draw    (224,213) -- (278.33,248.6) ;
\draw [shift={(280,249.7)}, rotate = 213.24] [color={rgb, 255:red, 0; green, 0; blue, 0 }  ][line width=0.75]    (10.93,-3.29) .. controls (6.95,-1.4) and (3.31,-0.3) .. (0,0) .. controls (3.31,0.3) and (6.95,1.4) .. (10.93,3.29)   ;
%Straight Lines [id:da30186828099426233] 
\draw [color={rgb, 255:red, 208; green, 2; blue, 27 }  ,draw opacity=1 ] [dash pattern={on 0.84pt off 2.51pt}]  (293.5,61) -- (469,61.69) ;
\draw [shift={(471,61.7)}, rotate = 180.23] [color={rgb, 255:red, 208; green, 2; blue, 27 }  ,draw opacity=1 ][line width=0.75]    (10.93,-3.29) .. controls (6.95,-1.4) and (3.31,-0.3) .. (0,0) .. controls (3.31,0.3) and (6.95,1.4) .. (10.93,3.29)   ;
%Straight Lines [id:da03502847072336035] 
\draw [color={rgb, 255:red, 74; green, 78; blue, 226 }  ,draw opacity=1 ] [dash pattern={on 0.84pt off 2.51pt}]  (298.5,280.5) -- (474,281.19) ;
\draw [shift={(476,281.2)}, rotate = 180.23] [color={rgb, 255:red, 74; green, 78; blue, 226 }  ,draw opacity=1 ][line width=0.75]    (10.93,-3.29) .. controls (6.95,-1.4) and (3.31,-0.3) .. (0,0) .. controls (3.31,0.3) and (6.95,1.4) .. (10.93,3.29)   ;
%Straight Lines [id:da9175752563557061] 
\draw [color={rgb, 255:red, 208; green, 2; blue, 27 }  ,draw opacity=1 ] [dash pattern={on 0.84pt off 2.51pt}]  (204,211) -- (514,210.21) ;
\draw [shift={(516,210.2)}, rotate = 179.85] [color={rgb, 255:red, 208; green, 2; blue, 27 }  ,draw opacity=1 ][line width=0.75]    (10.93,-3.29) .. controls (6.95,-1.4) and (3.31,-0.3) .. (0,0) .. controls (3.31,0.3) and (6.95,1.4) .. (10.93,3.29)   ;
%Straight Lines [id:da4350353394218165] 
\draw [color={rgb, 255:red, 74; green, 78; blue, 226 }  ,draw opacity=1 ] [dash pattern={on 0.84pt off 2.51pt}]  (203.65,125.75) -- (515.65,127.44) ;
\draw [shift={(517.65,127.45)}, rotate = 180.31] [color={rgb, 255:red, 74; green, 78; blue, 226 }  ,draw opacity=1 ][line width=0.75]    (10.93,-3.29) .. controls (6.95,-1.4) and (3.31,-0.3) .. (0,0) .. controls (3.31,0.3) and (6.95,1.4) .. (10.93,3.29)   ;
%Straight Lines [id:da5459454960484118] 
\draw [color={rgb, 255:red, 126; green, 211; blue, 33 }  ,draw opacity=1 ] [dash pattern={on 4.5pt off 4.5pt}]  (252.4,137.33) -- (368.6,176.07) ;
\draw [shift={(370.5,176.7)}, rotate = 198.43] [color={rgb, 255:red, 126; green, 211; blue, 33 }  ,draw opacity=1 ][line width=0.75]    (10.93,-3.29) .. controls (6.95,-1.4) and (3.31,-0.3) .. (0,0) .. controls (3.31,0.3) and (6.95,1.4) .. (10.93,3.29)   ;
\draw [shift={(250.5,136.7)}, rotate = 18.43] [color={rgb, 255:red, 126; green, 211; blue, 33 }  ,draw opacity=1 ][line width=0.75]    (10.93,-3.29) .. controls (6.95,-1.4) and (3.31,-0.3) .. (0,0) .. controls (3.31,0.3) and (6.95,1.4) .. (10.93,3.29)   ;
%Straight Lines [id:da33628025317266896] 
\draw [color={rgb, 255:red, 126; green, 211; blue, 33 }  ,draw opacity=1 ] [dash pattern={on 4.5pt off 4.5pt}]  (368.6,137.33) -- (252.4,176.07) ;
\draw [shift={(250.5,176.7)}, rotate = 341.57] [color={rgb, 255:red, 126; green, 211; blue, 33 }  ,draw opacity=1 ][line width=0.75]    (10.93,-3.29) .. controls (6.95,-1.4) and (3.31,-0.3) .. (0,0) .. controls (3.31,0.3) and (6.95,1.4) .. (10.93,3.29)   ;
\draw [shift={(370.5,136.7)}, rotate = 161.57] [color={rgb, 255:red, 126; green, 211; blue, 33 }  ,draw opacity=1 ][line width=0.75]    (10.93,-3.29) .. controls (6.95,-1.4) and (3.31,-0.3) .. (0,0) .. controls (3.31,0.3) and (6.95,1.4) .. (10.93,3.29)   ;
%Straight Lines [id:da09760949787029904] 
\draw [color={rgb, 255:red, 126; green, 211; blue, 33 }  ,draw opacity=1 ] [dash pattern={on 4.5pt off 4.5pt}]  (310.5,133.7) -- (310.5,179.7) ;
\draw [shift={(310.5,181.7)}, rotate = 270] [color={rgb, 255:red, 126; green, 211; blue, 33 }  ,draw opacity=1 ][line width=0.75]    (10.93,-3.29) .. controls (6.95,-1.4) and (3.31,-0.3) .. (0,0) .. controls (3.31,0.3) and (6.95,1.4) .. (10.93,3.29)   ;
\draw [shift={(310.5,131.7)}, rotate = 90] [color={rgb, 255:red, 126; green, 211; blue, 33 }  ,draw opacity=1 ][line width=0.75]    (10.93,-3.29) .. controls (6.95,-1.4) and (3.31,-0.3) .. (0,0) .. controls (3.31,0.3) and (6.95,1.4) .. (10.93,3.29)   ;

% Text Node
\draw (294.5,35.4) node [anchor=north west][inner sep=0.75pt]    {$D_{123}^{n,m}$};
% Text Node
\draw (186.5,96.4) node [anchor=north west][inner sep=0.75pt]    {$D_{213}^{n,m}$};
% Text Node
\draw (187.8,181.7) node [anchor=north west][inner sep=0.75pt]    {$D_{231}^{n,m}$};
% Text Node
\draw (297.7,250.9) node [anchor=north west][inner sep=0.75pt]    {$D_{321}^{n,m}$};
% Text Node
\draw (406,96.4) node [anchor=north west][inner sep=0.75pt]    {$D_{132}^{n,m}$};
% Text Node
\draw (405.8,182) node [anchor=north west][inner sep=0.75pt]    {$D_{312}^{n,m}$};
% Text Node
\draw (379.5,41) node [anchor=north west][inner sep=0.75pt]   [align=left] {{\fontfamily{pcr}\selectfont {\scriptsize even permutations}}};
% Text Node
\draw (384.5,258) node [anchor=north west][inner sep=0.75pt]   [align=left] {{\fontfamily{pcr}\selectfont {\scriptsize odd permutations}}};
% Text Node
\draw (453,99) node [anchor=north west][inner sep=0.75pt]   [align=left] {{\fontfamily{pcr}\selectfont {\scriptsize odd permutations}}};
% Text Node
\draw (449,186) node [anchor=north west][inner sep=0.75pt]   [align=left] {{\fontfamily{pcr}\selectfont {\scriptsize even permutations}}};
% Text Node
\draw (287.5,146) node [anchor=north west][inner sep=0.75pt]   [align=left] {{\scriptsize {\fontfamily{pcr}\selectfont reflections}}};

\end{tikzpicture}

       \caption{\textit{We organize the permutations of position-dependent corrections according to the order of scattering events. Starting from the identity permutation, $(123)$ we permute the indices by transpositions. In this case the relations between the various position-dependent corrections coming from the solution of one-variable generating functions are marked by colored arrows. Each color refers to a distinct element of the broken permutation group. These particular transformations also appear in Figure \ref{fig:sol-one-gen}. }}
       \label{fig:nice-relations}
   \end{figure}
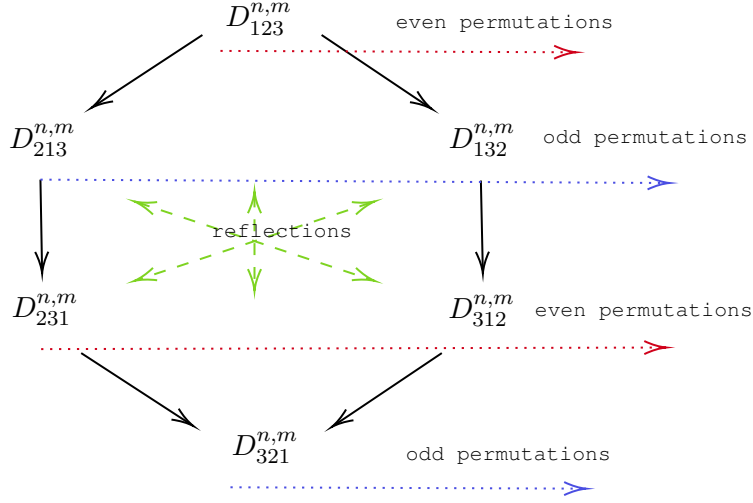

   Note that the expressions for the functions $g_{i,\sigma}$ do not spoil the analytical properties of the overall generating function \eqref{eq:g-gamma} at the origin.
   Because the denominator does not vanish for the chosen ranges of the momentum parameters. Therefore, we can insert the solution of the one-variable generating functions which are listed in \eqref{eq:one-var-sol-right} and \eqref{eq:one-var-sol-left} to the corresponding generating functions of the form \eqref{eq:generating-non-int} and obtain a final form of the generating functions for each permutation which solve the second class of equations. As depicted in Figure \ref{fig:sol-one-gen}, the final form of the solution of the second class of equations coming from the eigenvalue problem, \eqref{eq:secondclass}, admits a symmetric solution which is exactly the relation observed between the right and the left two-magnon interacting functions, \eqref{eq:two-body-rel}. However, this means that the particular relations between the permutations are not as straight forward as the usual CBA solution of rational or trigonometric models.\footnote{In CBA, we observe that the scattering coefficients factorize in terms of two-body scattering coefficients. Therefore, we can understand the set of scattering coefficients as a representation of the symmetric group $\mathcal{S}_N$.} Instead, we observe that the six permutations are divided into two sets. Each of the even (odd) permutations can be described by a common form for the one-variable generating functions, and the odd (even) permutations are obtained from reflections together with the transformations given in \eqref{eq:two-body-rel}. Hence, this division can be described in a unique way, such that knowing the solution of the one-variable generating function for one permutation is enough to generate all. The resulting form of the underlying permutation symmetry can be seen in Figure \ref{fig:nice-relations}.

   At this point, we can count the number of unknown coefficients in terms of the one-variable generating functions. After solving the non-interacting equations we obtain the generating function, \eqref{eq:generating-non-int} which has four undetermined one-variable functions for each of the six permutations. Furthermore, the analyticity condition given in \eqref{eq:sol-analyticity} brings six relations. Finally, the recurrence relations coming from the two-body interacting equations, \eqref{eq:two-body-rec1} and \eqref{eq:two-body-rec2} bring six more relations. Combining all these relations reduces the number of one-variable functions from twenty-four to twelve.

   After solving the first class of equations and using them as the initial conditions that determine the functions, $C_{ij}$ and $\tilde C_{ij}$ from equations \eqref{eq:const12} and \eqref{eq:const21} we end up with a generating function which encodes all the position-dependent corrections that solve the three-magnon eigenvalue problem. In terms of the final form of the generating functions each position-dependent correction is given in integral form,
   \begin{align}
       D_\sigma^{n,m}=-\oint \frac{dx dy}{4\pi^2}\frac{G_\sigma(x,y)}{x^{n+1}y^{m+1}}\;. \label{eq:integral}
   \end{align}
   We proceed to discuss the solution of the first class of equations which completes the solution of the three-magnon relation.
   \subsubsection{Solving the Equations for Minimum Separation}
   
   The last piece of the solution of the three-magnon eigenvalue problem is obtained by solving the equations given in \eqref{eq:firstclass}. These expressions contain the scattering coefficients as well as the position-dependent corrections with the shortest separation of magnons. The two-body interacting equations, when the third magnon is just one $\phi$ state away, are given by the following expressions and their permutations. We write the equations,
   \begin{align}
       \beta_{l,(ij)}(1)=0\;,\quad\text{and}\quad \beta_{r,(ij)}(1)=0\;,\label{eq:beta1vanish}
   \end{align}
   by isolating one of the scattering coefficients. The following equations correspond to the solution of the two-body interacting equations for which the interacting pair of magnons are on the left of the third magnon, separated by one $\phi$. Already from the first set of equations we can explicitly observe how the relations \eqref{eq:naive-sol-left} coming from the CBA are corrected,
   \begin{align}
       &A_{213}=S_\kappa(p_1,p_2)A_{123}+\frac{1}{a(p_1,p_2,\kappa)}\left[\left(e^{ip_1}\tilde D_{213}^{0,2} +e^{ip_2}\tilde D_{123}^{0,2}\right)+\left(e^{ip_1}\tilde D_{213}^{1,1}+e^{ip_2}\tilde D_{123}^{1,1}\right)\right.\nonumber\\&\quad\left.-\left(\EE-2\kappa \right)\left(e^{ip_1}\tilde D_{213}^{0,1} +e^{ip_2}\tilde D_{123}^{0,1}\right)+\PP \left(e^{ip_1}\tilde D_{213}^{1,0}+\tilde D_{123}^{1,0}\right)\right]\;,\label{eq:A213}\\
       &A_{312}=S_\kappa(p_1,p_3)A_{132}+\frac{1}{a(p_1,p_3,\kappa)}\left[\left(e^{ip_1}\tilde D_{312}^{0,2} +e^{ip_3}\tilde D_{132}^{0,2}\right)+\left(e^{ip_1}\tilde D_{312}^{1,1}+e^{ip_3}\tilde D_{132}^{1,1}\right)\right.\nonumber\\&\quad\left.-\left(\EE-2\kappa \right)\left(e^{ip_1}\tilde D_{312}^{0,1} +e^{ip_3}\tilde D_{132}^{0,1}\right)+\PP \left(e^{ip_1}\tilde D_{312}^{1,0}+\tilde D_{132}^{1,0}\right)\right]\;,\label{eq:A312}\\
       &A_{321}^L=S_\kappa(p_2,p_3)A_{231}+\frac{1}{a(p_2,p_3,\kappa)}\left[\left(e^{ip_2}\tilde D_{321}^{0,2} +e^{ip_3}\tilde D_{231}^{0,2}\right)+\left(e^{ip_2}\tilde D_{321}^{1,1}+e^{ip_3}\tilde D_{321}^{1,1}\right)\right.\nonumber\\&\quad\left.-\left(\EE-2\kappa \right)\left(e^{ip_2}\tilde D_{321}^{0,1} +e^{ip_3}\tilde D_{231}^{0,1}\right)+\PP \left(e^{ip_2}\tilde D_{321}^{1,0}+\tilde D_{231}^{1,0}\right)\right]\;.\label{eq:A321L}
   \end{align}
   Here, the function, $a(p_1,p_2,\kappa)=1+e^{ip_1+ip_2}-2\kappa e^{ip_1}$ is the numerator of the scattering coefficient given in \eqref{eq:scatdef}.  Furthermore, independent of the left two-body interacting equations, we also solve the right two-body interacting equations. Notice that these two sets of equations include the same scattering coefficients and position-dependent corrections but paired in two different ways. Under the assumption that $A_{321}^L$ is different from $A_{321}^R$ we obtain,
   \begin{align}
       &A_{132}=S_{1/\kappa}(p_2,p_3)A_{123}+\frac{1}{a(p_2,p_3,\kappa^{-1})}\left[\left(e^{ip_2}\tilde D_{132}^{2,0}+e^{ip_3}\tilde D_{123}^{2,0}\right)+\left(e^{ip_2}\tilde D_{132}^{1,1}+e^{ip_3}\tilde D_{123}^{1,1}\right)\right.\nonumber\\&-\left.\left(\EE-2\kappa^{-1}\right)\left(e^{ip_2}\tilde D_{132}^{1,0}+e^{ip_3}\tilde D_{123}^{1,0}\right)+\invPP\left(e^{ip_2}\tilde D_{132}^{0,1}+e^{ip_3}\tilde D_{123}^{0,1}\right)\right]\;,\label{eq:A132}\\
       &A_{231}=S_{1/\kappa}(p_1,p_3)A_{213}+\frac{1}{a(p_1,p_3,\kappa^{-1})}\left[\left(e^{ip_1}\tilde D_{231}^{2,0}+e^{ip_3}\tilde D_{213}^{2,0}\right)+\left(e^{ip_1}\tilde D_{231}^{1,1}+e^{ip_3}\tilde D_{213}^{1,1}\right)\right.\nonumber\\&-\left.\left(\EE-2\kappa^{-1}\right)\left(e^{ip_1}\tilde D_{231}^{1,0}+e^{ip_3}\tilde D_{213}^{1,0}\right)+\invPP\left(e^{ip_1}\tilde D_{231}^{0,1}+e^{ip_3}\tilde D_{213}^{0,1}\right)\right]\;,\label{eq:A231}\\
       &A_{321}^R=S_{1/\kappa}(p_1,p_2)A_{312}+\frac{1}{a(p_1,p_2,\kappa^{-1})}\left[\left(e^{ip_1}\tilde D_{321}^{2,0}+e^{ip_2}\tilde D_{312}^{2,0}\right)+\left(e^{ip_1}\tilde D_{321}^{1,1}+e^{ip_2}\tilde D_{312}^{1,1}\right)\right.\nonumber\\&-\left.\left(\EE-2\kappa^{-1}\right)\left(e^{ip_1}\tilde D_{321}^{1,0}+e^{ip_2}\tilde D_{312}^{1,0}\right)+\invPP\left(e^{ip_1}\tilde D_{321}^{0,1}+e^{ip_2}\tilde D_{312}^{0,1}\right)\right]\;.\label{eq:A321R}
   \end{align}
   We specifically labeled the scattering coefficients $A_{321}$ coming from the left and the right equations as $A_{321}^L$ and $A_{321}^R$ to emphasize that at least one of the position-dependent corrections in these equations must solve the equation,
   \begin{align}
       A_{321}^L=A_{321}^R\;,\label{eq:leftright}
   \end{align}
   for the two sets of equations to become compatible. It is important to realize that a YBE is hidden in this particular equation. More explicitly, the above equation can be brought into the form,
   \begin{align}
       \frac{A_{321}^L}{A_{231}}\frac{A_{231}}{A_{213}}\frac{A_{213}}{A_{123}}A_{123}=\frac{A_{321}^R}{A_{312}}\frac{A_{312}}{A_{132}}\frac{A_{132}}{A_{123}}A_{123}\;.\label{eq:firstYBE}
   \end{align}
   Since the two-magnon interacting equations are deformed in a non-trivial way, we can find solutions to this equation.  However, this is not the only YBE that arises in the solution of the long-range ansatz. We obtain a tower of YBEs for the position-dependent corrections for any configuration. We will postpone the discussion on this topic until we revisit the three-magnon state from the context of Yang operators in Section \ref{sec:Yang}.

\bigskip
   
   Finally, the three-body interacting equation brings one more constraint on the closest position-dependent corrections,
   \begin{align}
       \alpha(p_1,p_2,p_3)=0 \, .\quad\;\label{eq:alpha_vanish}
   \end{align}
   In the usual integrable spin chains which admit CBA, solving the two-magnon interacting equations automatically solves the three-magnon equation. However, after substituting the solutions \eqref{eq:A213}-\eqref{eq:leftright} into $\alpha(p_1,p_2,p_3)$, it does not vanish in our case. Since in both equations \eqref{eq:leftright} and \eqref{eq:alpha_vanish} the position-dependent corrections $D_\sigma^{1,0}$, $D_\sigma^{0,1}$, $D_\sigma^{2,0}$, $D_\sigma^{0,2}$, $D_\sigma^{1,1}$, for all $\sigma\in\mathcal{S}_3$, appear, we can just use them to solve for any two of these coefficients. This would be a highly non-symmetric way of solving these equations, but it would also be the most general way to obtain a solution of long-range ansatz. Here, we avoid giving the most general solution explicitly because we think that it does not offer any particular insight. Solving these equations in a $\mathcal{S}_3$ symmetric way in its whole generality is beyond our reach. Even more importantly, we do not want to restrict ourselves by looking for a fully manifest permutation symmetry, as it might be broken, twisted, or deformed. Instead, in the next section we present a special sub-set of three-magnon solutions for which we impose extra conditions to solve the eigenvalue problem fully and symmetrically.

   The undetermined position-dependent corrections can be packaged into twelve of the one-variable generating functions given in \eqref{eq:one-variable1}-\eqref{eq:one-variable4}. This freedom comes from the fact that we did not specify any boundary condition for our open spin chains and considered infinite length throughout this work. However, the counting of the coefficients suggests that imposing boundary conditions and considering finite-length spin chains would not cause the linear system of equations to be over-determined. Since the eigenstate with three magnons is not closable, it does not give us a gauge invariant operator in the gauge theory side, as discussed in \eqref{eq:gaugeop1}-\eqref{eq:gaugeop3}. Here, we only briefly discuss the possible ways to use this solution for finite open spin chains on the level of generating functions.

   The possible boundary conditions for open spin chains with three excited states can be summarized with the following analytic functions as these contain the undetermined coefficients and this leftover freedom would naturally be fixed after imposing boundary conditions,
   \begin{align}
       &g_{i,\sigma}(y)\quad\text{for}\quad \sigma\in\{(123),(231),(312)\}\quad\text{and}\quad i=3,4\label{eq;boundary1}\\
       &g_{i,\sigma}(x)\quad\text{for}\quad \sigma\in\{(213),(132),(321)\}\quad\text{and}\quad i=1,2\label{eq;boundary2}
   \end{align}
   
   Moreover, if we want to preserve the relations among the generating functions which are summarized in Figure \ref{fig:nice-relations}, then we only need to fix two of the generating functions, for example, $g_{i,123}(y)$ for $i=3,4$. Then the rest of them, which are listed in \eqref{eq;boundary1} and \eqref{eq;boundary2}, are determined as is prescribed for the solution of the two-magnon interacting equations and analyticity in the previous section.

  As we discuss in Section \ref{sec:intro-model}, our spin chain captures the spectral problem of single trace operators in the gauge theory side and thus the appropriate boundary conditions are periodic (or twisted).
Given the fact that the magnons, $Q_{12}$ and $Q_{21}$, are in the bifundamental representation of the color group,
 the three magnons eigenvectors cannot be closed and thus we cannot impose physically meaningful boundary conditions on the three-magnon states. That is possible only for an even number of magnons.
  To deal with this, we use our intuition coming from the symmetries of the physical problem to further fix
  this extra freedom and to obtain interesting special solutions of the three-magnon problem.

\section{Special Solutions}
\label{sec:special}

In this section, we present three special solutions to the three-magnon problem contained in the general solution presented in the previous section. Our approach deviates slightly from the one detailed in subsection \ref{sec:solution}. Rather than determining the scattering coefficients $A_\sigma$ through the equations \eqref{eq:firstclass}, we fix the scattering coefficients of the long-range ansatz according to the scattering coefficients arise from the usual CBA, in particular, we use either \eqref{eq:naive-sol-left} or \eqref{eq:naive-sol-right}, and use the position-dependent corrections to compensate for our choices. We fix all undetermined position-dependent corrections, as listed in \eqref{eq;boundary1} and \eqref{eq;boundary2}, to obtain solutions with factorized scattering coefficients.  The existence of these solutions makes the connection between the Temperly-Lieb model and our dynamical spin chain clearer, as it allows us to explicitly identify the corrections required for the scattering coefficients for each magnon configuration. Collectively, the position-dependent corrections should be the analog of the shifts of the dynamical parameter in a DYBE \cite{Felder:1994be,Felder:1996xym}.

We also consider starting from the scattering coefficients of the XXX model and compute the explicit corrections needed to solve this model. This particular solution not only indicates yet another direct connection between our model and the XXX model, it also enables comparison between two possible ways to obtain the dynamical spin chain model. Furthermore, in all of these special solutions, it is easy to see that the corrections always come with overall $(1-\kappa)$ factors which imply that the orbifold point limit, $\kappa\to1$ takes us back to the CBA for the $\mathbb Z_2$ orbifold of $\mathcal{N}=4$ SYM. Although these particular choices we present in the next three subsections, explicitly break the combination of parity and $\mathbb{Z}_2$ symmetry, we can easily obtain a solution that is invariant under the $\mathbb Z_2\mathcal{P}$ operator by combining these solutions we present.

By explicitly computing these corrections, we reveal the intricate relationship between the scattering processes and the underlying algebraic structures. Combining the special solutions that were presented here with the general solution of the second class of equations given in the previous section \eqref{eq:secondclass}, reveals the overall structure of the solution. Then we can interpret the position-dependent corrections as the corrections to plane wave basis more explicitly.

   \subsection{$S_{\kappa}$ Solution}
   \label{subsec:skappa}

For the first special solution, we consider scattering coefficients that are factorized in terms of $S_\kappa(p_1,p_2)$, given in \eqref{eq:scat12}. Choosing this form for the scattering coefficients implies a preference for the left-interacting equations of the naive CBA, as given in \eqref{eq:naive-sol-left},
\begin{align}
    &\frac{A_{213}}{A_{123}}=S_{\kappa}(p_1,p_2)\;,\quad  \frac{A_{312}}{A_{132}}=S_{\kappa}(p_1,p_3)\;,\quad\frac{A_{321}}{A_{231}}=S_{\kappa}(p_2,p_3)\label{eq:skappafac} \\ &\frac{A_{132}}{A_{123}}=S_{\kappa}(p_2,p_3)\;,\quad
    \frac{A_{231}}{A_{213}}=S_{\kappa}(p_1,p_3)\;,\quad 
    \frac{A_{321}}{A_{312}}=S_{\kappa}(p_1,p_2)\;.\label{eq:skappafac-corr}
\end{align}
The scattering coefficients \eqref{eq:skappafac} make the scattering factors disappear from the left two-magnon interacting equations, $\beta_l (1)$, but the scattering coefficients \eqref{eq:skappafac-corr} do not simplify the ones from the right, $\beta_r (1)$.

Before solving the equations \eqref{eq:firstclass}, we impose additional relations between the position-dependent corrections inspired by the CBA. In particular, for any $n,m\geq 0$, we have
\begin{align}
    &\tilde D_{213}^{n,m}= S_{\kappa}(p_1,p_2)  \tilde{D}_{123}^{n,m} \; ,\label{partial-fac2}  \\
    &\tilde D_{312}^{n,m}= S_{\kappa}(p_1,p_3)  \tilde{D}_{132}^{n,m}\label{partial-fac4} \; ,\\
    &\tilde D_{321}^{n,m}= S_{\kappa}(p_2,p_3)\tilde{D}_{231}^{n,m} \label{partial-fac6}\;. 
\end{align}
This implies that the position-dependent corrections are related by the appropriate scattering coefficients, $S_\kappa$, if their indices are related by a transposition of the first two indices, $ijk$ and $jik$. The fact that now the position-dependent corrections are \emph{partially-factorized} helps uncover the physics of this problem. In addition, we also fix the normalization of the state by setting
\begin{align}
    A_{123}=1\;.\label{eq:normalization}
\end{align}
Then, we first consider the three-magnon interacting equation, given in \eqref{eq:alpha}. We take advantage of the observation that the scattering coefficients and the position-dependent corrections share the same label and thus have a common exponential of momenta factor, such as $e^{ip_i+2ip_j}$. This allows us to solve each of these parts separately,
\begin{align}
    \left(\tilde D_{123}^{0,1}+\tilde D_{123}^{1,0}\right)&=2\left(\kappa-\frac{1}{\kappa}\right)\;.\label{eq:Dk10011}\\
    \left(\tilde D_{132}^{0,1}+\tilde D_{132}^{1,0}\right)&=2\left(\kappa-\frac{1}{\kappa}\right)S_\kappa(p_2,p_3)\;,\label{eq:Dk10012}\\
   \left(\tilde D_{231}^{0,1}+\tilde D_{231}^{1,0}\right)&=2\left(\kappa-\frac{1}{\kappa}\right)S_\kappa(p_1,p_3)S_\kappa(p_1,p_2)\;.\label{eq:Dk10013}
\end{align}
We obtain the same structure of scattering coefficients as one might infer from the naive CBA. However, it is important to emphasize that there is no intrinsic reason to justify this choice of coefficients beyond inspiration from the CBA and the fact that they satisfy the three-magnon interacting equation \eqref{eq:alpha}. Before moving to the two-magnon interacting equations for the shortest separation, we fix all the remaining freedom unconstrained by \eqref{eq:alpha} by choosing
\begin{align}
    \tilde D_\sigma^{1,1}=0\;,\quad\quad \tilde D_\sigma^{1,0}=\tilde D_\sigma^{0,1}\;.\label{eq:morechoices}
\end{align}
The two-body interacting equations from the left can be used to fix the coefficients $D^{0,2}_\sigma$,
\begin{align}
       \tilde D_{123}^{0,2}&=\left(\kappa-\frac{1}{\kappa}\right)\left(\EE-\PP -2\kappa\right) \;,\label{eq:Dk021}\\
       \tilde D_{132}^{0,2}&= S_\kappa(p_2,p_3) \left(\kappa-\frac{1}{\kappa}\right)\left(\EE-\PP -2\kappa\right) \;,\label{eq:Dk022}\\
       \tilde D_{231}^{0,2} &=S_\kappa(p_1,p_3)S_\kappa(p_1,p_2) \left(\kappa-\frac{1}{\kappa}\right)\left(\EE-\PP -2\kappa\right)  \label{eq:Dk023}\;.
   \end{align}
    Similarly, the two-body interacting equations from the right can be used to fix the coefficients $D^{2,0}_\sigma$. These equations give us the same structure with scattering coefficients as in  $D^{0,2}_\sigma$, but with a more involved overall factor. 
   \begin{align}
       \tilde D_{123}^{2,0} &=\frac{1-\kappa^2}{\kappa^2} f_{20}(p_3) \;,\label{eq:Dk201}\\
       \tilde D_{132}^{2,0} &=\frac{1-\kappa^2}{\kappa^2} S_\kappa(p_2,p_3) f_{20}(p_2)\;,\label{eq:Dk202} \\
       \tilde D_{231}^{2,0}&=\frac{1-\kappa^2}{\kappa^2} S_\kappa(p_1,p_3)S_\kappa(p_1,p_2) f_{20}(p_1)\;,\label{eq:Dk203} 
   \end{align}
   such that
   \begin{align}
       f_{20}(p_3)=\left[e^{-i p_3} \left(\EE+\PP +\invPP -2 \kappa \right) -\kappa \left(3\EE-2 \invPP  -4 \kappa -\frac{2}{\kappa} \right)\right]\;.
   \end{align}
   The reason why the solutions for the two-body interacting equations from the right are more complicated is that we fix the freedom of our coefficients using $S_\kappa$ factors, which are suited for solving the structure of the two-body interacting equations from the left but not for the two-body interacting equations from the right.

   We have now extracted all the information we can from the first class of equations, and we proceed to consider the second class of equations. As in the previous section, they are solved by imposing all the recurrence relations to the generating function \eqref{eq:generating-non-int} and ensuring analyticity. This last step can be performed by computing the resulting form of the functions, \eqref{eq:const12} and \eqref{eq:const21} under the special choices we made in this subsection. Using \eqref{partial-fac2}, we get
   \begin{equation}
       \left(e^{ip_1}\tilde D_{213}^{n,m}+e^{ip_2}\tilde D_{123}^{n,m}\right)=\omega(p_1,p_2,\kappa)\tilde D^{n,m}_{123} \;,
   \end{equation}
   where we define
   \begin{align}
       \omega(p_1,p_2,\kappa)=\frac{ (e^{ip_2}-e^{ip_1})(1+e^{ip_1+ip_2})}{1+e^{ip_1+ip_2}-2\kappa e^{ip_1}} \;.\label{eq:omega}
   \end{align}
   We combine this piece of observation with the resulting form of the functions which are defined in \eqref{eq:const12},
    \begin{align}
       &C_{12}(y)=\left(\kappa-\frac{1}{\kappa}\right)\left(1-y\PP\right)\omega(p_1,p_2,\kappa)\;,\label{eq:skappaC12}
   \end{align}
    and \eqref{eq:const21},
   \begin{align}
       &\tilde C_{12}(x)=\left(\kappa-\frac{1}{\kappa}\right)\left(1- \frac{x(\EE - \frac{2}{\kappa}-xe^{-ip_3}\invPP+\frac{f_{20}(p_3)}{\kappa})}{1-e^{-ip_3}x}\right)S_\kappa(p_2,p_3)S_\kappa(p_1,p_3)\omega(p_1,p_2,\kappa)\;.\label{eq:skappaCC12}
   \end{align}
   
   The common form of these expressions including factorized scattering coefficients and the permutation relations derived in Section \ref{subsubsec:two-magnons} and depicted in Figure \ref{fig:sol-one-gen} are enough to show that the factorization will continue to hold for higher position-dependent corrections. In addition to the factorized scattering coefficients, we will observe the corrections which are related to each other as observed in Section \ref{subsubsec:two-magnons}. Hence, for any $n,m\geq1$, we have
\begin{align}
\label{eqn:SpecialDs}
    D_{123}^{n,m}&=f(p_1,p_2,p_3;n,m;\kappa)\\
    D_{132}^{n,m}&=S_\kappa(p_2,p_3)f(p_1,p_3,p_2;n,m;\kappa)\\
    D_{213}^{n,m}&=S_\kappa(p_1,p_2)f(p_2,p_1,p_3;n,m;\kappa)\\
    D_{231}^{n,m}&=S_\kappa(p_1,p_3)S_\kappa(p_1,p_2)f(p_2,p_3,p_1;n,m;\kappa)\\
    D_{312}^{n,m}&=S_\kappa(p_1,p_3)S(p_2,p_3)f(p_3,p_1,p_2;n,m;\kappa)\\
    D_{321}^{n,m}&=S_{\kappa}(p_2,p_3)S_\kappa(p_1,p_3)S_\kappa(p_1,p_2)f(p_3,p_2,p_1;n,m;\kappa)\label{eqn:SpecialDs1}
\end{align}
such that $f(p_1,p_2,p_3;n,m;\kappa)$ is defined from the final form of the generating function $G_{123}(x,y)$,
\begin{align}
    f(p_1,p_2,p_3;n,m,\kappa)=-\oint_\Sigma \frac{dx dy}{4\pi^2}\frac{G_{123}(x,y)}{x^{n+1}y^{m+1}}\;,
\end{align}
where $\Sigma$ is a small ball around the origin. The relation between the various permutations of this function is given in Figure \ref{fig:nice-relations}. Under these choices of scattering coefficients and the position-dependent corrections for the shortest separation, we obtain a solution of the three-magnon state that satisfies the eigenvalue problem \eqref{eq:eigthree}. We are still left with three one-variable generating functions that codify many undetermined coefficients. We can absorb undetermined coefficients into $g_{2,123}$, $g_{2,132}$ and $g_{2,231}$, then setting all these coefficients to zero would give us a non-trivial solution with the structure laid down in equations \eqref{eqn:SpecialDs}-\eqref{eqn:SpecialDs1}. In Appendix \ref{sec:G123}, we elaborate on the details of combining the special solution of the first class of equations with the second class of equations. We label this solution as $\ket{\xi(p_1,p_2,p_3)}_\kappa$ and it has the following form,
   \begin{align}
     \ket{\xi(p_1,p_2,p_3)}_\kappa=\sum_{l_1<l_2<l_3}\left(\sum_{\sigma\in\mathcal{S}_3}\;A_\sigma\;\left(1+f(p_{\sigma(1)},p_{\sigma(2)},p_{\sigma(3)};n,m;\kappa)\right) e^{i\Vec{p}_{\sigma}\cdot \Vec{l}}\right)\ket{l_1,l_2,l_3}
   \end{align}
   Here, the separation between the scattering coefficients and position-dependent corrections becomes clearer. We can interpret the terms accompanying the plane wave corrections as explicit modifications to the plane wave basis. It is important to note that this eigenvector, by itself, does not remain invariant under the operator $\mathbb Z_2\mathcal{P}$. However, by combining it with the special solution presented in the next section, we obtain states that satisfy the relations given in  \eqref{eq:parity-identity}.

   \subsection{$S_{1/\kappa}$ Solution}
   \label{subsec:skappainverse}

   Although the solution we have presented in the previous subsection is a complete solution of the eigenvalue problem \eqref{eq:eigthree}, it is not invariant under the combination $\mathbb Z_2\mathcal{P}$ because it does not fulfill the conditions presented in \eqref{eq:parity-identity}. Here we construct another special solution with scattering coefficients that are factorized in terms of $S_{1/\kappa}(p_i,p_j)$ with the purpose of finding the $\mathbb Z_2\mathcal{P}$ conjugate of the previous solution.

    Similarly to the previous special solution, we can construct a special solution where the scattering coefficients are factorised in terms of $S_{1/\kappa}$ and agree with the ones predicted by the naive extension of the CBA \eqref{eq:naive-sol-left},
\begin{align}
    &\frac{A_{213}}{A_{123}}=S_{1/\kappa}(p_1,p_2)\;,\quad  \frac{A_{312}}{A_{132}}=S_{1/\kappa}(p_1,p_3)\;,\quad\frac{A_{321}}{A_{231}}=S_{1/\kappa}(p_2,p_3)\label{eq:sinversekappafac} \\ &\frac{A_{132}}{A_{123}}=S_{1/\kappa}(p_2,p_3)\;,\quad
    \frac{A_{231}}{A_{213}}=S_{1/\kappa}(p_1,p_3)\;,\quad 
    \frac{A_{321}}{A_{312}}=S_{1/\kappa}(p_1,p_2)\;.\label{eq:sinversekappafac-corr}
\end{align}
In this case, the scattering coefficients \eqref{eq:sinversekappafac} make the scattering factors disappear from the right two-magnon interacting equations, $\beta_r (1)$, but the scattering coefficients \eqref{eq:sinversekappafac-corr}, do not simplify the ones from the left, $\beta_l (1)$.  
%In this case we make the opposite choice compared to the previous special solution. The factors of the scattering coefficient \eqref{eq:sinversekappafac} agree with the scattering coefficients predicted by the CBA, given in \eqref{eq:naive-sol-right}. However, the factors that appear in \eqref{eq:sinversekappafac-corr} disagree with the equations \eqref{eq:naive-sol-left}. Again, the difference between this set of scattering coefficients and the ones predicted by the CBA will be compensated by the contact term contributions.

Before solving the equations coming from the three-body eigenvalue problem, we impose relations between the position-dependent corrections in the following way. For any $n,m\geq 0$ we impose
\begin{align}
    &\tilde D_{123}^{n,m}= S_{1/\kappa}(p_3,p_2)  \tilde{D}_{132}^{n,m} \;, \label{eq:1/kpartial-fac1} \\
    &\tilde D_{213}^{n,m}= S_{1/\kappa}(p_3,p_1) \tilde{D}_{231}^{n,m} \;,\label{eq:1/kpartial-fac2}  \\
    &\tilde D_{312}^{n,m}= S_{1/\kappa}(p_2,p_1)  \tilde{D}_{321}^{n,m}\label{eq:1/kpartial-fac3} \;,
\end{align}
where we have used that $S_{1/\kappa}(p,q)=1/S_{1/\kappa}(q,p)$, as we can see from \eqref{eq:scatdef}. In contrast to the previous choice, here we impose the partial factorization relations with respect to scattering coefficients $S_{1/\kappa}$ between two position-dependent corrections which are related to each other by an action of the transpositions to the last two indices, $ijk$ and $ikj$. In addition, we also fix the normalization of the state by setting
\begin{align}
    A_{321}=1\;,\label{eq:normalization1}
\end{align}
to make the relation with the previous solution more immediate. Following the same steps as for the previous solution, the three-body interacting equation is solved by
\begin{align}
   \left(\tilde D_{132}^{0,1}+\tilde D_{132}^{1,0}\right)&=-2\left(\kappa-\frac{1}{\kappa}\right) S_{1/\kappa}(p_3,p_1)S_{1/\kappa}(p_2,p_1) \;,\label{eq:D1/k10013}\\
    \left(\tilde D_{231}^{0,1}+\tilde D_{231}^{1,0}\right)&=-2\left(\kappa-\frac{1}{\kappa}\right) S_{1/\kappa}(p_3,p_2)\;,\label{eq:D1/k10012}\\
    \left(\tilde D_{321}^{0,1}+\tilde D_{321}^{1,0}\right)&=-2\left(\kappa-\frac{1}{\kappa}\right)\;.\label{eq:D1/k10011}
\end{align}
We can easily check that the position-dependent coefficients we get here are the same as the ones described in the previous subsection but with $\kappa$ replaced by $1/\kappa$ and the permutation, indices are written backward. Furthermore, if we use the two-body interacting equations to compute the coefficients $\tilde{D}_\sigma^{2,0}$ and $\tilde{D}_\sigma^{0,2}$, we get the same relation. After this point, we follow exactly the same steps as the $S_\kappa$ solution and insert this particular special solution of the first class of equations into the second class of equations as initial conditions and set the leftover freedom to zero, finding that all $D^{n,m}_\sigma$ can be similarly related to the ones of the previous solution. Indeed, this solution can be obtained from $\ket{\xi(p_1,p_2,p_3)}_\kappa$ by applying the combination of parity and $\mathbb{Z}_2$ described in Section \ref{subsec:symmetry}, and we denote it by $\ket{\xi(p_1,p_2,p_3)}_{1/\kappa}$. Therefore, the wave function obtained by the sum of these two solutions will solve the eigenvalue problem, as all our equations are linear in the position-dependent coefficients. Furthermore, it will be invariant under the combination of parity and $\mathbb{Z}_2$,

\begin{align}
    \mathbb Z_2\mathcal{P} \left(\ket{\xi(p_1,p_2,p_3)}_\kappa\pm\ket{\xi(p_1,p_2,p_3)}_{1/\kappa}\right)=\pm \left(\ket{\xi(p_1,p_2,p_3)}_\kappa\pm\ket{\xi(p_1,p_2,p_3)}_{1/\kappa}\right)
    \label{eq:sumparity}
\end{align}
The sum of these two solutions is even under $\mathbb Z_2\mathcal{P}$, while the difference is odd. These two wavefunctions correspond to untwisted and twisted open string states of the gravity dual, respectively.

   \subsection{$S_{\kappa=1}$ Solution}
   \label{subsec:skappa1}
   We present one more special solution, which is 
   obtained as a deformation of
   the two copies of the Heisenberg XXX spin chain. In this case, eventhough the full solution will have a non-trivial $\kappa$ dependence, we choose the  scattering coefficients to have the same form as in the orbifold point $\kappa=1$,
   \begin{align}
 &A_{213}=S(p_1,p_2)\;,\quad  \frac{A_{312}}{A_{132}}=S(p_1,p_3)\;,\quad\frac{A_{321}}{A_{231}}=S(p_2,p_3)\label{eq:skappa1fac} \\ &A_{132}=S(p_2,p_3)\;,\quad
    \frac{A_{231}}{A_{213}}=S(p_1,p_3)\;,\quad 
    \frac{A_{321}}{A_{312}}=S(p_1,p_2)\;.\label{eq:skappa1fac-corr}
\end{align}
In these equations, the scattering coefficient for $\kappa=1$ is such that $S_{\kappa=1}(p_i,p_j)=S(p_i,p_j)$. For simplicity, we also set the normalization of the eigenstate by fixing $A_{123}=1$. Additionally, the partial factorization of the position-dependent corrections for any $n,m\geq0$ should be given in terms of the same scatting coefficients,
\begin{align}
    &\tilde D_{213}^{n,m}= S(p_1,p_2)  \tilde{D}_{123}^{n,m} \;,\label{partial-fac12}  \\
    &\tilde D_{312}^{n,m}= S(p_1,p_3)  \tilde{D}_{132}^{n,m} \;,\label{partial-fac14} \\
    &\tilde D_{321}^{n,m}= S(p_2,p_3)  \tilde{D}_{231}^{n,m} \;.\label{partial-fac16}  
\end{align}
In this case, the three-magnon interacting equation is solved by the position-dependent corrections,
\begin{align}
    &\left(\tilde D_{123}^{0,1}+\tilde D_{123}^{1,0}\right)=- 2 \left(\sqrt{\kappa}-\frac{1}{\sqrt{\kappa}}\right)^2 \;,\label{Dk110110}\\
    &\left(\tilde D_{132}^{0,1}+\tilde D_{132}^{1,0}\right)=- 2\left(\sqrt{\kappa}-\frac{1}{\sqrt{\kappa}}\right)^2 S(p_2,p_3)\;,\label{Dk120110}\\
    &\left(\tilde D_{231}^{0,1}+\tilde D_{231}^{1,0}\right)=- 2\left(\sqrt{\kappa}-\frac{1}{\sqrt{\kappa}}\right)^2 S(p_1,p_3)S(p_1,p_2)\;.\label{Dk130110}
\end{align}
We observe exactly the same structure that we encountered in the previous special solutions. Then we fix the following relations among the position-dependent corrections of the shortest separation,
\begin{align}
    D_\sigma^{1,1}=0\;,\quad\quad D_\sigma^{1,0}=D_\sigma^{0,1}\;.
\end{align}
% Concerning these choices, the solution of the solution of the left two-magnon interacting equations gives the solution of the position-dependent corrections $D_\sigma^{0,2}$,
With respect to these choices, the left two-magnon interacting equations give the solution of the position-dependent corrections,
\begin{align}
    \tilde D_{123}^{0,2} &=2\frac{1-\kappa}{\kappa}\left(\kappa e^{ip_3}-(1-\kappa)(\EE -2\kappa ) \right)\;,\\
    \tilde D_{132}^{0,2} &=2\frac{1-\kappa}{\kappa}S(p_2,p_3) \left(\kappa e^{i p_2}-(1-\kappa) (\EE -2\kappa )  \right)\;, \\
    \tilde D_{231}^{0,2} &=2\frac{1-\kappa}{\kappa}S(p_1,p_3)S(p_1,p_2)\left(\kappa e^{ip_1}- (1-\kappa) (\EE -2\kappa )\right)\;.
\end{align}
Furthermore, the solution of the right two-magnon interacting equations for the shortest non-adjacent separation gives,
\begin{align}
    \tilde D_{123}^{2,0} &=\frac{\kappa -1}{\kappa} \left( e^{-ip_3}  (\EE + \PP + \invPP -4 ) -2 ( \EE - e^{ip_3} - e^{-ip_3} -2+\kappa\invPP) \right)\;,\\
    \tilde D_{132}^{2,0} &= \frac{\kappa -1}{\kappa} S(p_2,p_3)\left( e^{-ip_2}  (\EE + \PP + \invPP -4 ) -2 ( \EE - e^{ip_2} - e^{-ip_2} -2+\kappa\invPP) \right)\;,\\
    \tilde D_{231}^{2,0}&= \frac{\kappa -1}{\kappa}S(p_1,p_2)S(p_1,p_3) \left( e^{-ip_1}  (\EE + \PP + \invPP -4 )\right.\nonumber\\&\quad\quad\quad\quad\quad\quad\quad\quad\quad\quad\quad\quad\quad\quad\quad\quad\quad\left. -2 ( \EE - e^{ip_1} - e^{-ip_1} -2+\kappa\invPP) \right)\;,
\end{align}
We complete this special solution exactly as the previous ones. In particular, similarly to the previous ones, it admits a common description of its position-dependent corrections that admits the same form as \eqref{eqn:SpecialDs}-\eqref{eqn:SpecialDs1} when we replace the scattering coefficients with $S_{\kappa=1}$ ones. Even though the scattering coefficients are invariant under the $\kappa\to1/\kappa$ careful analysis shows that this solution cannot be parity invariant as given in \eqref{eq:restoredparity}. However, it is still an interesting sub-set of solutions and we expect there exists another special solution such that the linear combination of $S_{\kappa=1}$ solution with that one is parity invariant as in \eqref{eq:sumparity}.

\section{Yang Operator Formalism}
\label{sec:Yang}

To reveal the structure of the special solutions that we present in the previous section, we use the Yang Operator formalism \cite{YangPhysRev.168.1920}. If a quantum many-body system of $N$ particles has $N$ linearly independent conserved charges, the coordinate Bethe ansatz provides a universal form for the wave function. This form decomposes into plane wave contributions modulated by scattering coefficients. Implicitly, this structure precludes any intrinsically three-body events and diffraction. In this case, the Yang operator formalism is used to reveal the consequences of the large amount of conserved charges, for a detailed review see \cite{arutyunov2019elements}.

We consider a quantum many-body problem that admits $N$ many conserved charges for $N$-bodies. For a kinematic domain characterized by the positions of the magnons $x_1<\cdots<x_N$, the solution of the Schrödinger equation reads
\begin{align}
    \Psi(x_1,\cdots,x_N)=\sum_{\sigma\in\mathcal{S}_N}\mathcal{A}_{\sigma}(p_1,\cdots,p_N)e^{i\Vec{p}_{\sigma}\cdot\Vec{x}}\;,\label{eq:integrable-wave}
\end{align}
such that the coefficients are labeled by the permutation indices. We can generalize this wave function to any kinematic domain by specifying the order of the magnons using $\tau\in\mathcal{S}_N$. This means that, with respect to the initial set of coordinates, the following wave function is the solution of the stationary Schrödinger equations for $x_{\tau(1)}<\cdots<x_{\tau(N)}$,
\begin{align}
    \Psi(x_1,\cdots,x_N|\tau)=\sum_{\sigma\in\mathcal{S}_N}\mathcal{A}_{\tau|\sigma}(p_1,\cdots,p_N)e^{i\Vec{p}_{\sigma}\cdot\Vec{x}_{\tau}}\;.\label{eq:integrable-wave-gen}
\end{align}
In other words, we identify the $N$ magnons with their coordinates and permute them by changing the form of the incoming plane wave 
\begin{align}
    e^{ip_1x_1+\cdots +p_Nx_N}\longrightarrow e^{ip_1x_{\tau(1)}+\cdots +p_Nx_{\tau(N)}}\;.
\end{align}
We encode the scattering coefficients of the wave function, for any kinematic domain into a vector constructed from the coefficients next to the plane wave contributions,
\begin{align}
    \Phi(\tau)=\{\mathcal{A}_{\tau|\sigma},\quad \forall \sigma\in\mathcal{S}_N\} \;.
\end{align}
The \emph{Yang Operator} is defined as the map that relates two vectors in different kinematic domains
\begin{align}
    \Phi(\alpha_j\tau)=Y_j(\alpha_j)\Phi(\tau)\;, \label{eq:YangDef}
\end{align}
for any $\alpha_j,\tau\in\mathcal{S}_N$. For an integrable system that admits a non-diffracting solution to the Schrödinger equation, the coefficients of the wave function satisfy the following relations
\begin{align}
    \mathcal{A}_{\tau|\alpha_j\sigma}&=A(p_{\tau(j)},p_{\tau(j+1)})\mathcal{A}_{\tau|\sigma} \;,\\
    \mathcal{A}_{\alpha_j\tau|\alpha_j\sigma}&=B(p_{\tau(j)},p_{\tau(j+1)})\mathcal{A}_{\tau|\sigma} \;.
\end{align}
Here, $\alpha_j\in\mathcal{S}_N$ are the transpositions which exchange the $j$th element with $j+1$th. In this case, the Yang Operators include the \emph{reflection coefficient}s on the diagonal and the \emph{transmission coefficient}s on the off-diagonal,
\begin{align}
    Y_j(\alpha_j)=A(p_{\tau(j)},p_{\tau(j+1)})\mathbf{1}+B(p_{\tau(j)},p_{\tau(j+1)})\pi(\alpha_j)\;,
\end{align}
where $\mathbf{1}$ is the identity matrix for $N$-dimensional vector space, and the $\pi(\alpha_j)$ is permutation operator in the corresponding basis. Note that once we specify the representation of the symmetric group and compute the Yang operator for the transpositions for $(j,j+1)$, we can generate all other permutations by applying the Yang Operators successively. Furthermore, the Yang operators obey the YBE,
\begin{align}
    Y_j(p_2,p_3)Y_{j+1}(p_1,p_3)Y_j(p_1,p_2)=Y_{j+1}(p_1,p_2)Y_j(p_1,p_3)Y_{j+1}(p_2,p_3)\;.\label{eq:yang-ybe}
\end{align}
However, for our model, we encounter a form of the Yang operator which is more involved than this one.

\subsection{The Yang Operators of Our Model}
\label{sec:yangXZ}
We start by computing the Yang operator for the wave function of two magnons, as given in \eqref{eq:two-wave}. We identify the magnons by their positions, considering the excitation $Q_{12}$ located at $x_1$ and the excitation $Q_{21}$ located at $x_2$. This implies that the kinematic domain $x_1<x_2$ corresponds to the wave function associated with the state $|\Psi(p_1,p_2) \rangle_{12}$, while the kinematic domain $x_2<x_1$ corresponds to the wave function associated with the state $|\Psi(p_1,p_2) \rangle_{21}$. We can express both wave functions together as
\begin{align}
    \begin{pmatrix}
        \braket{x_1,x_2|\Psi(p_1,p_2)}_{12}\\
        \braket{x_2,x_1|\Psi(p_1,p_2)}_{21}
    \end{pmatrix}=\begin{pmatrix}
        1&S_\kappa(p_1,p_2)\\
        S_{1/\kappa}(p_1,p_2)&1
    \end{pmatrix}\begin{pmatrix}
        \exp(ip_1x_1+ip_2x_2)\\
        \exp(ip_2x_1+ip_1x_2)
    \end{pmatrix}\;.
\end{align}
Here, the vectors of coefficients are given by the first and the second columns of the $2\times2$ matrix in the right-hand side of the equation,
\begin{equation}
    \Phi(12)=\begin{pmatrix}
        1\\
        S_{1/\kappa}(p_1,p_2)
    \end{pmatrix}\;, \qquad \Phi(21)=\begin{pmatrix}
        S_\kappa(p_1,p_2)\\
        1
    \end{pmatrix}\;.
\end{equation}
The definition \eqref{eq:YangDef} of the Yang operator implies that
\begin{align}
    \Phi(21)&=Y_1(p_1 , p_2) \Phi (12) \;, & \Phi(12)&=Y_1 (p_2,p_1) \Phi(21) \;.
\end{align}
Therefore, the resulting Yang operator takes the form
\begin{align}
    Y_1(p_1,p_2)=\begin{pmatrix}
        S_\kappa(p_1,p_2)&0\\0&S_{1/\kappa}(p_2,p_1)
    \end{pmatrix}\;, \label{eq:Yangtwomagnons1}
\end{align}
and $Y_1(p_1,p_2)^{-1}=Y_1(p_2,p_1)$. As anticipated, the transmission coefficients vanish, indicating that the model permits only the reflection of magnons.

Let us consider now the case of an arbitrary number of magnons. If the magnon associated to the position $x_j$ is $Q_{12}$, then $Y_j (p_1, p_2)$ has the above form. From the form of the Hamiltonian and the fact that excitation $j+1$ has to be $Q_{21}$, it is clear that
\begin{align}
    Y_{j+1}(p_1,p_2)=\begin{pmatrix}
        S_{1/\kappa}(p_1,p_2)&0\\0&S_{\kappa}(p_2,p_1)
    \end{pmatrix}\;. \label{eq:Yangtwomagnons2}
\end{align}
Then, the YBE for the Yang operators, \eqref{eq:yang-ybe}, is not satisfied, as it suffers from the same mismatch as \eqref{eq:neq}
\begin{align}
    Y_{j+1}(p_1,p_2)Y_j(p_1,p_3)Y_{j+1}(p_2,p_3)\neq Y_{j}(p_2,p_3)Y_{j+1}(p_1,p_3)Y_{j}(p_1,p_2) \;.
\end{align}
Instead of directly generalizing the Yang operators to many body solutions we explicitly define the new form of the Yang operators for the long-range ansatz solution of the three-magnon problem.

\subsection{The Yang Operators of the Long-Range Solution}

When we explicitly derive the Yang operators for the long-range ansatz \eqref{eq:infocontwave}, we realize that Yang operators need to incorporate the additional position dependence on the coefficients,
%\begin{multline}
%    \mathcal{A}_{\tau|\sigma}(p_1,p_2,p_3;x_2-x_1,x_3-x_2)\\=A_\sigma(p_1,p_2,p_3)-\oint_\Sigma\frac{dx dy}{4\pi^2}\frac{G_\sigma(x,y)}{x^{x_{\tau(2)}-x_{\tau(1)}}y^{x_{\tau(3)}-x_{\tau(2)}}}\;,\label{eq:three-coeff}
%\end{multline}
%where we use the final form of the generating function which involves the position-dependent corrections that satisfy all the three-magnon eigenvalue equations.
\begin{equation}
   \mathcal{A}_{\tau|\sigma}(p_1,p_2,p_3;x_2-x_1,x_3-x_2) =A_\sigma(p_1,p_2,p_3)-D^{n_\tau,m_\tau}_\sigma (p_1,p_2,p_3) \;,
\end{equation}
where $n_\tau=x_{\tau(2)}-x_{\tau(1)}$ and $m_\tau=x_{\tau(3)}-x_{\tau(2)}$.

%Furthermore, because of the difference between the excited states $Q_{12}$ and $Q_{21}$, we are forced to restrict ourselves to two kinematic domains for the three-body problem instead of six,
The fact that we do not have a tensor product Hilbert space forbids some of the kinematic domains. In particular, given an allowed kinematic domain, only the one related by a permutation of the first and last magnon is also possible. Thus, out of the six kinematic domains that naturally arise by considering all the permutations, only the domains
\begin{align}
    x_1<x_2<x_3\quad\quad\text{and}\quad\quad x_3<x_2<x_1 \;,
\end{align}
are allowed for $\ket{\Psi(p_1,p_2,p_3)}_{12}$. This, together with the fact that the model does not have transmissions, implies that the Yang operator splits into two separate blocks related by a parity transformation. For the Yang operators coming from the two-magnon solution, the relation between the diagonal factors in \eqref{eq:Yangtwomagnons1} and \eqref{eq:Yangtwomagnons2} are a combination of parity and $\mathbb Z_2$ however here we only observe the parity relation as we need to use the wave functions coming from the $\mathbb Z_2\mathcal{P}$ invariant vectors, \eqref{eq:sumparity} to observe the full relation.
%This happens because, given an allowed kinematic domain appearing in $\ket{\Psi(p_1,p_2,p_3)}_{12}$, only the one related by a permutation of the first and last magnon is allowed. A similar situation happens for $\ket{\Psi(p_1,p_2,p_3)}_{21}$. Because of that and because the model does not have transmission, the Yang operator splits into two separated blocks related by a $\mathbb Z_2$ transformation, as it already can be seen in \eqref{eq:Yangtwomagnons}.

To construct the Yang operator, we first write the plane wave basis for the case of three magnons as
\begin{align}
    \textbf{v}=\begin{pmatrix}
        \exp(ip_1x_1+ip_2x_3+ip_3x_3)&&\cdots&&\exp(ip_3x_1+ip_2x_3+ip_1x_3)
    \end{pmatrix}^t \;,
\end{align}
which is the basis vectors of a six-dimensional vector space. Then, the matrix that contains the coefficients of the wave function is given as
\begin{align}
    \begin{pmatrix}
        \braket{x_1,x_2,x_3|\Psi(p_1,p_2,p_3)}_{12}\\
        \braket{x_3,x_2,x_1|\Psi(p_1,p_2,p_3)}_{12}
    \end{pmatrix}=\begin{pmatrix}
        \Phi(123)&&\cdots&&\Phi(321)
    \end{pmatrix}\textbf{v} \;,
\end{align}
where
\begin{equation}
    \Phi(\sigma)=\begin{pmatrix}
        A_{\sigma} + D^{n,m}_\sigma \\
        A_{\sigma^r} + D^{m,n}_{\sigma^r}
    \end{pmatrix} \;.
\end{equation}
Here $\sigma^r$ is the permutation $\sigma$ read backwards. From these expressions, the definition \eqref{eq:YangDef} of the Yang operator gives us
%In its most general form, the Yang operator that maps a column of the coefficient matrix takes the form
\begin{align}
    Y^{n,m}_j(p_{\tau(j)},p_{\tau(j+1)},p_k)=\begin{pmatrix}
        \frac{A_\sigma+D^{n,m}_\sigma}{A_{\alpha_j\sigma}+D^{n,m}_{\alpha_j\sigma}}&0\\0&\frac{A_{\sigma^r}+D^{m,n}_{\sigma^r}}{A_{\alpha_{j+1}\sigma^r}+D^{m,n}_{\alpha_{j+1}\sigma^r}} \;, \label{eq:generalyang}
    \end{pmatrix}
\end{align}
where $k$ is the third index, which labels the momentum parameter for $k\neq\tau(j)$ and $k\neq\tau(j+1)$.

To understand better the Yang operator, we write the form it takes for the $S_\kappa$ solution presented in Section \ref{subsec:skappa}. Substituting the rewriting \eqref{partial-fac2}-\eqref{partial-fac6} and the solutions \eqref{eqn:SpecialDs}-\eqref{eqn:SpecialDs1}, it reduces to
%We focus on one of these Yang operators for the $S_\kappa$ solution presented in Section \ref{subsec:skappa}. Note that the other special solution would have the same structure but different scattering coefficients,
%\begin{align}
%   Y_j^{n,m}(p_1,p_2,p_3)= \begin{pmatrix}
%        S_\kappa(p_1,p_2)\frac{1+e^{inp_2-imp_3}\tilde D_{123}^{n,m}}{1+e^{inp_1-imp_3}\tilde D_{123}^{n,m}} & 0 \\ 0 & S_\kappa(p_2,p_1)\frac{1+e^{inp_3-imp_2}\tilde D_{132}^{m,n}/S_\kappa(23)}{1+e^{inp_3-imp_1}D_{231}^{m,n}/S_\kappa(13)S_\kappa(12)}\end{pmatrix}\label{eq:yangskappa}
%\end{align}
%And, according to the common structure of the position-dependent corrections summarized in \eqref{eq:mainobs1}-\eqref{eq:mainobs3} and the fact that the second class of equations are solved under the permutational relations summarized in Figure \ref{fig:sol-one-gen}, we implicitly obtain a common definition of the position-dependent corrections for any separation, $n,m$. With respect to this common form of the position-dependent corrections, given in \eqref{eq:integral}, we get rid of the extra scattering coefficients that divide the position-dependent corrections. After simplifying those scattering coefficients, we obtain the following final form of the position-dependent corrections
\begin{align}
   Y_j^{n,m}(p_1,p_2,p_3)= \begin{pmatrix}
        S_\kappa(p_1,p_2)\frac{1+f(p_2,p_1,p_3;n,m;\kappa)}{1+f(p_1,p_2,p_3;n,m;\kappa)} & 0 \\ 0 & S_\kappa(p_2,p_1)\frac{1+f(p_3,p_1,p_2;m,n;\kappa)}{1+f(p_3,p_2,p_1;m,n;\kappa)}\end{pmatrix}\label{eq:yangskappa}
\end{align}
Note that the Yang operator for the $S_{1/\kappa}$ as well as the $S_{\kappa=1}$ solutions presented in Section \ref{subsec:skappainverse} have similar structures.

\subsection{An Infinite Tower of Yang-Baxter Equations}
At this point, it is easy to check that the following identity holds for any $n,m\geq0$
%Furthermore, for any $n,m\geq0$ we obtain the following identity,
\begin{multline}
    Y_j^{n,m}(p_2,p_3,p_1)Y_{j+1}^{n,m}(p_1,p_3,p_2)Y_j^{n,m}(p_1,p_2,p_3)\\
    =Y_{j+1}^{n,m}(p_1,p_2,p_3)Y_j^{n,m}(p_1,p_3,p_2)Y_{j+1}^{n,m}(p_2,p_3,p_1)\label{eq:towerofYBE}\;,
\end{multline}
which can be identified as an infinite tower of YBEs. This is a direct consequence of the initial form of the long-range ansatz, given in the equation \eqref{eq:infcontansatz}, which includes six plane wave contributions with respect to the permutations of the momentum variables. However, the fact that there is a common description of the Yang operators for permutation with any two indices is a highly non-trivial feature of this model which was revealed by special solutions. This was revealed by establishing the relations between the solution of the second class of equations in subsection \ref{subsubsec:two-magnons}. And further developed by choosing symmetric solutions to the first class of equations in subsection \ref{subsec:skappa}. In Figure \ref{fig:twoYBE} we present the tower of YBEs together with the inconsistency of the naive three-magnon CBA.

\bigskip

\section{Conclusions}
\label{sec:conclusions}

\begin{figure}[t!]
    \centering
    \includegraphics[scale=0.32]{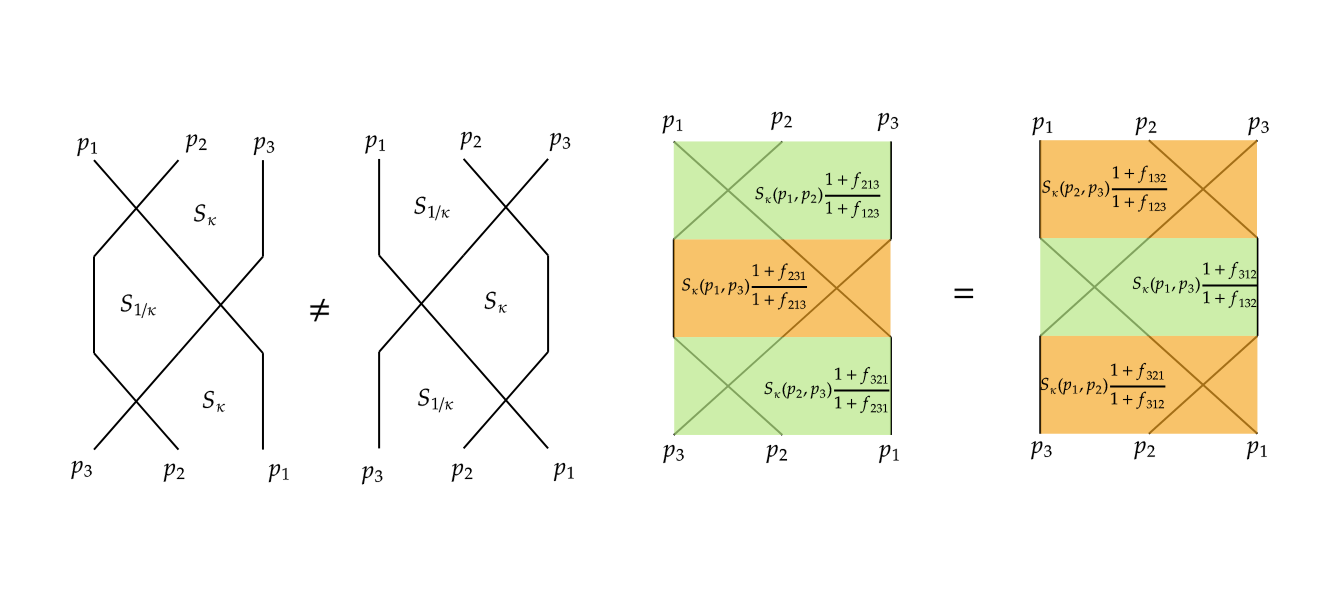}
    \caption{\it We compare the result of the naive CBA, which leads to inequality of two-sides of the YBE, as it was given in \eqref{eq:neq}. Only when we introduce the long-range ansatz with infinitely many position-dependent corrections we obtain a tower of YBEs which depend on the coordinates of the magnons and we attain a solution to the three-magnon problem.}
    \label{fig:twoYBE}
\end{figure}

With this paper, we took an important step in the study of spin chains capturing the spectral problem of gauge theories with less than maximal supersymmetry, in particular $\mathcal{N}=2$ SCFTs.
We attacked the up to now unsolved three-magnon problem (three bifundamental $Q$s in the sea of the adjoint $\phi$s) 
without which the question of integrability cannot be meaningfully addressed.

Our solutions have 
 additive  energy eigenvalue  
$E_3(p_1,p_2,p_3)=E_1(p_1)+E_1(p_2)+E_1(p_3)$
and
several novel features, the discussion of which is in order. Firstly,  they necessarily have the form of novel long-range wave functions.
Secondly, the general solution, as presented in Section \ref{sec:infinite}, contains an infinite number of unfixed parameters. This is due to not imposing boundary conditions as three magnon states are non-closable, {\it i.e.} cannot lead to single trace operators (as we review in the Subsection \ref{subsec:SpinChain}).
Nonetheless, using the symmetry of the problem and our intuition,  we can brute force fix all of them and get special solutions in Section \ref{sec:special}.
For the special solutions,  the position-dependent corrections to the scattering coefficients modify the plane wave basis as
\begin{align}
\label{eq:newbasis}
e^{ip_1l_1+ip_2l_2+ip_3l_3} 
\longrightarrow
\left(1+f(p_1,p_2,p_3;n,m;\kappa)\right) e^{ip_1l_1+ip_2l_2+ip_3l_3}
\end{align}
with the modification encoded in the
generating function $G_{123}(x,y)$ 
\begin{align}
\label{eq:fcon}
f(p_1,p_2,p_3;n,m,\kappa)=-\oint_\Sigma \frac{dx dy}{4\pi^2}\frac{G_{123}(x,y)}{x^{n+1}y^{m+1}}\;,
\end{align}
the form of which is still quite baroque, but can be explicitly written as in Appendix \ref{sec:G123}.
This allows us to perform a non-trivial change of  basis for the wave function in which the is solution, as in the case of the usual CBA, is given by permutations,
\begin{align}
\label{eq:newCBA}
\Psi(p_1,p_2,p_3;l_1,l_2,l_3)&=\sum_{\sigma\in\mathcal{S}_3}  A_\sigma \left(
1+f(p_{\sigma(1)},p_{\sigma(2)},p_{\sigma(3)};l_2-l_1,l_3-l_2;\kappa)\right)
e^{i\Vec{p}_\sigma\cdot\Vec{l}} 
        \;.
\end{align}
Interestingly, the generating function $G_{123}(x,y)$ is defined
on the elliptic curve $Q(x,y)=0$ \eqref{eq:ellipticCurve}. The later is parameterised only in terms of the total energy and the total momentum of the three magnons.
What is more, another feature of the special solutions is that they can be rendered in odd and even eigenvectors \eqref{eq:sumparity} of the combined $\mathbb{Z}_2$ plus parity symmetry, as expected from the orbifold construction.

Importantly, as discussed in Section \ref{sec:Yang}, our long-range solutions imply that we can define an infinite tower of Yang operators labeled by two integers that parameterize the distances between the three magnons. These  Yang operators obey an infinite tower of Yang-Baxter equations parameterized by the same two integers. As we take $\kappa\to1$, the contributions of the position-dependent corrections vanish and the tower of YBEs collapse to the single YBE of orbifold point, which is expected from \cite{Beisert:2005he}.

In \cite{Pomoni:2021pbj} it was argued that the model is elliptic. In this work, we find further evidence for this elliptic nature. In Section \ref{sec:solution} we saw that the corrections to the usual Bethe Ansatz are encoded in a generating function  \eqref{eq:GDefinition} that lives on an elliptic spectral curve \eqref{eq:ellipticCurve}. We believe that this fact is not just technical but deep. This fact reoccurs in the solution of the four magnons, which is a work in progress \cite{fourmagnon}.

The long-range correction to the usual Bethe Ansatz captures the fact that the permutation symmetry of the three-magnon states is naively broken because the magnons are bifundamental. 
For models where the permutation symmetry is present,
 the solution should always be possible to write in terms of the usual Bethe ansatz.
In our case, because of the $\mathbb{Z}_2$ orbifold in tandem with the marginal deformation $\kappa$, the permutation group of the three magnons breaks into two subsets, as we discuss in Section \ref{subsubsec:two-magnons}. 
The position-dependent corrections of our solution eventually restore the
 permutation symmetry \eqref{eq:newCBA}. This must be a very explicit manifestation of the dynamical nature of the model \cite{Pomoni:2021pbj}  which deserves further investigation.
This  must also be related to the fact that even though naively in three of the entries of the YBE we were supposed to have
$S_{1/\kappa}$ according to the naive Bethe Ansatz, our long-range solution has  $S_{\kappa}$ vividly reminding us of Felder's DYBE.
An attempt to visually capture this is presented in Figure \ref{fig:twoYBE}.

\bigskip 

Interesting future directions include: 

\begin{itemize}

\item To address the question of integrability and also naturally determine the unfixed parameters of our general solutions, the most imperative next step is the study of four magnons, which we are already pursuing \cite{fourmagnon}.

\item 

As discussed in \cite{Pomoni:2021pbj}, we expect this spin chain model to be elliptic. We expect that there is an elliptic parameterization in which the higher magnon wave functions can be beautifully repackaged. An important future step will be to explicitly find this parameterization. This can go in tandem with attempting to find a three-magnon solution in the alternating Q vacuum following \cite{Pomoni:2021pbj}.

 \item Our current approach needs to be combined with the algebraic approach of \cite{Hanno} such that we will understand what the novel symmetry of this spin chain is and how we can obtain the R-matrix that obeys $RTT$ as a function of the momentum/rapidity.

\item 
$\mathbb{Z}_2$ is the simplest orbifold in the ADE classification. We expect that our work is easily generalizable to $\mathbb{Z}_n$  orbifolds with $n>2$. Pursuing this direction, we expect to 
enlarge the landscape of $\mathcal{N}=2$ SCFTs
we have access to. What is more important, is that the spin chains of these $\mathbb{Z}_n$  orbifold SCFTs, as already discussed in \cite{Pomoni:2021pbj}, correspond to hyper-elliptic models with their 
spectral parameter living on a higher genus surface.

\item 

Having solved the three-magnon problem, we now have enough data to be able to attempt to find the R-matrix by at least brute force.
Especially combining this data together with the solutions in \cite{Pomoni:2021pbj}, where the $\phi$ excitations around the alternating Q vacuum were considered, it should be possible
to fix the form of the R-matrix.

\item 
Clearly, a very important strand of research is to look for higher conserved charges for this model, an approach that will answer the question of integrability. We have spent some time on this and we know that these cannot be constructed as in usual rational and trigonometric models using the boost operator \cite{deLeeuw:2020ahe,deLeeuw:2020xrw,DeLeeuw:2019gxe}.
To succeed in this endeavor
we need to come up with a dynamical generalization of this formalism.

\item After this very hopeful result, we feel a renewed interest in looking at the string theory side of the AdS/CFT correspondence. In light of the recent paper of \cite{Skrzypek:2023fkr}, it is time to study the dual 2D worldsheet QFT using the solution \cite{Skrzypek:2023fkr} of type IIB supergravity which is a resolution of the $AdS_5 \times S^5/\mathbb{Z}_2$ orbifold singularity.

\end{itemize}

\section*{Acknowledgements}
 We are grateful to Gleb  Arutyunov and Costas Zoubos for very useful discussions throughout the course of this work.
We also wish to thank  Marius de Leeuw, Jules Lamers, Ingo Runkel and Volker Schomerus. The authors have benefited from the German Research Foundation DFG under Germany’s Excellence Strategy – EXC 2121 Quantum Universe – 390833306 as well as from the CRC 1624 Higher Structures, Moduli Spaces and Integrability.
  EP is supported by ERC-2021-CoG - BrokenSymmetries 101044226.
DB presented this work for the first time at ``Integrability, Dualities, and Deformations 2024'' at
Swansea (UK).

\appendix
\section{Contact Term Ansatz and Kinematic Limit}
\label{sec:contact-terms}
We generalize the contact term ansatz given for the two-magnon state in equation \eqref{eq:twobody-cont} to three-magnon states and try to obtain a solution of the three-body problem, \eqref{eq:eigthree}. The resulting ansatz for three-magnons is referred as \emph{contact term Bethe ansatz},
\begin{align}
    \ket{\Psi^c(p_1,p_2,p_3)}_{12}=\sum_{l_1<l_2<l_3}\sum_{\sigma\in\mathcal{S}_3}\left(A_\sigma+B_\sigma \delta(l_2,l_1-1)+C_\sigma \delta(l_3,l_2-1)\right)e^{i\Vec{p}_\sigma\cdot\Vec{l}}\ket{l_1,l_2,l_3}\label{eq:threecont}
\end{align}
Both scattering coefficients, $A_\sigma$, and two types of contact term coefficients, $B_\sigma$, and $C_\sigma$, are functions of the momentum variables and $\kappa$. Note that we can also add a third type of contact term that only plays a role when three of the magnons are next to each other. This term would be added as $D_\sigma\delta(l_2,l_1-1)\delta(l_3,l_2-1)$ to the contact-term Bethe Ansatz. However, we conclude that it does not have a significant effect on the overall solution. Therefore, we decided to omit that contribution for simplicity.

The three-magnon eigenvalue problem, given in \eqref{eq:eigthree}, can be split into four types of equations. When we project the eigenvalue problem to a position space state with well-separated magnons, the contact term Bethe ansatz still automatically gives a vanishing expression,
\begin{align}
		\braket{l_1,l_2,l_3|\mathcal{H}-E_3|\Psi^c(p_1,p_2,p_3)}=0\;,\quad\text{for all}\quad l_2-l_1>2\;l_3-l_2>2\label{eq:cont-non-int}\;.
	\end{align}
 In contrast, the configurations with at least two of the magnons are separated by only one $\phi$ state does not. It gives the condition
 \begin{align}
		0=-e^{-il_1(p_1+p_2+p_3)}&\braket{l_1,l_1+2,l_3|\mathcal{H}-E_3|\Psi^c(p_1,p_2,p_3)}\nonumber\\&=\left(1+e^{-i (p_1+p_2)}\right) \left(B_{123} e^{i p_2}+B_{213} e^{i p_1}\right)e^{i(l_3-l_1)p_3}\nonumber\\&+\left(1+e^{-i (p_1+p_3)}\right) \left(B_{132} e^{i p_3}+B_{312} e^{i p_1}\right)e^{i(l_3-l_1)p_2}\nonumber\\&+\left(1+e^{-i (p_2+p_3)}\right) \left(B_{231} e^{i p_3}+B_{321} e^{i p_2}\right)e^{i(l_3-l_1)p_1}\;,\label{close-left}
	\end{align}
 where $l_3-l_1>4$. Similarly, for the last two magnons, we have the expressions 
 \begin{align}
		0=-e^{-il_1(p_1+p_2+p_3)}&\braket{l_1,l_3-2,l_3|\mathcal{H}-E_3|\Psi^c(p_1,p_2,p_3)}\nonumber\\&=\left(1+e^{-i (p_1+p_2)}\right) \left(C_{321} e^{i p_2}+C_{312} e^{i p_1}\right)e^{i(l_3-l_1)p_3}\nonumber\\&+\left(1+e^{-i (p_1+p_3)}\right) \left(C_{231} e^{i p_3}+B_{213} e^{i p_1}\right)e^{i(l_3-l_1)p_2}\nonumber\\&+\left(1+e^{-i (p_2+p_3)}\right) \left(C_{132} e^{i p_3}+B_{123} e^{i p_2}\right)e^{i(l_3-l_1)p_1}\;,\label{close-right}
	\end{align}
 Since we initially assumed that the contact term coefficients are only functions of momentum variables and $\kappa$, we can separate each of these two expressions into three parts that have to vanish separately. From them, we obtain the following solution for the contact terms,
 \begin{align}
     B_{jik}&=-e^{ip_j-ip_i}B_{ijk}\label{eq:cont-B-sol}\\
     C_{kji}&=-e^{ip_j-ip_i}B_{kij}\label{eq:cont-C-sol}
 \end{align}
 for any permutation of $B_\sigma$ and $C_\sigma$. Notice that these relations closely resemble the relations we derived for the two-magnon contact term relations, \eqref{eq: two-contact-rel}. Moreover, we observe that the relations given above imply,
 \begin{align}
     \braket{l_1,l_1+2,l_1+4|\mathcal{H}-E_3|\Psi^c(p_1,p_2,p_3)}=0\;.
 \end{align}
 Furthermore, the contribution of the contact terms \eqref{eq:cont-B-sol} and \eqref{eq:cont-C-sol} in the wave function, is completely washed out after inserting these relations into the contact term ansatz, \eqref{eq:threecont}. Therefore, we are left with exactly the same expressions as the three-magnon interacting equations of the naive CBA, \eqref{eq:twobodyl-naive} and \eqref{eq:twobodyr-naive}. This leads to the failure of the contact term Bethe ansatz due to the inconsistency coming from the failure of the YBE, \eqref{eq:neq}.

 We can avoid the trivial relations between the contact term coefficients by imposing special kinematics. Since the contact term coefficients that appear in the equations, \eqref{close-left} and \eqref{close-right} have the common factors, $1+e^{ip_i+ip_j}$ for any $i,j\in\{1,2,3\}$, by constraining two of the momentum variables we are able to lift one of the relations that trivialize the contact term contributions for each $B$ and $C$ type contact terms.

 Without loss of generality, we present the solution of the three-magnon problem with contact term ansatz under the constraint,
 \begin{align}
     p_2+p_3=\pi\;.\label{eq:mom-const}
 \end{align}
 Then the solution of the equations given for the magnons which are separated by one $\phi$ state is only
 \begin{align}
     B_{213}&=-e^{ip_2-ip_1}B_{123}\label{eq:cont1}\;,\\
     B_{312}&=-e^{ip_3-ip_1}B_{132}\label{eq:cont2}\;,\\
     C_{231}&=-e^{ip_3-ip_1}B_{213}\label{eq:cont3}\;,\\
     C_{321}&=-e^{ip_2-ip_1}B_{312}\label{eq:cont4}\;.
 \end{align}
 These four relations between the contact terms make the following expressions vanish.
 \begin{align}
      \braket{l_1,l_1+2,l_3|\mathcal{H}-E_3|\Psi^c(p_1,p_2,p_3)}=0\\
      \braket{l_1,l_3-3,l_3|\mathcal{H}-E_3|\Psi^c(p_1,p_2,p_3)}=0\\
      \braket{l_1,l_1+2,l_1+4|\mathcal{H}-E_3|\Psi^c(p_1,p_2,p_3)}=0
 \end{align}
 Furthermore, the interacting equations are solved by the following scattering coefficients
\begin{align}
    A_{213}&=S_\kappa(p_1,p_2)\label{s213}\;,\\
    A_{231}&=S_\kappa(p_1,p_2)S_{1/\kappa}(p_1,p_3)\;, \label{s231}\\
    A_{132}&=\mathcal{S}(p;\frac{1}{\kappa})S_{1/\kappa}(p_2,p_3)\label{s132}\;,\\
    A_{312}&=\mathcal{S}(p;\frac{1}{\kappa})S_{1/\kappa}(p_2,p_3)S_\kappa(p_1,p_3)\label{s312}\;,\\
    A_{321}&=\mathcal{S}(p;\frac{1}{\kappa})S_{1/\kappa}(p_2,p_3)S_\kappa(p_1,p_3)S_{1/\kappa}(p_1,p_2)\;. \label{s321}
\end{align}
together with the additional relations between contact terms that are needed for the interacting equations,
 \begin{align}
    B_{321}=&B_{231} e^{-2 i p_2}+e^{-2 i p_2}S_\kappa(p_1,p_2)S_{1/\kappa}(p_1,p_3)\nonumber\\&\quad\quad\quad\quad-\mathcal{S}(p;\frac{1}{\kappa})S_{1/\kappa}(p_2,p_3)S_{1/\kappa}(p_1,p_2)S_\kappa(p_1,p_3),\label{b1}\;,\\
    C_{132}= &C_{123} e^{-2 i p_2}+e^{-2 i p_2}-\mathcal{S}(p;\frac{1}{\kappa})S_{1/\kappa}(p_2,p_3)\;. \label{c1}
\end{align}
Where we fix the overall normalization of the state by setting $A_{123}=1$. In these expressions appears an additional type of scattering coefficient that takes the following form,
\begin{equation}
    \resizebox{.9\hsize}{!}{$
    \mathcal{S}(p_1,p_2,p_3;\frac{1}{\kappa})=\frac{e^{-i p_2+ip_3} \left(1+e^{i p_1+i p_3}-2 \kappa e^{i p_1}\right) \left(1+e^{i p_1+i p_2}-2 \kappa^{-1} e^{i p_1}\right) \left(e^{i p_2} \left(-\kappa +e^{i p_1}\right)^2-e^{i p_1}\left(\kappa ^2-1\right) \right)}{\left(1+e^{i p_1+i p_2}-2 \kappa e^{i p_1}\right) \left(1+e^{i p_1+i p_3}-2 \kappa^{-1} e^{i p_1}\right) \left(e^{i p_3}\left(-\kappa +e^{i p_1}\right)^2-e^{i p_1}\left(\kappa ^2-1\right) \right)}\label{S32}$}\;.
\end{equation}
This additional scattering coefficient contains the three momentum variables, and it describes the process when two magnons carry momenta $p_2$ and $p_3$. scatter These two variables are constrained by the condition, \eqref{eq:mom-const}. Then the third momentum variable carried by the third magnon, which is far away from the interacting magnons, plays the role of an \emph{observer}. We notice that the non-trivial contact terms given in equations \eqref{b1} and \eqref{c1} exchange the global form of the scattering coefficients to the one predicted by the naive form of the CBA, \eqref{eq:naive-sol-left} and \eqref{eq:naive-sol-right} under the kinematic limit. Here, the observation \eqref{eq:pi-relation} plays a crucial role\footnote{Although we are writing $S_\kappa(p_2,p_3)$ to keep track of the factorization of the scattering coefficient this particular coefficient is equal to $e^{-2ip_2}$ under the momentum constraint, $p_2+p_3=\pi$.} and, together with the contributions of the contact terms the contact term Bethe ansatz, is able to capture a part of the dynamical nature of the spin chain model. However, the special kinematics highly constrain the part of the spectrum we can access with this solution. The form of the total energy eigenvalue, \eqref{eq:threeeigv} is even more restricted than the eigenvectors because the momentum constraint reduces the eigenvalue to be only a function of $p_1$,
\begin{align}
    E_3(p_1,p_2,\pi-p_2)=3\left(\sqrt{\kappa}-\frac{1}{\sqrt{\kappa}}\right)^2+4\sin^2\left(\frac{p_1}{2}\right)\;,
\end{align}
which is not very desirable.

Remarkably, the contact term Bethe ansatz under the momentum constraint can be generalized to any number of excitations. This allows us to compute these solutions of the eigenvalue problem for an even number of excitations and get an idea about the physical spectrum,
\begin{align}
    \ket{\Psi^c(p_1,\cdots,p_N)}_{12}=\sum_{l_1<\cdots<l_N}\sum_{\sigma\in\mathcal{S}_3}\left(A_\sigma+\sum_{i=1}^{N-1}B^{(i)}_\sigma\delta(l_{i+1},l_i+1)\right)e^{i\Vec{p}_\sigma\cdot\Vec{l}}\ket{l_1,\cdots,l_N}_{12}\;.\label{eq:gen-cont-ansatz}
\end{align}
Here we denote the contact term contribution between the magnon located at the position $l_i$ and $l_{i+1}$ as $B^{(i)}$. Therefore, we obtain $N-1$ many distinct sets of contact term coefficients, which are also labeled by the permutation indices. The number of momentum constraints should also be increased to $N-2$. We can choose to set
\begin{align}
    p_2+p_3=p_3+p_4=\cdots=p_{N-1}+p_N=\pi\label{eq:general-mom-const}\;.
\end{align}
Under these constraints, we obtain an eigenvector of the $N$-magnon eigenvalue problem with factorized scattering coefficients. Every time the factor of the scattering coefficient includes two momentum variables which are constrained by the kinematic limit \eqref{eq:general-mom-const}, we need to add the $\mathcal{S}(p,\kappa)$ or $\mathcal{S}(p,1/\kappa)$ factor depending on whether the scattering event initially would be described by a factor of $S_\kappa$ or $S_{1 / \kappa}$. This is apparent from the order of the magnons that enter into the scattering process.  Additionally, the observer momentum variable can be identified by tracing back the sequence of scattering events. The observer momentum should be the one resulting from the scattering event immediately preceding this one. Including the three-magnon state, $N$-magnon contact term Bethe ansatz is given in terms of these two types of scattering coefficients, \eqref{eq:scatdef} and \eqref{S32}.

\section{$G_{123}$ for the $S_{\kappa}$ solution}
\label{sec:G123}
   Here, we revisit the special solution presented in Section \ref{subsec:skappa}. Combining the solution of the position-dependent coefficients for the shortest separation of the magnons to the solution of position-dependent corrections which are contained in the second class of equations, \eqref{eq:secondclass} requires some effort. We follow the steps which are given in Section \ref{subsubsec:non-int} and \ref{subsubsec:two-magnons}. We give the pieces of the solution for the position-dependent corrections with identity permutation. Recall the form of the generating function that makes the non-interacting equation for the identity permutation, given in \eqref{eq:gamma123} vanish,
   \begin{align}
   G_{123}(e^{-ip_{1}}x,e^{ip_{3}}y)&=g_{1,123}(x)+g_{3,123}(x)-\frac{\left(\invPP x^2 y+\PP x y^2\right)\left(\kappa-1/\kappa\right)}{Q(x,y)}  \notag \\
       &+\frac{\PP y^2\left(1 +\invPP x \right) }{Q(x,y)} g_{1,123}(x)+\frac{\invPP x^2\left( 1 +\PP y\right) }{Q(x,y)} g_{3,123}(y)\nonumber \\
       &-\frac{xy\left(1+\invPP x\right)}{Q(x,y)} g_{2,123}(x)-\frac{xy\left(1+\PP y\right)}{Q(x,y)} g_{4,123}(y)\; , %\label{eq:generating-non-int}
   \end{align}
   where the common denominator $Q(x,y)$ is a third-order polynomial of two variables without any constant term, introduced in \eqref{eq:ellipticCurve}. We only replaced the position-dependent corrections, $D_{123}^{1,0}$ and $D_{123}^{0,1}$ with their solution coming from \eqref{eq:Dk10011} and \eqref{eq:morechoices}. Furthermore, we solve the right two-magnon interacting equations, given in \eqref{eq:two-body-rec21}, but we set the one-variable generating functions $g_{2,123}(x)$, $g_{2,132}(x)$ and $g_{2,231}(x)$ to zero as they would be undetermined even after all the steps of second class of equations. We obtain,
 \begin{align}
     g_{1,123}(x)&=\frac{\kappa  e^{-i p_3} x (-\tilde C_{12}(x)+S_\kappa(p_2,p_3) \tilde C_{13}(x)+S_\kappa(p_1,p_3) \tilde C_{23}(x))}{\left(x (2-\EE \kappa )+\kappa +\kappa  x^2\right) [S_\kappa(p_1,p_2)S_\kappa(p_2,p_3)+S_\kappa(p_1,p_3)]} \; .\label{eq:g1123} 
 \end{align}
   Note that the resulting form of the functions $C_{ij}$ and $\tilde C_{ij}$ is given in \eqref{eq:skappaC12} and \eqref{eq:skappaCC12} and the rest can be obtained by permuting the indices of $\omega(p_1,p_2)$ which is defined in \eqref{eq:omega}, and adding the necessary scattering coefficients with respect to the structure of the solution. Finally, we solve the left two-magnon equations, given in \eqref{eq:two-body-rec12} together with the analyticity conditions of the form \eqref{eq:sol-analyticity}. After substituting the vanishing $g_{2,123}(x)$, we are left with
  \begin{align}
      g_{3,123}(y)=&\frac{y\; C_{12}(y)}{\omega(p_1,p_2,\kappa) h_{12}(y,p)}+\frac{y \left(\PP y+\invPP\rho_+(y,p)\right)\left(\kappa-1/\kappa\right)}{ h_{12}(y,p)}\nonumber\\&-\frac{ y^2 g_{1,123}(\rho_+(y,p)) (1+\PP/\rho_+(y,p))}{ h_{12}(y,p)}\;,
  \end{align}
 and 
 \begin{align}
     g_{4,123}(y)=&\frac{\invPP\rho_+(y,p) C_{12}(y)}{\omega(p_1,p_2,\kappa)h_{12}(y,p)}-\frac{\left(1-\EE y+y^2+2 \kappa  y\right) \left(\PP y+\invPP\rho_+(y,p)\right)\left(\kappa-1/\kappa\right)}{\left(1+\PP y\right)h_{12}(y,p)}\nonumber\\&+\frac{\left(1-\EE y+y^2+2 \kappa  y\right)\left((1+\PP/\rho_+(y,p)\right)y g_{1,123}(\rho_+(y,p)) }{\left(1+\PP y\right)h_{12}(y,p)}\;,
 \end{align}
 such that $h_{12}(y,p)$ is given in equation \eqref{eq:h12}. The solution of the one-variable generating functions completes the solution of the position-dependent corrections for the identity permutation.  Then we observe that the remaining parts of the solution are directly related to the solution of the position-dependent corrections, $D_{123}^{n,m}$ in a rather simple way. We denote the generating function which includes the solution for all first-class equations and the one-variable generating functions which are given above for the identity permutation as $G(p_1,p_2,p_3;\kappa)$ and we suppress the complex variables $x$ and $y$. In this case, rest of the generating functions can be given just by using $G(p_1,p_2,p_3;\kappa)$,
\begin{align}
    G_{123}(e^{-ip_1}x,e^{ip_3}y)&=G(p_1,p_2,p_3;\kappa) \;,\\
    G_{132}(e^{-ip_1}x,e^{ip_2}y)&=S_\kappa(p_2,p_3)G(p_1,p_3,p_2;\kappa) \;,\\
    G_{213}(e^{-ip_2}x,e^{ip_3}y)&=S_\kappa(p_1,p_2)G(p_2,p_1,p_3;\kappa) \;,\\
    G_{231}(e^{-ip_2}x,e^{ip_1}y)&=S_\kappa(p_1,p_3)S_\kappa(p_1,p_2)G(p_2,p_3,p_1;\kappa) \;,\\
    G_{312}(e^{-ip_3}x,e^{ip_2}y)&=S_\kappa(p_1,p_3)S(p_2,p_3)G(p_3,p_1,p_2;\kappa) \;,\\
    G_{321}(e^{-ip_3}x,e^{ip_1}y)&=S_{\kappa}(p_2,p_3)S_\kappa(p_1,p_3)S_\kappa(p_1,p_2)G(p_3,p_2,p_1;\kappa) \;.
\end{align}
Let us now define the function
\begin{align}
    f(p_1,p_2,p_3;n,m,\kappa)=-\oint_\Sigma \frac{dx dy}{4\pi^2}\frac{G_{123}(x,y)}{x^{n+1}y^{m+1}}\;,
\end{align}
where $\Sigma$ is a small ball around the origin. Then, for any $n,m\geq1$, we obtain the description of the position-dependent corrections as it is given in equations \eqref{eqn:SpecialDs}-\eqref{eqn:SpecialDs1}.

\newpage
  %%%%%Bibliography%%%%%
	 \bibliographystyle{JHEP}
	 \bibliography{1db}

\providecommand{\href}[2]{#2}\begingroup\raggedright\begin{thebibliography}{10}

\bibitem{Beisert:2010jr}
N.~Beisert et~al., \emph{{Review of AdS/CFT Integrability: An Overview}}, \href{https://doi.org/10.1007/s11005-011-0529-2}{\emph{Lett. Math. Phys.} {\bfseries 99} (2012) 3} [\href{https://arxiv.org/abs/1012.3982}{{\ttfamily 1012.3982}}].

\bibitem{Minahan:2002ve}
J.A.~Minahan and K.~Zarembo, \emph{{The Bethe ansatz for $\mathcal{N}=4$ superYang-Mills}}, \href{https://doi.org/10.1088/1126-6708/2003/03/013}{\emph{JHEP} {\bfseries 03} (2003) 013} [\href{https://arxiv.org/abs/hep-th/0212208}{{\ttfamily hep-th/0212208}}].

\bibitem{Korchemsky:2010kj}
G.P.~Korchemsky, \emph{{Review of AdS/CFT Integrability, Chapter IV.4: Integrability in QCD and $N\ensuremath{<}4$ SYM}}, \href{https://doi.org/10.1007/s11005-011-0516-7}{\emph{Lett. Math. Phys.} {\bfseries 99} (2012) 425} [\href{https://arxiv.org/abs/1012.4000}{{\ttfamily 1012.4000}}].

\bibitem{Zoubos:2010kh}
K.~Zoubos, \emph{{Review of AdS/CFT Integrability, Chapter IV.2: Deformations, Orbifolds and Open Boundaries}}, \href{https://doi.org/10.1007/s11005-011-0515-8}{\emph{Lett. Math. Phys.} {\bfseries 99} (2012) 375} [\href{https://arxiv.org/abs/1012.3998}{{\ttfamily 1012.3998}}].

\bibitem{Pomoni:2019oib}
E.~Pomoni, \emph{{4D $\mathcal{N}=2$ SCFTs and spin chains}}, \href{https://doi.org/10.1088/1751-8121/ab7f66}{\emph{J. Phys. A} {\bfseries 53} (2020) 283005} [\href{https://arxiv.org/abs/1912.00870}{{\ttfamily 1912.00870}}].

\bibitem{Gadde:2010zi}
A.~Gadde, E.~Pomoni and L.~Rastelli, \emph{{Spin Chains in $\mathcal{N}$=2 Superconformal Theories: From the $\mathbb{Z}_{2}$ Quiver to Superconformal QCD}}, \href{https://doi.org/10.1007/JHEP06(2012)107}{\emph{JHEP} {\bfseries 06} (2012) 107} [\href{https://arxiv.org/abs/1006.0015}{{\ttfamily 1006.0015}}].

\bibitem{Beisert:2005he}
N.~Beisert and R.~Roiban, \emph{{The Bethe ansatz for $Z_S$ orbifolds of $\mathcal{N}=4$ super Yang-Mills theory}}, \href{https://doi.org/10.1088/1126-6708/2005/11/037}{\emph{JHEP} {\bfseries 11} (2005) 037} [\href{https://arxiv.org/abs/hep-th/0510209}{{\ttfamily hep-th/0510209}}].

\bibitem{Beisert:2005if}
N.~Beisert and R.~Roiban, \emph{{Beauty and the twist: The Bethe ansatz for twisted $\mathcal{N}=4$ SYM}}, \href{https://doi.org/10.1088/1126-6708/2005/08/039}{\emph{JHEP} {\bfseries 08} (2005) 039} [\href{https://arxiv.org/abs/hep-th/0505187}{{\ttfamily hep-th/0505187}}].

\bibitem{Gurdogan:2015csr}
O.~G\"urdo\u{g}an and V.~Kazakov, \emph{{New Integrable 4D Quantum Field Theories from Strongly Deformed Planar $\mathcal{N}=4$ Supersymmetric Yang-Mills Theory}}, \href{https://doi.org/10.1103/PhysRevLett.117.201602}{\emph{Phys. Rev. Lett.} {\bfseries 117} (2016) 201602} [\href{https://arxiv.org/abs/1512.06704}{{\ttfamily 1512.06704}}].

\bibitem{Pomoni:2013poa}
E.~Pomoni, \emph{{Integrability in $\mathcal{N}=2$ superconformal gauge theories}}, \href{https://doi.org/10.1016/j.nuclphysb.2015.01.006}{\emph{Nucl. Phys. B} {\bfseries 893} (2015) 21} [\href{https://arxiv.org/abs/1310.5709}{{\ttfamily 1310.5709}}].

\bibitem{Pomoni:2021pbj}
E.~Pomoni, R.~Rabe and K.~Zoubos, \emph{{Dynamical spin chains in 4D $ \mathcal{N} $ = 2 SCFTs}}, \href{https://doi.org/10.1007/JHEP08(2021)127}{\emph{JHEP} {\bfseries 08} (2021) 127} [\href{https://arxiv.org/abs/2106.08449}{{\ttfamily 2106.08449}}].

\bibitem{Hanno}
H.~Bertle, E.~Pomoni, X.~Zhang and K.~Zoubos, \emph{{Hidden Symmetries in Four-dimensional Gauge Theories}}, .

\bibitem{Baxter:1982zz}
R.J.~Baxter, \emph{{Exactly solved models in statistical mechanics}} (1982), \href{https://doi.org/$10.1142/9789814415255\_0002$}{$10.1142/9789814415255\_0002$}.

\bibitem{Belavin:1981ix}
A.A.~Belavin, \emph{{Dynamical Symmetry of Integrable Quantum Systems}}, \href{https://doi.org/10.1016/0550-3213(81)90414-4}{\emph{Nucl. Phys. B} {\bfseries 180} (1981) 189}.

\bibitem{Felder:1994be}
G.~Felder, \emph{{Elliptic quantum groups}},  in \emph{{11th International Conference on Mathematical Physics (ICMP-11) (Satellite colloquia: New Problems in the General Theory of Fields and Particles, Paris, France, 25-28 Jul 1994)}}, pp.~211--218, 7, 1994 [\href{https://arxiv.org/abs/hep-th/9412207}{{\ttfamily hep-th/9412207}}].

\bibitem{Felder:1996xym}
G.~Felder and A.~Varchenko, \emph{{Algebraic Bethe ansatz for the elliptic quantum group $E_{\ensuremath{\tau},\ensuremath{\eta}}( sl(2))$}}, \href{https://doi.org/10.1016/S0550-3213(96)00461-0}{\emph{Nucl. Phys. B} {\bfseries 480} (1996) 485} [\href{https://arxiv.org/abs/q-alg/9605024}{{\ttfamily q-alg/9605024}}].

\bibitem{konnobook}
H.~Konno, \emph{Elliptic Quantum Groups}, Springer Singapore (2020), \href{https://doi.org/https://doi.org/10.1007/978-981-15-7387-3}{https://doi.org/10.1007/978-981-15-7387-3}.

\bibitem{Sklyanin:1982tf}
E.K.~Sklyanin, \emph{{Some algebraic structures connected with the Yang-Baxter equation}}, \href{https://doi.org/10.1007/BF01077848}{\emph{Funct. Anal. Appl.} {\bfseries 16} (1982) 263}.

\bibitem{Douglas:1996sw}
M.R.~Douglas and G.W.~Moore, \emph{{D-branes, quivers, and ALE instantons}},  \href{https://arxiv.org/abs/hep-th/9603167}{{\ttfamily hep-th/9603167}}.

\bibitem{Gadde:2009dj}
A.~Gadde, E.~Pomoni and L.~Rastelli, \emph{{The Veneziano Limit of $\mathcal{N}= 2$ Superconformal QCD: Towards the String Dual of $\mathcal{N}= 2$ $SU(N_c)$ SYM with $N_f=2N_c$}},  \href{https://arxiv.org/abs/0912.4918}{{\ttfamily 0912.4918}}.

\bibitem{Liendo:2011xb}
P.~Liendo, E.~Pomoni and L.~Rastelli, \emph{{The Complete One-Loop Dilation Operator of $\mathcal{N}=2$ SuperConformal QCD}}, \href{https://doi.org/10.1007/JHEP07(2012)003}{\emph{JHEP} {\bfseries 07} (2012) 003} [\href{https://arxiv.org/abs/1105.3972}{{\ttfamily 1105.3972}}].

\bibitem{fourmagnon}
D.~Bozkurt, Z.~Kong, J.M.~Nieto~Garcia and E.~Pomoni.

\bibitem{Bibikov2016}
P.N.~Bibikov, \emph{Three magnons in an isotropic $s=1$ ferromagnetic chain as an exactly solvable non-integrable system}, \href{https://doi.org/10.1088/1742-5468/2016/03/033109}{\emph{Journal of Statistical Mechanics: Theory and Experiment} {\bfseries 2016} (2016) 033109}.

\bibitem{YangPhysRev.168.1920}
C.N.~Yang, \emph{$s$ matrix for the one-dimensional $n$-body problem with repulsive or attractive $\ensuremath{\delta}$-function interaction}, \href{https://doi.org/10.1103/PhysRev.168.1920}{\emph{Phys. Rev.} {\bfseries 168} (1968) 19201923}.

\bibitem{arutyunov2019elements}
G.~Arutyunov, \emph{Elements of Classical and Quantum Integrable Systems}, UNITEXT for Physics, Springer International Publishing (2019).

\bibitem{deLeeuw:2020ahe}
M.~de~Leeuw, C.~Paletta, A.~Pribytok, A.L.~Retore and P.~Ryan, \emph{{Classifying Nearest-Neighbor Interactions and Deformations of AdS}}, \href{https://doi.org/10.1103/PhysRevLett.125.031604}{\emph{Phys. Rev. Lett.} {\bfseries 125} (2020) 031604} [\href{https://arxiv.org/abs/2003.04332}{{\ttfamily 2003.04332}}].

\bibitem{deLeeuw:2020xrw}
M.~de~Leeuw, C.~Paletta, A.~Pribytok, A.L.~Retore and P.~Ryan, \emph{{Yang-Baxter and the Boost: splitting the difference}}, \href{https://doi.org/10.21468/SciPostPhys.11.3.069}{\emph{SciPost Phys.} {\bfseries 11} (2021) 069} [\href{https://arxiv.org/abs/2010.11231}{{\ttfamily 2010.11231}}].

\bibitem{DeLeeuw:2019gxe}
M.~de~Leeuw, A.~Pribytok and P.~Ryan, \emph{{Classifying two-dimensional integrable spin chains}}, \href{https://doi.org/10.1088/1751-8121/ab529f}{\emph{J. Phys. A} {\bfseries 52} (2019) 505201} [\href{https://arxiv.org/abs/1904.12005}{{\ttfamily 1904.12005}}].

\bibitem{Skrzypek:2023fkr}
T.~Skrzypek and A.A.~Tseytlin, \emph{{On AdS/CFT duality in the twisted sector of string theory on $AdS_{5}\times S^{5}/\ensuremath{\mathbb{Z}}_{2}$ orbifold background}}, \href{https://doi.org/10.1007/JHEP03(2024)045}{\emph{JHEP} {\bfseries 03} (2024) 045} [\href{https://arxiv.org/abs/2312.13850}{{\ttfamily 2312.13850}}].

\end{thebibliography}\endgroup
	 %%%%%%%%%%%%%%%%%%%%%%
\end{document}